# Some Novel Thought Experiments Involving Foundations of Quantum Mechanics and Quantum Information

Dissertation
submitted in partial fulfillment of the requirements for the degree of
Doctor of Philosophy in Physics

**Omid Akhavan**
*Department of Physics, Sharif University of Technology*

Tehran, July 2003

*To*

*my mother, who has been with me every day of my life*

*and to*

*my wife Eli, who gave me l♡ve*

In the memory of
our beloved friend and colleague the late Dr. Majid Abolhasani,
who gave me some useful comments on the foundations of quantum mechanics,
and was my initial encourager for working on quantum information theory.

## Acknowledgements

It is a great pleasure to thank the many people who have contributed to this dissertation. My deepest thanks go to Dr. Mehdi Golshani, my professor, for his moral and financial support through the years of my PhD, for his unfailing positive attitude which remounted my morale more than once, for his understanding and sympathy for my problems with my hands, and for being my guide through the maze of quantum world. He has been a valued teacher, and I hope my seven years at Sharif University have given me even a few of his qualities.

Special thanks go to Ali T. Rezakhani, my close friend, collaborator and colleague, for his considerable influence on this dissertation. Much of the view point mentioned here was worked out in valuable conservations with him.

Warm thanks to Dr. Alireza Z. Moshfegh for introducing me to experimental physics, for our common works which are not part of this thesis, and for his moral and financial supports through all years of my presence at Sharif University.

Thanks also go to Dr. Vahid Karimipour for his stimulating discussions on quantum information theory, for reading this thesis and for his illuminating comments.

I am thankful to Drs. Mohammad Akhavan, Mohammadreza Hedayati and Majid Rahnama for reading this dissertation and for their valuable comments.

I would like to thank all my teachers, colleagues and friends for many useful and instructive discussions on physics and life. I am grateful also to those who are not mentioned by name in the following. In particular let me thank Drs.: Hesam Arfaie, Farhad Ardalan, Reza Mansouri, Jalal Samimi, and Hamid Salamati as teachers, and Saman Moghimi, Masoud M. Shafiee, Ahmad Ghodsi, Ali Talebi, Parviz Kameli, Saeed Parvizi, Husein Sarbolouki, Mohammad Kazemi, Alireza Noiee, Nima Hamedani, Farhad Shahbazi, Mahdi Saadat, Hamid Molavian, Mohammad Mardani, Ali A. Shokri, Masoud Borhani, Javad Hashemifar, Rouhollah Azimirad, Afshin Shafiee, Ali Shojaie, Fatimah Shojaie, Mohammad M. Khakian, Abolfazl Ramezanpour, Sohrab Rahvar, Parvaneh Sangpour, Ali Tabeie, Hashem H. Vafa, Ahmad Mashaie, Akbar Jafari, Alireza Bahraminasab, Akbar Fahmi, Mohammad R. Mohammadizadeh, Sima Ghasemi, Omid Saremi, Davoud Pourmohammad, Fredric Faure and Ahmad Mohammadi as colleagues and friends.

I would also like to thank my teachers at Physics Department of Uroumieh University who encouraged me to continue physics. Particularly, I thank Drs.: Rasoul Sedghi, Mohammadreza Behforouz, Rasoul Khodabakhsh, Mostafa Poshtkouhi, Mir Maqsoud Golzan, Jalal Pesteh, Shahriar Afshar, and Mohammad Talebian.

There are also many people to whom I feel grateful and whom I would like to thank at this occasion. Each of the following have in one way or another affected this dissertation, even if only by prompting an explanation or turn of phrase. I thank Drs.: Partha Ghose, Louis Marchildon, Ward Struyve, Willy De Baere, Marco Genovese, Adan Cabello, Hrvoje Nikolic, Jean-Francois Van Huele, Edward R. Floyd, Farhan Saif, Manzoor Ikram, Seth Lloyd, Vladimir E.

Kravtsov, Antonio Falci, Ehud Shapiro, Vlatko Vedral, Denis Feinberg, Massimo Palma, Irinel Chiorescu, Jonathan Friedman, Ignacio Cirac and Paolo Zanardi.

I would like to thank Institute for Studies in Theoretical Physics and Mathematics (IPM) for financial support of this thesis.

I also appreciate hospitality of the the abdus salam international centre for theoretical physics (ICTP, Italy) where some part of this work was completed.

Thanks also go to the following people for a lot of beer: Parisa Yaqoubi, Edris Bagheri, Khosro Orami, Vaseghinia, Yahyavi, Beheshti and Nicoletta Ivanissevich.

**Omid Akhavan**

*Sharif University of Technology*
*July 2003*

# Some Novel Thought Experiments Involving Foundations of Quantum Mechanics and Quantum Information

by

Omid Akhavan

B. Sc., Physics, Uroumieh University, Uroumieh, 1996

M. Sc., Physics, Sharif University of Technology, Tehran, 1998

PhD, Physics, Sharif University of Technology, Tehran, 2003

## *Abstract*

In this thesis, we have proposed some novel thought experiments involving foundations of quantum mechanics and quantum information theory, using quantum entanglement property. Concerning foundations of quantum mechanics, we have suggested some typical systems including two correlated particles which can distinguish between the two famous theories of quantum mechanics, i.e. the standard and Bohmian quantum mechanics, at the individual level of pair of particles. Meantime, the two theories present the same predictions at the ensemble level of particles. Regarding quantum information theory, two theoretical quantum communication schemes including quantum dense coding and quantum teleportation schemes have been proposed by using entangled spatial states of two EPR particles shared between two parties. It is shown that the rate of classical information gain in our dense coding scheme is greater than some previously proposed multi-qubit protocols by a logarithmic factor dependent on the dimension of Hilbert space. The proposed teleportation scheme can provide a complete wave function teleportation of an object having other degrees of freedom in our three-dimensional space, for the first time. All required unitary operators which are necessary in our state preparation and Bell state measurement processes are designed using symmetric normalized Hadamard matrix, some basic gates and one typical conditional gate, which are introduced here for the first time.

CONTENTS

## PREFACE

The present dissertation consists of two parts which are mainly based on the following papers and manuscripts:

- *Bohmian prediction about a two double-slit experiment and its disagreement with standard quantum mechanics,* M. Golshani and O. Akhavan, *J. Phys. A* **34**, 5259 (2001); quant-ph/0103101.
- *Reply to: Comment on "Bohmian prediction about a two double-slit experiment and its disagreement with SQM"* O. Akhavan and M. Golshani, quant-ph/0305020.
- *A two-slit experiment which distinguishes between standard and Bohmian quantum mechanics,* M. Golshani and O. Akhavan, quant-ph/0009040.
- *Experiment can decide between standard and Bohmian quantum mechanics,* M. Golshani and O. Akhavan, quant-ph/0103100.
- *On the experimental incompatibility between standard and Bohmian quantum mechanics*, M. Golshani and O. Akhavan, quant-ph/0110123.
- *Quantum dense coding by spatial state entanglement,* O. Akhavan, A.T. Rezakhani, and M. Golshani, *Phys. Lett. A* **313**, 261 (2003); quant-ph/0305118.
- *Comment on "Dense coding in entangled states"*, O. Akhavan and A.T. Rezakhani, *Phys. Rev. A* **68**, 016302 (2003); quant-ph/0306148.
- *A scheme for spatial wave function teleportation in three dimensions*, O. Akhavan, A.T. Rezakhani, and M. Golshani, *J. Quant. Inf. Comp.*, submitted.

The first part of this dissertation includes three chapters. In chapter 1, an introduction about the foundations of quantum mechanics, which is mainly concentrated on explanations of; some problems in the standard quantum mechanics, the quantum theory of motion, some new insights presented by Bohmian quantum mechanics and noting some objections that have been advanced against this theory, has been presented. In chapter 2, by using position entanglement property of two particles in a symmetrical two-plane of double-slit system, we have shown that the standard and Bohmian quantum mechanics can predict different results at an individual level of entangled pairs. However, as expected, the two theories predict the same interference pattern at an ensemble level of the particles. In chapter 3, the predictions of the standard and Bohmain quantum mechanics have been compared using a double-slit system including two correlated particles. It has been shown that using a selective joint detection of the two particles at special conditions, the two theories can be distinguished at a statistical level of the selected particles. But, by considering all particles, the predictions of the two theories are still identical at the ensemble level of particles.

In the second part of the dissertation there are also three chapters. In its first chapter, i.e. chapter 4, an introduction about quantum information theory including quantum dense coding and teleportation has been presented. In chapter 5, using a two-particle source similar to that is applied in chapter 1, a more efficient quantum dense coding scheme has been proposed. In this regard, the suitable encoding and decoding unitary operators along with its corresponding Bell states have been studied. The rate of classical information gain of this scheme has been obtained and then compared with some other well-known protocols. Furthermore, possibility of designing of the required position operators using some basic gates and one conditional position gates has been investigated. Next, in chapter 6, using a system the same as the dense coding scheme, wave function teleportation of a three dimensional object having some other degrees of freedom has been studied. Concerning this, the required operators for performing Bell state measurement and reconstruction process have been designed using some position and momentum gates.

In appendix, which consists three sections, some more details on our considered EPR source, preparing and measuring processes utilized in some initial cases of the dense coding and teleportation schemes, and comparison of our dense coding scheme with some other ones can be found.

# Part I

# NEW SUGGESTED EXPERIMENTS RELATED TO THE FOUNDATIONS OF QUANTUM MECHANICS

# 1. INTRODUCTION-FOUNDATIONS OF QUANTUM MECHANICS

## 1.1 Standard quantum mechanics

The standard view of quantum mechanics (SQM), accepted almost universally by physicists, is commonly termed the Copenhagen interpretation. This interpretation requires complementarity, e.g. wave-particle duality, inherent indeterminism at the most fundamental level of quantum phenomena, and the impossibility of an event-by-event causal representation in a continuous space-time background [1]. In this regard, some problems embodied in this interpretation are concisely described in the following.

### 1.1.1 Some of the major problems of SQM

*Measurement*

As an example, consider a two-state microsystem whose eigenfunctions are labelled by $\psi_+$ and $\psi_-$. Furthermore, there is a macrosystem apparatus with eigenfunctions $\phi_+$ and $\phi_-$ corresponding to an output for the microsystem having been in the $\psi_+$ and $\psi_-$ states, respectively. Since prior to a measurement we do not know the state of the microsystem, it is a superposition state given by

$$\psi_0 = \alpha\psi_+ + \beta\psi_-, \qquad |\alpha|^2 + |\beta|^2 = 1. \tag{1.1}$$

Now, according to the linearity of Scrödinger's equation, the final state obtained after the interaction of the two systems is

$$\Psi_0 = (\alpha\psi_+ + \beta\psi_-)\phi_0 \longrightarrow \Psi_{out} = \alpha\psi_+\phi_+ + \beta\psi_-\phi_- \tag{1.2}$$

where it is assumed that initially the two systems are far apart and do not interact. It is obvious that, the state on the far right side of the last equation does not correspond to a definite state for a macrosystem apparatus. In fact, this result would say that the macroscopic apparatus is itself in a superposition of both plus and minus states. Nobody has observed such macroscopic superpositions. This is the so-called measurement problem, since the theory predicts results that are in clear conflict with all observations. It is at this point that the standard program to resolve this problem invokes the reduction of wave packet upon observation, that is,

$$\alpha\psi_+\phi_+ + \beta\psi_-\phi_- \longrightarrow \begin{cases} \psi_+\phi_+, & P_+ = |\alpha|^2; \\ \psi_-\phi_-, & P_- = |\beta|^2. \end{cases} \tag{1.3}$$

Various attempts to find reasonable explanation for this reduction are at the heart of the measurement problem.

*Schrödinger's cat*

Concerning the measurement problem, there is a paradox introduced by Schrödinger in 1935. He suggested the coupling of an uranium nucleus or atom as a microsystem and a live cat in a box as a macrosystem. The system is so arranged that, if the nucleus with a life time $\tau_0$ decays, it triggers a device that kills the cat. Now the point is to consider a quantum description of the time evolution of the system. If $\Psi(t)$, $\phi$ and $\psi$ represent the wave functions of the system, cat and atom, respectively, then the initial state of the system would be

$$\Psi(0) = \psi_{atom}\phi_{live}. \tag{1.4}$$

This initial state evolves into

$$\Psi(t) = \alpha(t)\psi_{atom}\phi_{live} + \beta(t)\psi_{decay}\phi_{dead} \tag{1.5}$$

and the probabilities of interest are

$$P_{live}(t) = |\langle\psi_{atom}\phi_{live}|\Psi(t)\rangle|^2 \sim e^{-t/\tau_0} \tag{1.6}$$

$$P_{dead}(t) = |\langle\psi_{decay}\phi_{dead}|\Psi(t)\rangle|^2 \sim 1 - e^{-t/\tau_0}. \tag{1.7}$$

As time goes on, chance looks less for the cat's survival. Before one observes the system, $\Psi(t)$ represents a superposition of a live and a dead cat. However, after observation the wave function is reduced to live or dead one. Now, the main question which was posed by Schrödinger is: what does $\Psi(t)$ represent? Possible answers are that, it represents (1) our state of knowledge, and so quantum mechanics is incomplete, and (2) the actual state of the system which beers a sudden change upon observation. If we choose (1) (which is what Schrödinger felt intuitively true), then quantum mechanics is incomplete, i.e., there are physically meaningful questions about the system that it cannot answer-surly the cat was either alive or dead before observation. On the other hand, Choice (2) faces us with the measurement problem, in which the actual collapse of the wave function must be explained.

*The classical limit*

It is well established that when a theory supersedes an earlier one, whose domain of validity has been determined, it must reduce to the old one in a proper limit. For example, in the special theory of relativity there is a parameter $\beta = v/c$ such that when $\beta \ll 1$, the equations of special relativity reduce to those of classical mechanics. In general relativity, also, the limit of weak gravitational fields or small space-time curvature leads to Newtonian gravitational theory. If quantum mechanics is to be a candidate for a fundamental physical theory that replaces classical mechanics, then we would expect that there is a suitable limit in which the equations of quantum mechanics approach those of classical mechanics. It is often claimed that the desired limit is $\hbar \longrightarrow 0$. But $\hbar$ is not a dimensionless constant and it is not possible for us to set it equal to zero. A more formal attempt at a classical limit is Ehrenfest's theorem, according to which expectation values satisfy Newton's second law as

$$\langle \mathbf{F} \rangle = m \frac{d^2 \langle \mathbf{r} \rangle}{dt^2}. \tag{1.8}$$

This really only implies that the centroid of the packet follows the classical trajectory. However, wave packets spread and the above equation is just not the same as $\mathbf{F} = m\mathbf{a}$. A similar formal attempt is the WKB (Wentzel-Kramers-Brillouin) approximation which is often advertised as a classical limit of the Schrödinger's equation. Again, there is not a well-defined limit (in terms of a dimensionless parameter) for which one can obtain exactly the equations of classical mechanics for all future of times. Therefore, if it is not possible to find a classical description for macroscopic objects in a suitable limit, then we do not have a complete theory that is applicable to both the micro and macro domains.

*Concept of the wave function*

The quantum theory that developed in the 1920s is related to its classical predecessor by the mathematical procedure of quantization, in which classical dynamics variables are replaced by operators. Hence, a new entity appeares on which the operators act, i.e., the wave function. For a single-body system this is a complex function, $\psi(\mathbf{x}, t)$, and for a field it is a complex functional, $\psi[\phi, t]$. In fact, the wave function introduces a new notion of the state of a physical system. But, in prosecuting their quantization procedure, the founding fathers introduced the new notion of the state not in addition to the classical variables but instead of them. They could not see, and finally did not want to see, even when presented with a consistent example, how to retain in some form the assumption that matter has substance and form, independently of whether or not it is observed. The wave function alone was adapted as the most complete characterizing the state of a system. Since there was no deterministic way to describe individual processes using just the wave function, it seemed natural to claim that these are indeterminate and unanalyzable in principle. Furthermore, quantum mechanics appears essentially as a set of working rules for computing the likely outcomes of certain as yet undefined processes called measurement. So, one might well ask what happened to the original program embodied in the old quantum theory of explaining the stability of atoms as objective structures in space-time. In fact, quantum mechanics leaves the primitive notion of system undefined; it contains no statement regarding the objective constitution of matter corresponding to the conception of particles and fields employed in classical physics. There are no electrons or atoms in the sense of distinct localized entities beyond the act of observation. These are simply names attributed to the mathematical symbols $\psi$ to distinguish one functional form from another one. So the original quest to comprehend atomic structure culminated in just a set of rules governing laboratory practice. To summarize, according to the completeness assumption of SQM, the wave function is associated with an individual physical system. It provides the most complete description of the system that is, in principle, possible. The nature of the description is statistical, and concerns the probabilities of the outcomes of all conceivable measurements that may be performed on the system. Therefore, in this view, quantum mechanics does not present a causal and deterministic theory for the universe.

## 1.2 The quantum theory of motion

We have seen that, the quantum world is inexplicable in classical terms. The predictions concerning the interaction of matter and light, embodied in Newtonian mechanics and Maxwell's equations, are inconsistent with the experimental facts at the microscopic level. An important feature of quantum effects is their apparent indeterminism, that individual atomic events are unpredictable, uncontrollable, and literally seem to have no cause. Regularities emerge only when one considers a large ensemble of such events. This indeed is generally considered to constitute the heart of the conceptual problem posed by quantum phenomena. A way of resolving this problem is that the wave function does not correspond to a single physical system but rather to an ensemble of systems. In this view, the wave function is admitted to be an incomplete representation of actual physical states and plays a role roughly analogous to the distribution function in classical statistical mechanics. Now, to understand experimental results as the outcome of a causally connected series of individual processes, one can seek further significance of the wave function (beyond its probabilistic aspect), and can introduce other concepts (hidden variables) in addition to the wave function. It was in this spirit that Bohm [2] in 1952 proposed his theory and showed how underlying quantum mechanics is a causal theory of the motion of waves and particles which is consistent with a probabilistic outlook, but does not require it. In fact, the additional element that he introduced apart from the wave function is just a particle, conceived in the classical sense of pursuing a definite continuous track in space-time. The basic postulates of Bohm's quantum mechanics (BQM) can be summarized as follows:
*1.* An individual physical system comprises a wave propagating in space-time together with a particle which moves continuously under the guidance of the wave.
*2.* The wave is mathematically described by $\psi(\mathbf{x},t)$ which is a solution to the Scrödinger's equation:

$$i\hbar\frac{\partial\psi}{\partial t}=(-\frac{\hbar^2}{2m}\nabla^2+V)\psi \tag{1.9}$$

*3.* The particle motion is determined by the solution $\mathbf{x}(t)$ to the guidance condition

$$\dot{\mathbf{x}} = \frac{1}{m}\nabla S(\mathbf{x},t)|_{\mathbf{x}=\mathbf{x}(t)} \tag{1.10}$$

where $S$ is the phase of $\psi$.
These three postulates on their own constitute a consistent theory of motion. Since BQM involves physical assumptions that are not usually made in quantum mechanics, it is preferred to consider it as a new theory of motion which is appropriately called the quantum theory of motion [3]. In order to ensure the compatibility of the motions of the ensemble of particles with the results of quantum mechanics, Bohm added the following further postulate:
*4.* The probability that a particle in the ensemble lies between the points $\mathbf{x}$ and $\mathbf{x} + d\mathbf{x}$ at time $t$ is given by

$$R^2(\mathbf{x},t)d^3x \tag{1.11}$$

where $R^2 = |\psi|^2$. This shows that the concept of probability in BQM only enters as a subsidiary condition on a causal theory of the motion of individuals, and the statistical meaning of the wave function is of secondary importance. Failure to recognize this has been the source of much confusion in understanding the causal interpretation.

Now, here, it is proper to compare and contrast Bohm's quantum theory with the standard one. It can be seen that, some of the most perplexing interpretational problems of SQM are simply solved in BQM.

### *1.2.1 Some new insights by BQM*

#### *No measurement problem*

One of the most elegant aspects of BQM is its treatment on the measurement problem, where it becomes a non-problem. In BQM, measurement is a dynamical and essentially many-body process. There is no collapse of the wave function, and so no measurement problem. The basic idea is that a particle always has a definite position before measurement. So there is no superposition of properties, and measurement or observation is just an attempt to discover this position.

To clarify the subject, consider, as an example, an inhomogeneous magnetic field which produces a spatial separation among the various angular momentum components of an incident beam of atoms. The incident wave packet $g(x)$ moves with a velocity $v_0$ along the $y$-axis. This function $g(x)$ (e.g., a Gaussian) is fairly sharply peaked about $x = 0$. The initial quantum state of the atom is a superposition of angular momentum eigenstates $\psi_n(\xi)$ of the atom. Thus, the initial wave function for the system before the atom has entered the region of the magnetic field can be written as

$$\Psi_0(x,\xi,t) = g(x - v_0 t)\sum_n C_n\psi_n(\xi). \tag{1.12}$$

The interaction between the inhomogeneous magnetic field and the magnetic moment of the atom exerts a net force on the atom in the $z$-direction. Once the packet emerges from the field, the $n$ components of the packet diverge along separate paths. After that sufficient time has elapsed, the $n$ component packets no longer overlap and have essentially disjoint supports. Then the wave function has evolved into

$$\Psi(x, \xi, t) = \sum_n C_n g_n[x - x_n(t)] e^{i\varphi_n} \psi_n(\xi) \tag{1.13}$$

where $x_n(t)$ show the particle trajectories and the $\varphi_n$ are simply constant phases. The description given sofar is similar to an account of a measurement in SQM frame. So, the next step would be to apply the projection postulate once an atom has been observed. One would simply erase the other packets. In BQM, however, the situation is different. The probability of finding the atom at some particular position is

$$P(x, \xi, t) = \sum_n |C_n|^2 |g_n[x - x_n(t)]|^2 |\psi_n(\xi)|^2. \tag{1.14}$$

There are no interference or cross terms here, because the various $g_n$ no longer overlap. After the particle has been found in one packet, it cannot be in any of the others and has negligible probability of crossing to other ones (because $P$ effectively vanishes between the packets). Now, it is necessary that the microsystem interact, effectively irreversibly, with a macroscopic measuring device that has many degrees of freedom to make it practically impossible (i.e., overwhelmingly improbable) for these lost wave packets to interfere once again with the one actually containing the particle. Thus, the process of measurement is a two-step one in which (1) the quantum states of the microsystem are separated into nonoverlapping parts by an, in principle, reversible interaction and (2) a practically irreversible interaction with a macroscopic apparatus registers the final results.

### *The classical limit*

By using the guidance condition along with the Schrödinger's equation, the quantum dynamical equation for the motion of a particle with mass $m$ is given by

$$\frac{dp}{dt} = -\nabla(V + Q) \tag{1.15}$$

where $V$ is the usual classical potential energy and $Q$ is the so-called quantum potential that is given in terms of the wave function as

$$Q = -\frac{\hbar^2}{2m} \frac{\nabla^2 R}{R}. \tag{1.16}$$

This $Q$ has the classically unexpected feature that its value depends sensitively on the shape, but not necessarily strongly on the magnitude of $R$, so that $Q$ need not falloff with distance as $V$ does. Now, it is evident that there are no problems in obtaining the classical equations of motion from BQM, because the above dynamical equation has the form of Newton's second law. In fact, when $(Q/V) \ll 1$ and $(\nabla Q/\nabla V) \ll 1$ (dimensionless parameters) the quantum dynamical equation becomes just the classical equation of motion. So the suitable limit is $Q \longrightarrow 0$ (in the sense of $(Q/V) \longrightarrow 0$ and also $(\nabla Q/\nabla V) \longrightarrow 0$), rather than anything like $\hbar \longrightarrow 0$. It is interesting to know that there are solutions to the Schrödinger's equation with no classical limit (quantum system with no classical analogue). Thus, one cannot exclude a priori the possibility that there be a class of solutions to the classical equations of motion which do not correspond to the limit of some class of quantum solutions (classical systems with no quantum analogue). Therefore, it seems reasonable to conceive classical mechanics as a special case of quantum mechanics in the sense that the latter has new elements ($\hbar$ and $Q$) not anticipated in the former. However, the possibility that the classical theory admits more general types of ensemble which cannot be described using the limit of quantum ensembles, because the latter corresponds to a specific type of linear wave equation and satisfy special conditions such as being built from single-valued conserved pure states, suggests that the two statistical theories can be considered independent while having a common domain of application. This domain is characterized by $Q \longrightarrow 0$ in the quantum theory. Now, there is a well-defined conceptual and formal connection between the classical and quantum domains but, as a new result, they merely intersect rather than are being contained in the other.

*The uncertainty relations*

One of the basic features of quantum mechanics is the association of Hermitian operators with physical observables, and the consequent appearance of noncommutation relations between the operators. For example, whatever the interpretation, from SQM or BQM one can obtain the Heisenberg uncertainty relation

$$\triangle x_i \triangle p_j \geq (\hbar/2)\delta_{ij} \tag{1.17}$$

for operators $x$ and $p$ that satisfy $[x_i, p_j] = i\hbar\delta_{ij}$ [3]. As a result, a wave function cannot be simultaneously an eigenfunction of $x$ and $p$. Since measuring of an observable involves the transformation of the wave function into an eigenfunction of the associated operator, it appears that a system cannot simultaneously be in a state by which its position and momentum are precisely known. How may one reconcile the uncertainty relation with the assumption that a particle can be ascribed simultaneously well-defined position and momentum variables as properties that exist during all interactions, including measurements? To answer this, we note that our knowledge of the state of a system should not be confused with what the state actually is. Quantum mechanics is constructed so that we cannot observe position and momentum simultaneously, but this fact does not prevent us to think of a particle having a well-defined track in reality. Bohm's discussion shows how the act of measurement, through the influence of the quantum potential, can disturb the microsystem and thus produce an uncertainty in the outcome of a measurement [2]. In other words, we can interpret the uncertainty relations as an expression of the different types of motion accessible to a particle when its wave undergoes the particular types of interaction appropriate to the measurement. In fact, the formal derivation of the uncertainty relations goes through as before, but now we have some understanding of how the spreads come about physically. According to BQM, the particle has a position and momentum prior to, during, and after the measurement, whether this be of position, momentum or any other observable. But in a measurement, we usually cannot observe the real value that an observable had prior to the measurement. In fact, as Bohm mentioned [2], in the suggested new interpretation, the so-called observables are not properties belonging to the observed system alone, but instead potentialities whose precise development depends just as much on the observing apparatus as on the observed system.

*Concept of the wave function*

As we have seen, to find a connection between the two aspects of matter, i.e. particle and wave, one can rewrite the complex Schrödinger's equation as a coupled system of equations for the real fields $R$ and $S$ which are defined by $\psi = Re^{iS}$. Then, in summary, these fields can play the following several roles simultaneously:
*1.* They are associated with two physical fields propagating in space-time and define, together with the particle, an individual physical system.
*2.* They act as actual agents in the particle motion, via the quantum potential.
*3.* They enter into the definition of properties associated with a particle (momentum, energy and angular momentum). These are not arbitrarily specified but are a specific combination of these fields, and are closely related to the associated quantum mechanical operators.
*4.* They have other meanings which ensure the consistency of BQM with SQM, and moreover, their connection with the classical mechanics.
Generally in BQM, the wave function plays two conceptually different roles. It determines (1) the influence of the environment on the quantum system and (2) the probability density by $P = |\psi|^2$. Now, since the guidance condition along with the Schrödinger's equation uniquely specify the future and past continuous evolution of the particle and field system, BQM forms the basis of a causal interpretation of quantum mechanics.

*Wave function of the universe*

By quantizing the Hamiltonian constraint of general relativity in the standard way one obtains the Wheeler-De Witt's equation, which is the Schrödinger's equation of the gravitational field. In this regard, there is an attempt to apply quantum mechanics to the universe as a whole in the so-called quantum cosmology. This has been widely interpreted according to the many-worlds picture of quantum mechanics. But there is no need for this, because acoording to many physicists, quantum cosmology deals with a single system - our universe. We have seen that, BQM is eminently suited to a description of systems that are essentially unique, such as the universe. Therefore, quantum cosmology is independent of any subsidiary probability interpretation one may like to attach to the wave function.

*Quantum potential as the origin of mass?*

In BQM it can be shown that the equation of motion of a bosonic massless quantum field is given by

$$\partial^2\psi(\mathbf{x},t) = -\frac{\delta Q[\psi(\mathbf{x}),t]}{\delta\psi(\mathbf{x})}|_{\psi(\mathbf{x})=\psi(\mathbf{x},t)} \tag{1.18}$$

which generally implies noncovariant and nonlocal properties of the field [3]. In fact, these features characterize the extremes of quantum behavior and, in principle, there exist states for which the right hand side of the above equation of motion is a scalar and local function of the space-time coordinate. The fact that this term is finite means that although the wave will be essentially nonclassical but will obey the type of equation we might postulate for a classical field, in which the scalar wave equation is equated to some function of the field. Here, the interesting point is that using quantization of a massless field it is possible to give mass to the field in the sense that the quantum wave obeys the classical massive Klein-Gordon equation

$$(\partial^2 + m^2)\psi = 0 \tag{1.19}$$

as a special case for the equation of motion of a massless quantum field, where $m$ is a real constant [3]. Therefore, the quantum potential acts so that the massless quantum field behaves as if it were a classical field with mass.

### 1.2.2 Some current objections to BQM

There are some of the typical objections that have been advanced against BQM. So, here, these objections are summarized and some preliminary answers are given to them.

*Predicting nothing new*

It is completely right that BQM was constructed so that its predictions are exactly the same as SQM's ones at the ensemble level. But, BQM permits more detailed predictions to be made pertaining to the individual processes. Whether this may be subjected to an experimental test is an open question, which is studied here using some examples.

*Nonlocality is the price to be paid*

Nonlocality is an intrinsic and clear feature of BQM. This property does not contradict special theory of relativity and the statistical predictions of relativistic quantum mechanics. But sometimes it is considered to be in some way a defect, because local theories are considered to be preferable. Yet nonlocality seems to be a small price to pay if the alternative is to forego any account of objective processes at all (including local ones). Furthermore, Aspect's experiment [4] established that, quantum mechanics is really a nonlocal theory without superluminal signalling [5]. Therefore, it is not necessary to worry about this property.

*Existence of trajectories cannot be proved*

BQM reproduces the assertion of SQM that one cannot simultaneously perform a precise measurement on both position and momentum. But this cannot be adduced as an evidence against the tenability of the trajectory concept. Science would not exist if ideas were only admitted when evidence for them already exists. For example, one cannot after all empirically prove the completeness postulate. The argument in favor of trajectory lies elsewhere, in its capacity to make intelligible a large amount of empirical facts.

*An attempt to return to classical physics*

BQM has been often objected for reintroducing the classical paradigm. But, as we mentioned in relation to BQM's classical limit, BQM is a more complete theory than SQM and classical mechanics, and includes both of them nearly independent theories in different domains, and also represents the connection between them. Therefore, BQM which apply the quantum states to guide the particle is, in principle, an intelligible quantum theory, not a classical one.

*No mutual action between the guidance wave and the particle*

Among the many nonclassical properties exhibited by BQM, one is that the particle does not react dynamically on the wave that is guided by. But, while it may be reasonable to require reciprocity of actions in classical theory, this cannot be regarded as a logical requirement of all theories that employ the particle and field concepts, particularly the ones involving a nonclassical field.

*More complicated than quantum mechanics*

Mathematically, BQM requires a reformulation of the quantum formalism, but not an alternation. The present reason is that SQM is not the one most appropriate to the physical interpretation. But, mathematically, the desirable theory, particularly at the ensemble level, can be considered quantum mechanics, because the quantum potential is implicit in the Schrödinger's equation.

In part I of this dissertation, we have concentrated on the first objection and studied some thought experiments in which BQM can predict different results from SQM, at the individual level.

# 2. TWO DOUBLE-SLIT EXPERIMENT USING POSITION ENTANGLEMENT OF EPR PAIR

## *2.1 Introduction*

According to the standard quantum mechanics (SQM), the complete description of a system of particles is provided by its wave function. The empirical predictions of SQM follow from a mathematical formalism which makes no use of the assumption that matter consists of particles pursuing definite tracks in space-time. It follows that the results of the experiments designed to test the predictions of the theory, do not permit us to infer any statement regarding the particle–not even its independent existence.

In the Bohmian quantum mechanics (BQM), however, the additional element that is introduced apart from the wave function is the particle position, conceived in the classical sense as pursuing a definite continuous track in space-time [1-3]. The detailed predictions made by this causal interpretation explains how the results of quantum experiments come about, but it is claimed that they are not tested by them. In fact, when Bohm [2] presented his theory in 1952, experiments could be done with an almost continuous beam of particles, but not with individual particles. Thus, Bohm constructed his theory in such a fashion that it would be impossible to distinguish observable predictions of his theory from SQM. This can be seen from Bell's comment about empirical equivalence of the two theories when he said: "*It* [the de Broglie-Bohm version of non-relativistic quantum mechanics] *is experimentally equivalent to the usual version insofar as the latter is unambiguous*" [5]. So, could it be that a certain class of phenomena might correspond to a well-posed problem in one theory but to none in the other? Or might definite trajectories of Bohm's theory lead to a prediction of an observable where SQM would just have no definite prediction to make?

To draw discrepancy from experiments involving the particle track, we have to argue in such a way that the observable predictions of the modified theory are in some way functions of the trajectory assumption. The question raised here is whether BQM's laws of motion can be made relevant to experiment. At first, it seems that definition of time spent by a particle within a classically forbidden barrier provides a good evidence for the preference of BQM. But, there are difficult technical questions, both theoretically and experimentally, that are still unsolved about this tunnelling times [1]. Furthermore, a recent work indicates that it is not practically feasible to use tunnelling effect to distinguish between the two theories [6]. In another proposal, Englert *et al.* [7] and Scully [8] have claimed that in some cases Bohm's approach gives results that disagree with those obtained from SQM and, in consequence, with experiment. However, Dewdney *et al.* [9] and then Hiley *et al.* [10] showed that the specific objections raised by them cannot be sustained. Meanwhile, Hiley believes that no experiment can decide between the standard and Bohm's interpretation. On the other hand, Vigier [11], in his recent work, has given a brief list of new experiments which suggests that the U(1) invariant massless photon, with properties of light within the standard interpretation, are too restrictive and that the O(3) invariant massive photon causal de Broglie-Bohm interpretation of quantum mechanics, is now supported by experiments. In addition, Leggett [12] considered some thought experiments involving macrosystems which can predict different results for SQM and BQM. [1] In other work, Ghose *et al.* [13, 14] indicated that although BQM is equivalent to SQM when averages of dynamical variables are taken over a Gibbs ensemble of Bohmian trajectories, the equivalence breaks down for ensembles built over clearly separated short intervals of time in specially entangled two-bosonic particle systems. Moreover, Ghose [15] showed that BQM is incompatible with SQM unless the Bohmian system corresponding to an SQM system is ergodic. Some other recent work in this regard can be also found in [16, 17, 18, 19].

Here, using an original EPR source [20] placed between two double-slit plane, we have suggested a thought experiment which can distinguish between the standard and Bohmian quantum mechanics [21, 22]. Some details on the considered EPR source have been examined to clarify the realizability of this experiment. Finally, an experimental effort for the realization of this thought experiment has been indicated [23].

## 2.2 Description of the proposed experiment

To distinguish between SQM and BQM we consider the following scheme. A pair of identical non-relativistic particles with total momentum zero, labelled by 1 and 2, originate from a source S that is placed exactly in the middle of a two double-slit screens, as shown in Fig. 2.1. We assume that the intensity of the beam is so low that during any individual experiment we have only a single pair of particles passing through the slits. In addition, we assume that the detection screens $S_1$ and $S_2$ register only those pairs of particles that reach the two screens simultaneously. Thus, we are sure that the registration of single particles is eliminated from final interference pattern. The detection process at the screens $S_1$ and $S_2$ may be nontrivial, but they play no causal role in the basic phenomenon of the interference of particles waves [3]. In the two-dimensional system of coordinates $(x, y)$ whose origin $S$ is shown, the center of slits lie at the points $(\pm d, \pm Y)$. Suppose that before the arrival of the two particles on the

[1] Leggett [12] assumes that the experimental predictions of SQM will continue to be realized under the extreme conditions specified, although a test of this hypothesis is part of the aim of the macroscopic quantum cohrence program. In addition, he considered BQM as another interpretation of the same theory rather than an alternative theory.

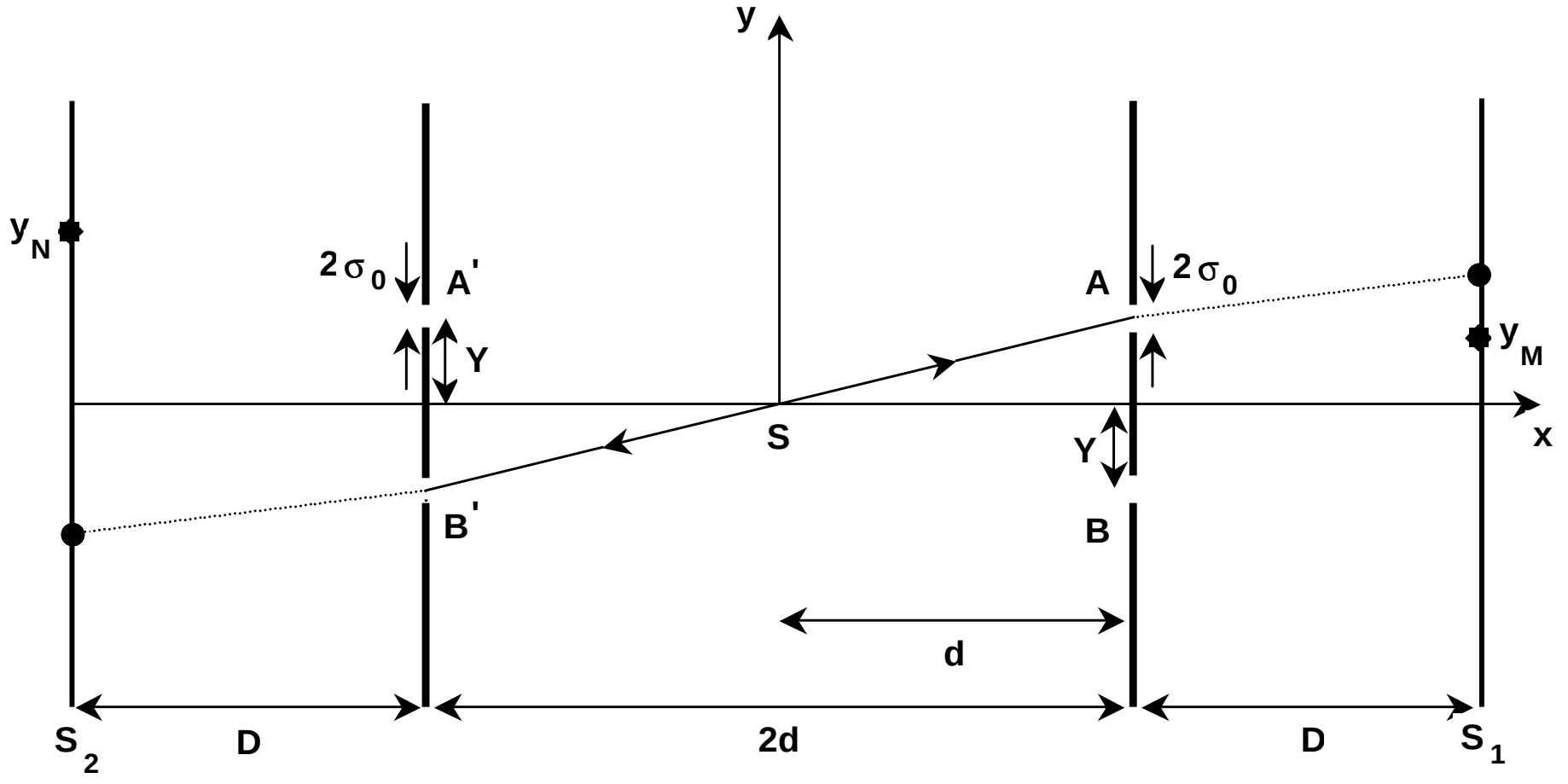

*Fig. 2.1:* A two double-slit experiment configuration. Two identical particles with zero total momentum are emitted from the source $S$ and then they pass through slits $A$ and $B^{'}$ or $B$ and $A^{'}$. Finally, they are detected on $S_1$ and $S_2$ screens, simultaneously. It is necessary to note that dotted lines are not real trajectories.

slits, the entangled wave function describing them is given by

$$\begin{aligned}\psi_{in}(x_1, y_1; x_2, y_2; t) &= \chi(x_1, x_2)\hbar \int_{-\infty}^{+\infty} exp[ik_y(y_1 - y_2)]dk_y e^{-iEt/\hbar} \\ &= 2\pi\hbar\chi(x_1, x_2)\delta(y_1 - y_2)e^{-iEt/\hbar} \qquad (2.1)\end{aligned}$$

where $E = E_1 + E_2 = \hbar^2(k_x^2 + k_y^2)/m$ is the total energy of the system of the two particles, and $\chi(x_1, x_2)$ is the $x$-component of the wave function that could have a form similar to the $y$-component. However, its form is not important for the present work. The wave function (2.1) is just the one represented in [20], and it shows that the two particles have vanishing total momentum in the $y$-direction, and their $y$-component of the center of mass is exactly located on the $x$-axis. This is not inconsistent with Heisenberg's uncertainty principle, because

$$[p_{y_1} + p_{y_2}, y_1 - y_2] = 0. \qquad (2.2)$$

The plane wave assumption comes from large distance between source $S$ and double-slit screens. To avoid the mathematical complexity of Fresnel diffraction at a sharp-edge slit, we suppose the slits have soft edges that generate waves having identical Gaussian profiles in the $y$-direction while the plane wave in the $x$-direction is unaffected [3]. The instant at which the packets are formed will be taken as our zero of time. Therefore, the four waves emerging from the slits $A$, $B$, $A^{'}$ and $B^{'}$ are initially

$$\begin{aligned} \psi_{A,B}(x,y) &= (2\pi\sigma_0^2)^{-1/4}e^{-(\pm y-Y)^2/4\sigma_0^2}e^{i[k_x(x-d)+k_y(\pm y-Y)]} \\ \psi_{A',B'}(x,y) &= (2\pi\sigma_0^2)^{-1/4}e^{-(\pm y+Y)^2/4\sigma_0^2}e^{i[-k_x(x+d)+k_y(\pm y+Y)]} \end{aligned} \tag{2.3}$$

where $\sigma_0$ is the half-width of each slit. At time $t$ the general total wave function at a space point $(x, y)$ of our considered system for bosonic and fermionic particles is given by

$$\begin{aligned} \psi(x_1, y_1; x_2, y_2; t) = N[&\psi_A(x_1, y_1, t)\psi_{B'}(x_2, y_2, t) \pm \psi_A(x_2, y_2, t)\psi_{B'}(x_1, y_1, t) \\ &+\psi_B(x_1, y_1, t)\psi_{A'}(x_2, y_2, t) \pm \psi_B(x_2, y_2, t)\psi_{A'}(x_1, y_1, t)] \end{aligned} \tag{2.4}$$

where $N$ is a reparametrization constant that its value is unimportant in this work and

$$\begin{aligned} \psi_{A,B}(x,y,t) &= (2\pi\sigma_t^2)^{-1/4}e^{-(\pm y-Y-\hbar k_y t/m)^2/4\sigma_0\sigma_t} \\ &\quad \times e^{i[k_x(x-d)+k_y(\pm y-Y)-Et/\hbar]} \\ \psi_{A',B'}(x,y,t) &= (2\pi\sigma_t^2)^{-1/4}e^{-(\pm y-Y-\hbar k_y t/m)^2/4\sigma_0\sigma_t} \\ &\quad \times e^{i[-k_x(x+d)+k_y(\pm y-Y)-Et/\hbar]} \end{aligned} \tag{2.5}$$

with

$$\sigma_t = \sigma_0(1 + \frac{i\hbar t}{2m\sigma_0^2}). \tag{2.6}$$

In addition, the upper and lower sings in the total wave function (2.4) refer to symmetric and anti-symmetric wave function under exchange of particle 1 to particle 2, corresponding to bosonic and fermionic property, while in Eq. (2.5) they refer to upper and lower slits, respectively. In the next section, we shall use BQM to derive some of the predictions of this proposed experiment.

## 2.3 Bohmian quantum mechanics prediction

In BQM, the complete description of a system is given by specifying the position of the particles in addition to their wave function which has the role of guiding the particles according to following guidance condition for $n$ particles, with masses $m_1, m_2, ..., m_n$

$$\dot{\mathbf{x}}_i(\mathbf{x}, t) = \frac{1}{m_i}\nabla_i S(\mathbf{x}, t) = \frac{\hbar}{m_i} Im\left(\frac{\nabla_i\psi(\mathbf{x}, t)}{\psi(\mathbf{x}, t)}\right) \tag{2.7}$$

where $\mathbf{x} = (\mathbf{x}_1, \mathbf{x}_2, \ldots, \mathbf{x}_n)$ and

$$\psi(\mathbf{x}, t) = R(\mathbf{x}, t)e^{iS(\mathbf{x},t)/\hbar} \tag{2.8}$$

is a solution of Schrödinger's wave equation. Thus, instead of SQM with indistinguishable particles, in BQM the path of particles or their individual histories distinguishes them and each one of them can be studied separately [3]. In addition, Belousek [25] in his recent work, concluded that the problem of Bohmian mechanical particles being statistically (in)distinguishable is a matter of theory choice underdetermined by logic and experiment, and that such particles are in any case physically distinguishable. For our proposed experiment, the speed of the particles 1 and 2 in the $y$-direction is given , respectively, by

$$\begin{aligned}\dot{y}_1(x_1,y_1;x_2,y_2;t) &= \frac{\hbar}{m}Im(\frac{\partial_{y_1}\psi(x_1,y_1;x_2,y_2;t)}{\psi(x_1,y_1;x_2,y_2;t)})\\ \dot{y}_2(x_1,y_1;x_2,y_2;t) &= \frac{\hbar}{m}Im(\frac{\partial_{y_2}\psi(x_1,y_1;x_2,y_2;t)}{\psi(x_1,y_1;x_2,y_2;t)}).\end{aligned} \tag{2.9}$$

With the replacement of $\psi(x_1,y_1;x_2,y_2;t)$ from Eqs. (2.4) and (2.5), we have

$$\begin{aligned}\dot{y}_1 = \quad N\tfrac{\hbar}{m} \quad & Im\{\frac{1}{\psi}[[-2(y_1-Y-\hbar k_y t/m)/4\sigma_0\sigma_t+ik_y]\psi_{A_1}\psi_{B'_2}\\ \pm \quad & [-2(y_1+Y+\hbar k_y t/m)/4\sigma_0\sigma_t-ik_y]\psi_{A_2}\psi_{B'_1}\\ + \quad & [-2(y_1+Y+\hbar k_y t/m)/4\sigma_0\sigma_t-ik_y]\psi_{B_1}\psi_{A'_2}\\ \pm \quad & [-2(y_1-Y-\hbar k_y t/m)/4\sigma_0\sigma_t+ik_y]\psi_{B_2}\psi_{A'_1}]\}\\ \dot{y}_2 = \quad N\tfrac{\hbar}{m} \quad & Im\{\frac{1}{\psi}[[-2(y_2+Y+\hbar k_y t/m)/4\sigma_0\sigma_t-ik_y]\psi_{A_1}\psi_{B'_2}\\ \pm \quad & [-2(y_2-Y-\hbar k_y t/m)/4\sigma_0\sigma_t+ik_y]\psi_{A_2}\psi_{B'_1}\\ + \quad & [-2(y_2-Y-\hbar k_y t/m)/4\sigma_0\sigma_t+ik_y]\psi_{B_1}\psi_{A'_2}\\ \pm \quad & [-2(y_2+Y+\hbar k_y t/m)/4\sigma_0\sigma_t-ik_y]\psi_{B_2}\psi_{A'_1}]\}\end{aligned} \tag{2.10}$$

where, for example, the short notation $\psi_A(x_1,y_1,t)=\psi_{A_1}$ is used. Furthermore, from Eq. (2.5) it is clear that

$$\begin{aligned}&\psi_A(x_1,y_1,t)=\psi_B(x_1,-y_1,t)\\ &\psi_A(x_2,y_2,t)=\psi_B(x_2,-y_2,t)\\ &\psi_{B'}(x_1,y_1,t)=\psi_{A'}(x_1,-y_1,t)\\ &\psi_{B'}(x_2,y_2,t)=\psi_{A'}(x_2,-y_2,t)\end{aligned} \tag{2.11}$$

which indicates the reflection symmetry of $\psi(x_1,y_1;x_2,y_2;t)$ with respect to the $x$-axis. Utilizing this symmetry in Eq. (2.10), we can see that

$$\begin{aligned}&\dot{y}_1(x_1,y_1;x_2,y_2;t)=-\dot{y}_1(x_1,-y_1;x_2,-y_2;t)\\ &\dot{y}_2(x_1,y_1;x_2,y_2;t)=-\dot{y}_2(x_1,-y_1;x_2,-y_2;t)\end{aligned} \tag{2.12}$$

which are valid for both bosonic and fermionic particles. Relations (2.12) show that if $y_1(t)=y_2(t)=0$, then the speed of each particles in the $y$-direction is zero. This means that none of the particles can cross the $x$-axis nor are they tangent to it, provided both of them are simultaneously on this axis. There is the same symmetry of the velocity about the $x$-axis as for an ordinary double-slit experiment [3].

If we consider $y = (y_1 + y_2)/2$ to be the vertical coordinate of the center of mass of the two particles [2], then we can write

$$\begin{aligned} \dot{y} &= (\dot{y}_1 + \dot{y}_2)/2 \\ &= -N\frac{\hbar}{2m} Im\{\frac{1}{\psi}(\frac{y_1 + y_2}{2\sigma_0\sigma_t})(\psi_{A_1}\psi_{B_1'} \pm \psi_{A_2}\psi_{B_1'} + \psi_{B_1}\psi_{A_2'} \pm \psi_{B_2}\psi_{A_1'})\} \\ &= \frac{(\hbar/2m0\sigma_0^2)^2}{1 + (\hbar/2m\sigma_0^2)^2 t^2} yt. \end{aligned} \tag{2.13}$$

Solving the equation of motion (2.13), we obtain the path of the $y$-coordinate of the center of mass

$$y(t) = y(0)\sqrt{1 + (\hbar/2m\sigma_0^2)^2 t^2}. \tag{2.14}$$

If it is assumed that, at $t = 0$ the center of mass of the two particles is exactly on the $x$-axis, that is $y(0) = 0$, then the center of mass of the particles will always remain on the $x$-axis. Thus, according to BQM, the two particles will be detected at points symmetric with respect to the $x$-axis, as shown in Fig. 2.1.

It seems that calculation of quantum potential can give us another perspective of this experiment. As we know, to see the connection between the wave and particle, the Schrödinger equation can be rewritten in the form of a generalized Hamilton-jacobi equation that has the form of the classical equation, apart from the extra term

$$Q(\mathbf{x}, t) = -\frac{\hbar^2}{2m}\frac{\nabla^2 R(\mathbf{x}, t)}{R(\mathbf{x}, t)} \tag{2.15}$$

where the function Q has been called quantum potential. However, it can be seen that the calculation and analysis of Q, by using our total wave function (2.4), is not very simple. On the other hand, we can use the form of Newton's second law, in which the particle is subject to a quantum force ($-\nabla Q$), in addition to the classical force ($-\nabla V$), namely

$$\mathbf{F} = -\nabla(Q + V). \tag{2.16}$$

Now, if we utilize the equation of motion of the center of mass $y$-coordinate (2.14) and Eq. (2.16), it is possible to obtain the quantum potential for the center of mass motion ($Q_{cm}$). Thus, we can write

$$-\frac{\partial Q}{\partial x} = m\ddot{x} = 0 \tag{2.17}$$

[2] Here, one may argue that this center of mass definition seems inconsistent with the previous definition introduced in the incident wave function (2.1). But, in appendix A, we have shown that these two definitions of the center of mass coordinate are two consistent representations.

$$-\frac{\partial Q}{\partial y}=m\ddot{y}=\frac{my(0)(\hbar/2m\sigma_0^2)^2}{(1+(\hbar t/2m\sigma_0^2)^2)^{3/2}}=\frac{my^4(0)}{y^3}(\frac{\hbar}{2m\sigma_0^2})^2 \tag{2.18}$$

where the result of Eq. (2.17) is clearly due to motion of plane wave in the $x$-direction. In addition, we assume that $\nabla V=0$ in our experiment. Thus, our effective quantum potential is only a function of the $y$-variable and it has the form

$$Q=\frac{my^4(0)}{2y^2}(\frac{\hbar}{2m\sigma_0^2})^2=\frac{1}{2}my^2(0)\frac{(\hbar/2m\sigma_0^2)^2}{1+(\hbar t/2m\sigma_0^2)^2}. \tag{2.19}$$

If it is assumed that $y(0)=0$, the quantum potential for the center of mass of the two particles is zero at all times and it remains on the $x$-axis. However, if $y(0)\neq 0$, then the center of mass can never touch or cross the $x$-axis. These conclusions are consistent with our earlier results (Eq. (2.14)).

## 2.4 Predictions of standard quantum mechanics

So far, we have been studying the results obtained from BQM at the individual level. Now it is well known from SQM that the probability of simultaneous detection of two particles at $y_M$ and $y_N$, at the screens $S_1$ and $S_2$, is equal to

$$P_{12}(y_M,y_N)=\int_{y_M}^{y_M+\triangle}dy_1\int_{y_N}^{y_N+\triangle}dy_2|\psi(x_1,y_1;x_2,y_2;t)|^2. \tag{2.20}$$

The parameter $\Delta$, which is taken to be small, is a measure of the size of the detectors. It is clear that the probabilistic prediction of SQM is in disagreement with the symmetrical prediction of BQM for the $y(0)=0$ condition, because SQM predicts that probability of asymmetrical detection, at the individual level of a pair of particles, can be different from zero, in opposition to BQM's symmetrical predictions. In the other words, based on SQM's prediction, the probability of finding the two particles on one side of the $x$-axis can be nonzero while we showed that BQM's prediction forbids such events in our scheme, and its probability is exactly zero. Thus, if necessary arrangements to perform this experiment are provided, one can choose one of the two theories as a more complete description of the quantum universe.

## 2.5 Statistical distribution of the center of mass coordinate around the $x$-axis

We have assumed that the two particles are entangled so that in spite of a position distribution for each particle, $y(0)$ can be always considered to be on the $x$-axis. However, one may argue that, it is necessary to consider a position distribution for $y(0)$, that is, $\triangle y(0)\neq 0$ while $\langle y(0)\rangle=0$. Therefore, it may seem that, not only symmetrical detection of the two particles is violated, but also they can be found on one side of the $x$-axis on the screens, because the majority of the pairs can not be simultaneously on the $x$-axis [26]. In this regard, Ghose [27] believes that the two entangled bosonic particles cannot cross the symmetry axis even if we have the situation $(y_1 + y_2)_{t=0} \neq 0$. However, even by accepting Marchildon's argument about this situation [26], this problem can be solved if we adjust $\triangle y(0)$ to be very small. We assume that, to keep symmetrical detection about the $x$-axis with reasonable approximation, the center of mass dispersion in the $y$-direction must be smaller than the distance between any two neighboring maxima on the screens, that is,

$$\triangle y(t) \ll \frac{\lambda D}{2Y} \simeq \frac{\pi\hbar t}{Ym} \tag{2.21}$$

where $\lambda$ is the de Broglie wavelength. By using Eq. (2.14), we can write

$$\triangle y(0) \ll \frac{2a\pi}{\sqrt{1+a^2}} \frac{\sigma_0^2}{Y} \tag{2.22}$$

where $a = \hbar t/2m\sigma_0^2$. Since considering $a \geq 1$ condition is a reasonable assumption in interferences experiments, we have

$$\triangle y(0) \ll 2\pi \frac{\sigma_0^2}{Y}. \tag{2.23}$$

This relation shows that $Y \ll \sigma_0$ condition can be considered as a suitable choice in our scheme. For usual condition of $Y \sim \sigma_0$, we have

$$\triangle y(0) \ll 2\pi\sigma_0. \tag{2.24}$$

Therefore, to prevent joint detection of the two particles on the one side of the $x$-axis, and also, to obtain an acceptable symmetrical joint detection around this axis, it is enough to adjust the $y$ dispersion of the center of mass position of the two particles very smaller than the width of slits. In this case, we only lose our information about the initial trajectory of bosonic particles. It is evident that, if one considers $\triangle y(0) \sim \sigma_0$, as was done in [26], the incompatibility between the two theories will disappear. But because of the entanglement of the two particles in the $y$-direction, it is possible to adjust $y(0)$ independent of $\sigma_0$, so that

$$0 \leq y(0) = \frac{1}{2}(y_1 + y_2)_{t=0} \ll \sigma_0. \tag{2.25}$$

Although it is obvious that $(\triangle y_1)_{t=0} = (\triangle y_2)_{t=0} \sim \sigma_0$, but the position entanglement of the two particles at the source $S$ in the $y$-direction makes them always satisfy Eq. (2.25), which is not feasible in the one-particle double-slit devices with $\triangle y(0) \sim \sigma_0$.

## 2.6 Comparison between SQM and BQM at the ensemble level

Now, one can compare the results of SQM and BQM at the ensemble level of the particles. To do this, we consider an ensemble of pairs of particles that have arrived at the detection screens at different times $t_i$. It is well known that, in order to ensure compatibility between SQM and BQM for ensemble of particles, Bohm added a further postulate to his three basic and consistent postulates [1-3]. Based on this further postulate, the probability density that a particle in the ensemble lies between $\mathbf{x}$ and $\mathbf{x} + d\mathbf{x}$, at time $t$, is given by

$$P(\mathbf{x}, t) = R^2(\mathbf{x}, t). \tag{2.26}$$

Thus, the joint probability of simultaneous detection for all pairs of particles of the ensemble, arriving on the two screens at different time $t_i$, with $y(0) = 0$, is

$$P_{12} = \lim_{N\to\infty} \sum_{i=1}^{N} R^2(y_1(t_i), -y_1(t_i), t_i) \equiv \int_{-\infty}^{+\infty} dy_1 \int_{-\infty}^{+\infty} dy_2 |\psi(y_1, y_2, t)|^2 = 1 \tag{2.27}$$

where every term in the sum shows only one pair arriving on the screens at the symmetrical points about the $x$-axis at time $t_i$, with the intensity of $R^2$. If all times $t_i$ in the sum are taken to be $t$, the summation on $i$ can be converted to an integral over all paths that cross the screens at that time. Now, one can consider that the joint detection of two points on the two screens at time $t$ is not symmetrical around the $x$-axis, but we know that they are not detected simultaneously. So, it is possible to consider the joint probability of detecting two particles at two arbitrary points $y_M$ and $y_N$ as follows

$$P_{12}(y_M, y_N, t) = \int_{y_M}^{y_M+\Delta} dy_1 \int_{y_N}^{y_N+\Delta} dy_2 |\psi(y_1, y_2, t)|^2, \tag{2.28}$$

which is similar to the prediction of SQM, but obtained in a Bohmian way. Thus, it appears that under such conditions, the possibility of distinguishing the two theories at the statistical level is impossible, as expected.

Here, to show equivalence of the two theories, we have assumed for simplicity that $y(0) = 0$. If one consider $y(0) \neq 0$ or $\triangle y(0) \neq 0$, the equivalence of the two theories is maintained, as it is argued by Marchildon [26]. But, using this special case, we show that assumption of $y(0) = 0$ is consistent with statistical results of SQM, and in consequence, finding such a source may not be impossible.

## 2.7 Quantum equilibrium hypothesis and our proposed experiment

In some of recent comments [28, 29, 30, 31], the quantum equilibrium hypothesis (QEH) is utilized in order to show that our proposed experiment cannot distinguish between SQM and BQM. In this section, we have presented some explanations to show that their argument may not be right and our basic conclusions about this scheme are still intact.

We have seen that, when the entangled particles pass through the slits, the transformation

$$\psi_{in}(x_1, y_1; x_2, y_2) \longrightarrow \psi(x_1, y_1; x_2, y_2) \tag{2.29}$$

occurs to the wave function describing the system. Now, it is interesting to know what can happen to the entanglement when the two particles emerge from the slits producing the Gaussian wave functions represented by Eq. (2.3). In the other words, there is a question as to whether the position entanglement property of the two particles is kept after this transformation. To answer this question, one can first examine the effect of the total momentum operator on the wave function of the system, $\psi(x_1, y_1; x_2, y_2; t)$, which yields

$$\begin{aligned}(p_{y_1} + p_{y_2})\psi(x_1, y_1; x_2, y_2; t) &= -i\hbar(\frac{\partial}{\partial y_1} + \frac{\partial}{\partial y_2})\psi(x_1, y_1; x_2, y_2; t) \\ &= i\hbar(\frac{y_1(t) + y_2(t)}{2\sigma_0\sigma_t})\psi(x_1, y_1; x_2, y_2; t)\end{aligned} \tag{2.30}$$

where one can see that the wave function is an eigenfunction of the total momentum operator. Now, if we can assume that the total momentum of the particles remains zero at all times (an assumption about which we shall elaborate later on), then it can be concluded that the center of mass of the two particles in the $y$-direction is always located on the $x$-axis. In other words, a momentum entanglement in the form $p_1 + p_2 = 0$ leads to the position entanglement in this experiment. However, Born's probability principle, i.e. $P = |\psi|^2$, which is a basic rule in SQM, shows that the probability of asymmetrical joint detection of the two particles can be non-zero on the screens. Thus, there is no position entanglement and consequently no momentum entanglement between the two particles. This compels us to believe that, according to SQM, the momentum entanglement of the two particles must be erased during their passage through the slits, and the center of mass position has to be distributed according to $|\psi|^2$.

In BQM, however, Born's probability principle is not so important as a primary rule and all particles follow well-defined tracks determined by the wave function $\psi(\mathbf{x}, t)$, using the guidance condition (2.7) with the unitary time development governed by Schrödinger's equation. However, to ensure the consistency of statistical results of BQM with SQM, Bohm [2] added QEH, i.e. $P = |\psi|^2$, to his self-consistent theory just as an additional assumption. Now, let us review the previous details, but this time in BQM frame. Based on our supposed EPR source, there are momentum and position entanglements between the two particles before they were arrived on the slits. Then, the wave function of the emerging particles from the slits suffers a transformation represented by Eq. (2.29). It is not necessary to know in details how this transformation acts, but what is important is that the two double-slit screens are considered to be completely identical. Thus, we expect that the two particles in the slits undergo the same transformation(s), and so the momentum entanglement, i.e. $p_1 + p_2 = 0$, must still be valid in BQM which is a deterministic theory, contrary to SQM. Then, according to Eq. (2.30), the validity of the momentum entanglement immediately leads to the position entanglement

$$y(t) = \frac{1}{2}(y_1(t) + y_2(t)) = 0. \tag{2.31}$$

We would like to point out that this entanglement is obtained by using the quantum wave function of the system. Therefore, this claim that the supposed position entanglement can not be understood by using the assumed wave function for the system is not correct. By the way, if we accept that the momentum entanglement is not kept and consequently $y(0)$ obeys QEH, then deterministic property of BQM, which is a main property of this theory, must be withdrawn. However, it is well known that Bohm [2] put QEH only as a subsidiary constraint to ensure the consistency of the motion of an ensemble of particles with SQM's results. Thus, although in this experiment, the center of mass position of the two entangled particles turns out to be a constant in BQM frame, the position of each particle is consistent with QEH so that the final interference pattern is identical to what is predicted by SQM. Therefore, Bohm's aim concerning QEH is still satisfied and the deterministic property of BQM is left intact. In addition, superluminal signals resulting from nonlocal conditions between our distant entangled particles are precisely masked by considering QEH for the distribution of each entangled particle.

So far, we have shown that in BQM frame, the center of mass position of the two entangled particles can be considered to be a constant, without any distribution. So, this property provides a way to make a discrepancy between SQM and BQM, even for an ensemble of entangled particles. For instance, suppose that we only consider those pairs one of which arrives at the upper half of the right screen. Thus, BQM predicts that only detectors located on the lower half of the left screen become ON and the other ones are always OFF. In fact, we obtain two identical interference patterns at the upper half of the right screen and the lower half of the left screen. Instead, SQM is either silent or predicts a diluted interference pattern at the left screen. Therefore, concerning the validity of the initial constraint $y(0) = 0$ in BQM, selecting of some pairs to obtain a desired pattern, which is called selective detection in [21], can be applied to evaluate the two theories, at the ensemble level of pairs.

## *2.8 Approaching realization of the experiment*

Based on a recent work on Bohmian trajectories for photons [14], the first effort for realization of a typical two-particle experiment was performed very recently by Brida *et al.* [23, 24], using correlated photons produced in type I parametric down conversion (PDC). In this realization a beam of a 351 nm pump laser of 0.4 W power with 1 mm in diameter is directed into a lithium iodate crystal, where correlated pairs of photons are generated by type I PDC [32]. The two photons are emitted at the same time (within femtoseconds, whilst correlation time is some orders of magnitude larger) in a well-defined direction for a specific frequency. By means of an optical condenser the produced photons, within two correlated directions corresponding to 702 nm emission (the degenerate emission for a 351 nm pump laser), are sent on a double slit (obtained by a metal deposition on a thin glass by a photolithographic process) placed just before the focus of the lens system. The two slits are separated by 100 $\mu$m and have a width of 10 $\mu$m. They lay in a plane orthogonal to the incident laser beam and are orthogonal to the table plane. Two single photon detectors are placed at a 1.21 and a 1.5 m distance after the slits. They are preceded by an interferential filter at 702 nm of 4 nm full width at half height and by a lens of 6 mm diameter and 25.4 mm focal length. The output signals from the detectors are routed to a two channel counter, in order to have the number of events on a single channel, and to a time to amplitude converter (TAC) circuit, followed by a single channel analyzer, for selecting and counting the coincidence events. Figure 2.2 illustrates this experimental set-up.

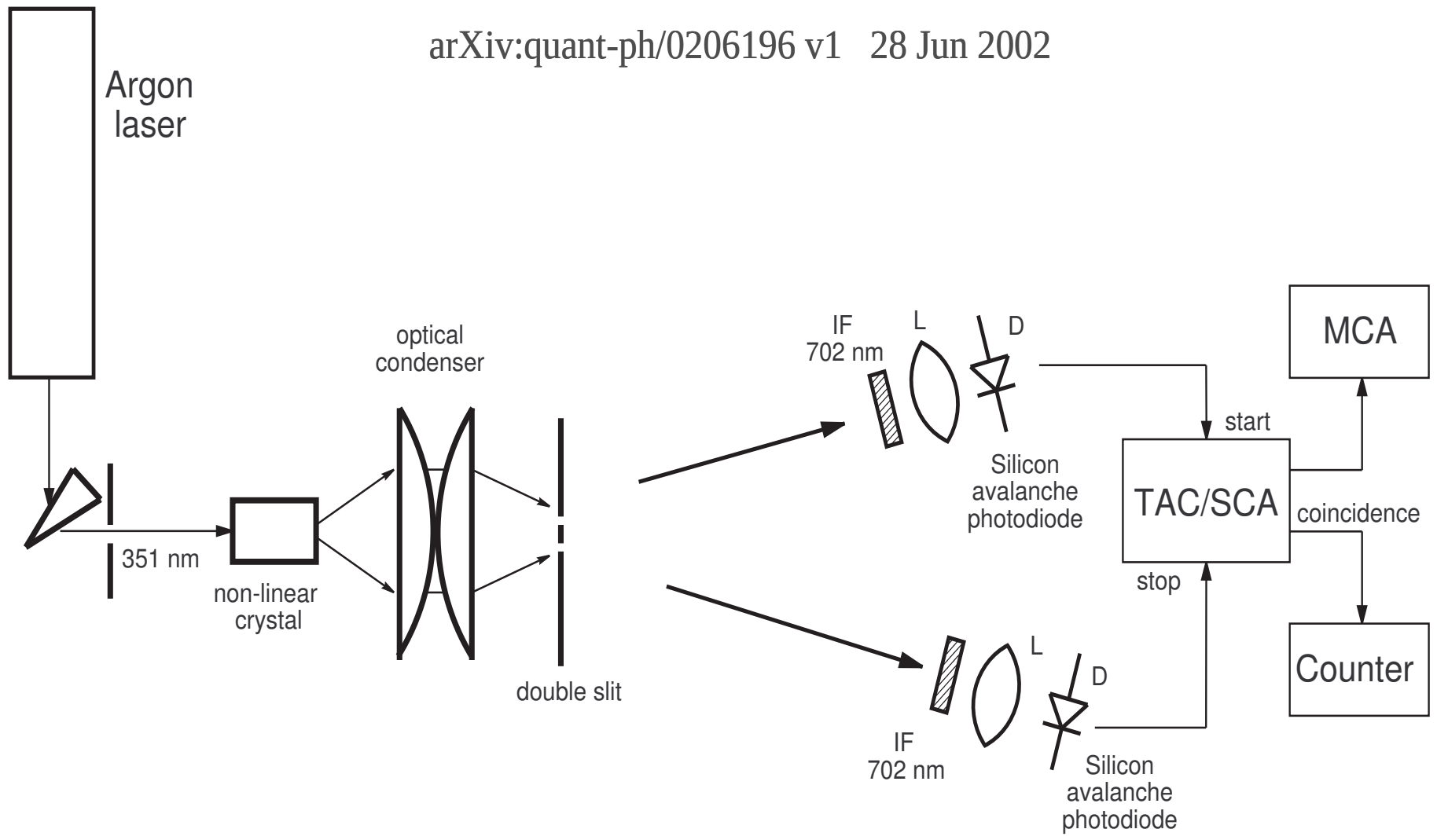

*Fig. 2.2:* The experimental apparatus. A pump laser at 351 nm generates parametric down conversion of type I in a lithium-iodate crystal. Conjugated photons at 702 nm are sent to a double-slit by a system of two piano-convex lenses in a way that each photon of the pair crosses a well defined slit. The first photodetector is placed at 1.21 m and the second one at 1.5 m from the slit. Both the single photon detectors (D) are preceded by an interferential filter at 702 nm (IF) and a lens (L) of 6 mm diameter and 25.4 mm focal length. Signals from detectors are sent to a time amplitude converter and then to the acquisition system (multi-channel analyzer and counters)[23].

By scanning the diffraction pattern using the first detector and leaving the second fixed at 55 mm from the symmetry axis, it is found that the coincidences pattern perfectly followed SQM's predictions, as Fig. 2.3 shows. The last ones are given by

$$\begin{aligned} C(\theta_1,\theta_2) &= g(\theta_1,\theta_i^A)^2 g(\theta_2,\theta_i^B)^2 + g(\theta_2,\theta_i^A)^2 g(\theta_1,\theta_i^B)^2 \\ &\quad +2g(\theta_1,\theta_i^A)g(\theta_2,\theta_i^B)g(\theta_2,\theta_i^A)g(\theta_1,\theta_i^B)cos[2kY(sin\theta_1 - sin\theta_2)] \end{aligned} \tag{2.32}$$

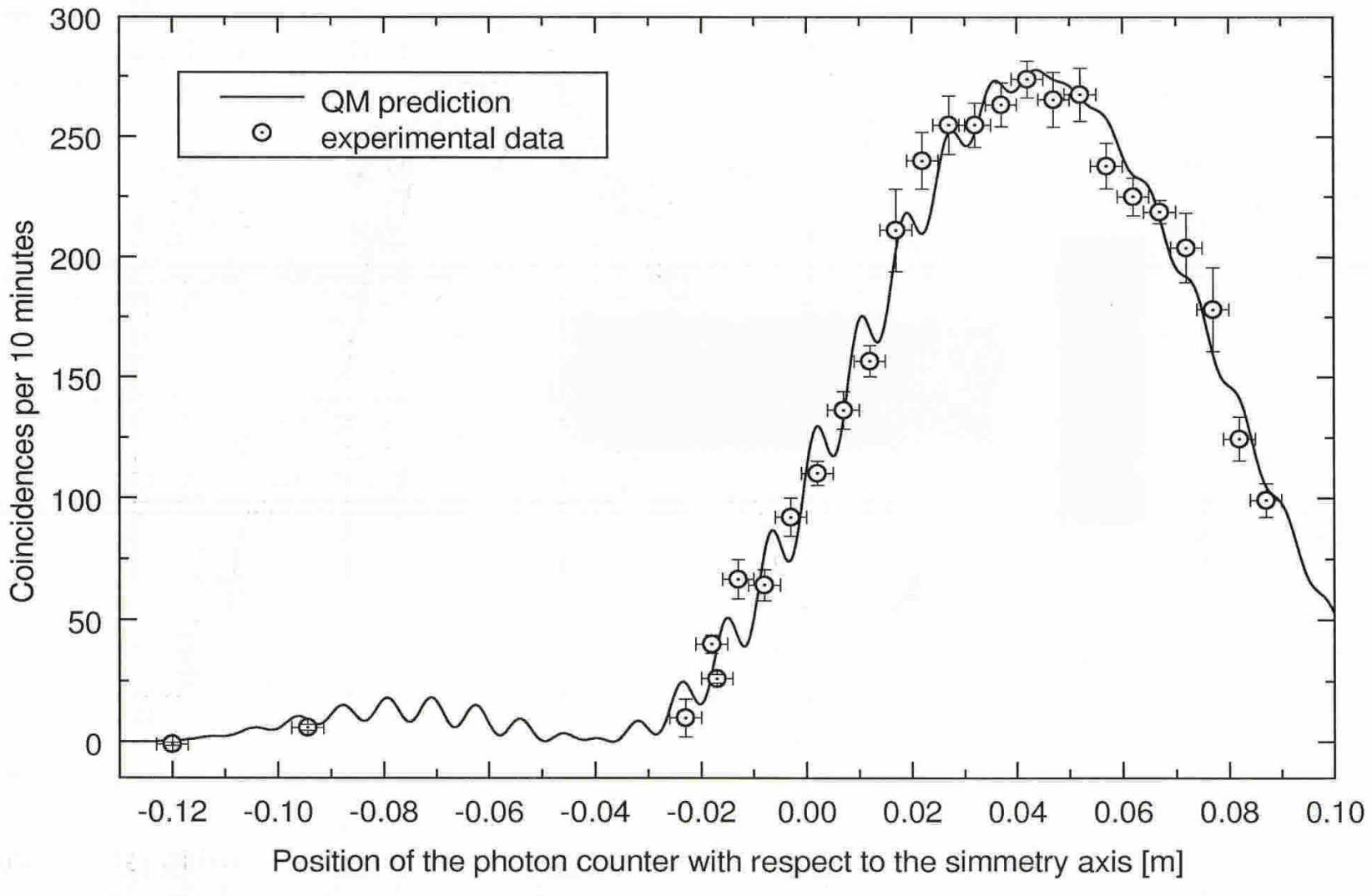

*Fig. 2.3:* Coincidences data in the region of interest compared with SQM's predictions. The second detector is kept fixed at -55 mm from the $x$-axis. The x errors bars represent the width of the lens before the detector [23].

in which

$$g(\theta, \theta_i^l) = \frac{sin(k\sigma_0(sin(\theta) - sin(\theta_i^l))}{k\sigma_0(sin(\theta) - sin(\theta_i^l))} \tag{2.33}$$

where $\theta_1$ ($\theta_2$) is the diffraction angle of the photon observed by detector 1 (2), and $\theta_i^l$ is the incidence angle of the photon on the slit $l$ (A or B).

The next result of this experiment is that a coincidence peak is observed (see Fig. 2.4) also when the first detector is placed inside enough the same semiplane of the second one. The coincidences acquisition with a temporal window of 2.6 ns is considered, and the background is evaluated shifting the delay between start and stop of TAC of 16 ns and acquiring data for the same time of the undelayed acquisition. When the center of the lens of the first detector is placed 17 mm after the median symmetry axis of the two slits and the second detector is kept at 55 mm, with 35 acquisitions of 30 minutes, it is obtained 78 $\pm$ 10 coincidences per 30 minutes after background subtraction, whilst in this situation BQM's prediction for coincidences is strictly zero. Furthermore, even when the two detectors were placed in the same semiplane, the first at 44.4 mm and the second at 117 mm from the symmetry axis, in correspondence of the second diffraction peak, a clear coincidence signal was still observed (albeit less evident than in the former case): in fact, after background subtraction, an average of 41 $\pm$ 14 coincidences per hour with 17 acquisitions of one hour (and a clear peak appeared on the multichannel analyzer) is obtained.

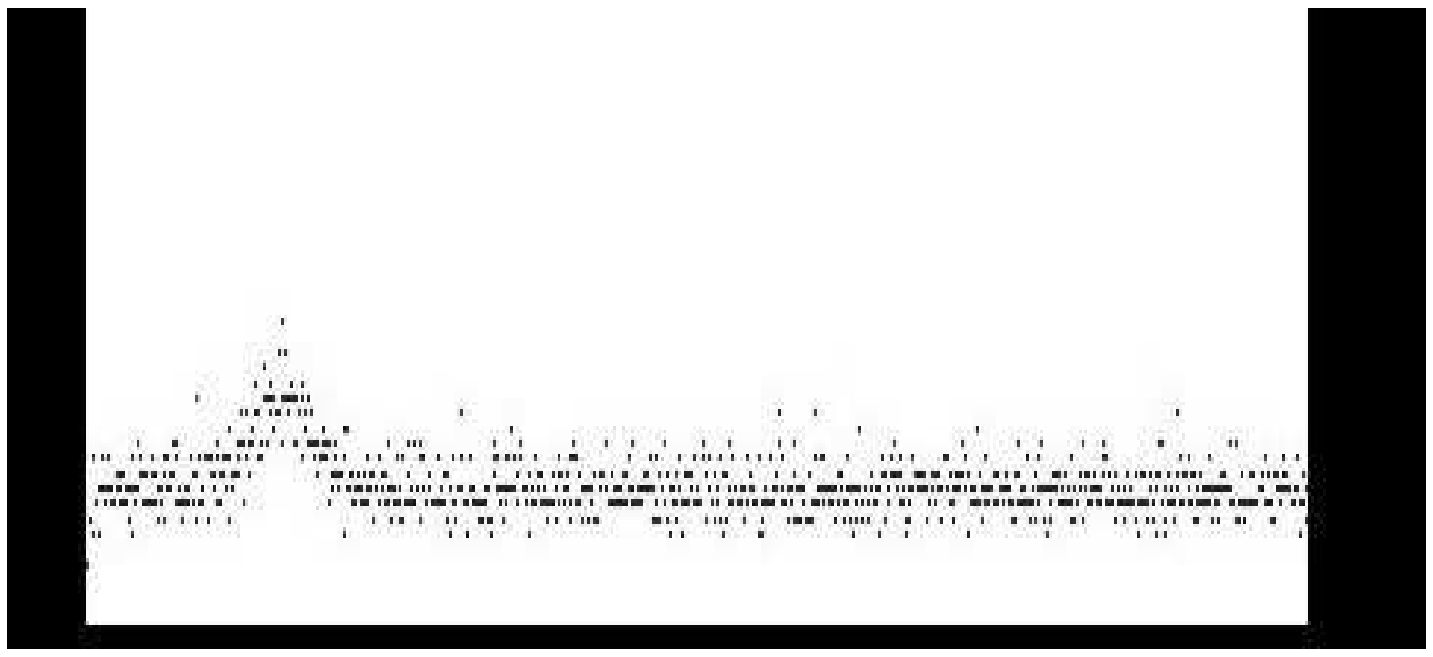

*Fig. 2.4:* Observed coincidence peak (output of the multi-channel analyzer) when the center of the lens of the first photodetector is placed 17 mm after the median symmetry axis of the double slit in the same semiplane of the other photodetector, which is kept at 55 mm after the median symmetry axis. Acquisition time lasts 17 hours. No background subtraction is done. A coincidence peak is clearly visible for a delay between start (first photodetector) and stop (second photodetector) of 9 ns (the delay inserted on the second line signal)[23].

Although performing of this experiment using photons, by Brida *et al.* [23], shows that it is feasible to realize the proposed thought experiment, however, their work does not satisfy all of our necessary conditions to enter into the region in which BQM's prediction is different from SQM's. For instance, in sec. 1.5, we have shown that the constraint $\triangle y \ll \lambda D/2Y$ is necessary to keep the symmetrical detection of the two particles in BQM frame of nonrelativistic domain. So, by considering $\hbar t/2m\sigma_0^2 \geq 1$, one obtains $\triangle y(0) \ll 2\pi\sigma_0^2/Y$. This roughly means that in the considered experimental set-up, with $\sigma_0 = 5\mu$m and $Y = 50\mu$m, we also should adjust $\triangle y(0) \ll 1\mu$m to observe a clear difference between the standard and Bohmian predictions. But in [23], the applied $\triangle y(0)$ in the laser beam is as much as 1 mm [33]. Therefore, it seems that we still need more elaborate efforts to complete the realization of this thought experiment.

## 2.9 Conclusions

In conclusion, we have suggested a two-particle system which can be adjusted to yield only symmetrical detections for the two entangled particles in BQM frame whilst according to SQM the probability for asymmetrical detections is not zero. The main reason for the existence of the mentioned differences between SQM and BQM in this thought experiment is that, in BQM as a deterministic theory, the position and momentum entanglements are kept at the slits, while in SQM, due to its probabilistic interpretation, we must inevitably accept that the entanglements of the two particles are erased when the two particles pass through the slits. Incidentally, the saved position entanglement in BQM, i.e. $y(0) = 0$, which is a result of the deterministic property of the thoery is not inconsistent with QEH, because we are still able to reproduce SQM's prediction for an ensemble of such particles, just as QEH requires. Therefore, our proposed experiment is a suitable candidate to distinguish between the standard and Bohmian quantum mechanics.

# 3. STUDY ON DOUBLE-SLIT DEVICE WITH TWO CORRELATED PARTICLES

## *3.1 Introduction*

The statistical interpretation of the wave function of the standard quantum mechanics (SQM) is consistent with all performed experiments. An interference pattern on a screen is built up by a series of apparently random events, and the wave function correctly predicts where the particle is most likely to land in an ensemble of trials. One may, however, take the view that the characteristic distribution of spots on a screen which builds up an interference pattern is an evidence for the fact that the wave function has a more potent physical role. If one attempts to understand the experimental results as the outcome of a causally connected series of individual process, then one is free to inquire about further significance of the wave function and to introduce other concepts in addition to it. Bohm [2], in 1952, showed that an individual physical system comprises a wave propagating in space-time together with a point particle which moves continuously under the guidance of the wave [1-3]. He applied his theory to a range of examples drawn from non-relativistic quantum mechanics and speculated on the possible alternations in the particle and field laws of motion such that the predictions of the modified theory continue to agree with those of SQM where this is tested, but it could disagree in as yet unexplored domains [3]. For instance, when Bohm presented his theory in 1952, experiments could be done with an almost continuous beam of particles. Thus, it was impossible to discriminate between the standard and the Bohmian quantum mechanics (BQM) at the individual levels. In fact, the two theories can be discriminated at this level, because SQM is a probabilistic theory while BQM is a precisely defined and deterministic theory.

In this chapter, we have studied entangled and disentangled wave functions that can be imputed to a two-particle interference device, using a Gaussian wave function as a real representation. Then, SQM and BQM predictions are compared at both the individual and the statistical levels [1][29-30].

[1] The individual level refers to our experiment with pairs of particles which are emitted in clearly separated short intervals of time, and by statistical level we mean our final interference pattern.

## 3.2 Description of the two-particle experiment

Consider the famous double-slit experiment. In the two-dimensional coordinate system, the centers of the two slits are located at $(0, \pm Y)$. Instead of the usual one-particle emitting source, consider a special source $S_1$, so that a pair of identical non-relativistic particles originate simultaneously from it. We assume that, the intensity of the beam is so low that at a time we have only a single pair of particles passing through the slits and the detection screen $S_2$ registers only those pairs of particles that reach it simultaneously, and so the interference effects of single particles will be eliminated. For mathematical simplicity, we avoid slits with sharp edges which produce the mathematical complexity of Fresnel diffraction, i.e., we assume that the slits have soft edges, so that the Gaussian wave packets are produced along the $y$-direction, and that the plane wave along the $x$-axis remain unchanged [3]. We take the time of the formation of the Gaussian wave to be $t = 0$. Then, the emerging wave packets from the slits $A$ and $B$ are respectively

$$\begin{aligned} \psi_A(x,y) &= (2\pi\sigma_0^2)^{-1/4} e^{-(y-Y)^2/4\sigma_0^2} e^{i[k_x x + k_y(y-Y)]} \\ \psi_B(x,y) &= (2\pi\sigma_0^2)^{-1/4} e^{-(y+Y)^2/4\sigma_0^2} e^{i[k_x x - k_y(y+Y)]} \end{aligned} \tag{3.1}$$

where $\sigma_0$ is the half-width of each slit. Moreover at time $t$ we can write

$$\begin{aligned} \psi_A(x,y,t) &= (2\pi\sigma_t^2)^{-1/4} e^{-(y-Y-\hbar k_y t/m)^2/4\sigma_0\sigma_t} e^{i[k_x x + k_y(y-Y) - Et/\hbar]} \\ \psi_B(x,y,t) &= (2\pi\sigma_t^2)^{-1/4} e^{-(y+Y+\hbar k_y t/m)^2/4\sigma_0\sigma_t} e^{i[k_x x - k_y(y+Y) - Et/\hbar]} \end{aligned} \tag{3.2}$$

where

$$\sigma_t = \sigma_0(1 + \frac{i\hbar t}{2m\sigma_0^2}). \tag{3.3}$$

Concerning the two-particle source, in general, we can have two alternatives:
*1.* The wave function describing the system (the two emitted particles+double slit screen) is so entangled that if one particle passes through the upper (lower) slit, the other particle must go through lower (upper) slit. In other words, the total momentum and the center of mass of the two particles in the $y$-direction is considered zero at the source.
*2.* The wave function describing the system is not entangled. In other words, the emission of each particle is done freely and the two particles can be considered independently. However, as a key point, the two correlated particles are still emitted simultaneously.

In the following, we shall study each one of the two alternatives, separately, and explain SQM's predictions. Then, Bohmian predictions have been compared with those of SQM.

## 3.3 Entangled wave function

We take the wave incident on the double-slit screen to be a plane wave of the form

$$\begin{aligned} \psi_{in}(x_1, y_1; x_2, y_2; t) &= \chi(x_1, x_2)\hbar \int_{-\infty}^{+\infty} exp[ik_y(y_1 - y_2)]dk_y e^{-iEt/\hbar} \\ &= 2\pi\hbar\chi(x_1, x_2)\delta(y_1 - y_2)e^{-iEt/\hbar} \end{aligned} \tag{3.4}$$

where $\chi(x_1, x_2)$ is the $x$-component of the wave function and $E = E_1 + E_2 = \hbar^2(k_x^2 + k_y^2)/m$ is the total energy of the system of the two identical particles. The parameter $m$ is the mass of each particle and $k_i$ is the wave number of particle in $i$-direction.

For this two-particle system, the total wave function after passing the two particles through the slits can be written as

$$\psi(x_1, y_1; x_2, y_2; t) = N[\psi_A(x_1, y_1, t)\psi_B(x_2, y_2, t) \pm \psi_A(x_2, y_2, t)\psi_B(x_1, y_1, t)] \tag{3.5}$$

where $N$ is a normalization constant which its value is not important here. Moreover, we consider that $\psi_A$ and $\psi_B$ are the Gaussian wave function introduced in Eq. (3.2). Also note that, the upper and lower signs in the total entangled wave function (3.5) are due to symmetric and anti-symmetric wave function under the exchange of particles 1 and 2, corresponding to bosonic and fermionic property, respectively.

## 3.4 Disentangled wave function

In this case, the incident plane wave can be considered to be

$$\tilde{\psi}_{in}(x_1, y_1; x_2, y_2; t) = ae^{i[k_x(x_1+x_2)+k_y(y_1+y_2)]}e^{-iEt/\hbar} \tag{3.6}$$

where $a$ is a constant. Now, for such a two-particle system, the total wave function after passing the two particles through the slits at time $t$ can be written as

$$\begin{aligned} &\tilde{\psi}(x_1, y_1; x_2, y_2; t) = \\ &\quad \widetilde{N}[\psi_A(x_1, y_1, t)\psi_B(x_2, y_2, t) + \psi_A(x_2, y_2, t)\psi_B(x_1, y_1, t) \\ &\quad +\psi_A(x_1, y_1, t)\psi_A(x_2, y_2, t) + \psi_B(x_1, y_1, t)\psi_B(x_2, y_2, t)] \\ &\quad = \widetilde{N}[\psi_A(x_1, y_1, t) + \psi_B(x_1, y_1, t)][\psi_A(x_2, y_2, t) + \psi_B(x_2, y_2, t)] \end{aligned} \tag{3.7}$$

where $\widetilde{N}$ is another normalization constant and $\psi_A$ as well as $\psi_B$ can be considered the Gaussian wave function represented in Eq. (3.2).

## 3.5 Standard quantum mechanics predictions

Based on SQM, the wave function can be associated with an individual physical system. It provides the most complete description of the system that is, in principle, possible. The nature of description is statistical, and concerns the probabilities of the outcomes of all conceivable measurements that may be performed on the system. It is well known from SQM that, the probability of simultaneous detection of the particles at $y_M$ and $y_N$, on the screen $S_2$, located at $x_1 = x_2 = D$ and $t = Dm/\hbar k_x$, is equal to

$$P_{12}(y_M, y_N, t) = \int_{y_M}^{y_M+\triangle} dy_1 \int_{y_N}^{y_N+\triangle} dy_2 |\psi(x_1, y_1; x_2, y_2; t)|^2. \tag{3.8}$$

The parameter $\Delta$, which is taken to be small, is a measure of the size of the detectors. We shall compare this prediction of SQM with that of BQM.

## 3.6 Bohmian predictions for the entangled case

Consider the entangled wave function (3.5). By substituting it in the guidance condition (2.7), one can obtain

$$\begin{aligned} \dot{y}_1 = \; & N\tfrac{\hbar}{m} \quad Im\{\frac{1}{\psi}[[-2(y_1 - Y - \hbar k_y t/m)/4\sigma_0\sigma_t + ik_y]\psi_{A_1}\psi_{B_2} \\ & \pm \quad [-2(y_1 + Y + \hbar k_y t/m)/4\sigma_0\sigma_t - ik_y]\psi_{A_2}\psi_{B_1}]\} \\ \dot{y}_2 = \; & N\tfrac{\hbar}{m} \quad Im\{\frac{1}{\psi}[[-2(y_2 + Y + \hbar k_y t/m)/4\sigma_0\sigma_t - ik_y]\psi_{A_1}\psi_{B_2} \\ & \pm \quad [-2(y_2 - Y - \hbar k_y t/m)/4\sigma_0\sigma_t + ik_y]\psi_{A_2}\psi_{B_1}]\}. \end{aligned} \tag{3.9}$$

On the other hand, from Eq. (3.2) one can see that,

$$\begin{aligned} \psi_A(x_1, y_1, t) = \psi_B(x_1, -y_1, t) \\ \psi_A(x_2, y_2, t) = \psi_B(x_2, -y_2, t) \end{aligned} \tag{3.10}$$

which indicate the reflection symmetry of $\psi(x_1, y_1; x_2, y_2; t)$ with respect to the $x$-axis. Using this symmetry in Eq. (3.9), we have

$$\begin{aligned} \dot{y}_1(x_1, y_1; x_2, y_2; t) = \mp\dot{y}_1(x_1, -y_1; x_2, -y_2; t) \\ \dot{y}_2(x_1, y_1; x_2, y_2; t) = \mp\dot{y}_2(x_1, -y_1; x_2, -y_2; t). \end{aligned} \tag{3.11}$$

These relations show that if $y_1(t) = y_2(t) = 0$, i.e., two particles are on the $x$-axis, simultaneously, then the speed of each bosonic particles in the $y$-direction is zero along the symmetry axis $x$, but we have no such constraint on fermionic particles, as was also mentioned by Ghose [13]. However, in the previous chapter, we have shown that, there is such a constraint on both bosonic and fermionic particles, using the two entangled particles in a two double-slit device.

If we consider $y = (y_1 + y_2)/2$ to be the vertical coordinate of the center of mass of the two particles, then we can write

$$\begin{aligned} \dot{y} \; &= \; (\dot{y}_1 + \dot{y}_2)/2 \\ &= \; N\frac{\hbar}{2m} Im\{\frac{1}{\psi}(-\frac{y_1 + y_2}{2\sigma_0\sigma_t})(\psi_{A_1}\psi_{B_2} \pm \psi_{A_2}\psi_{B_1})\} \end{aligned}$$

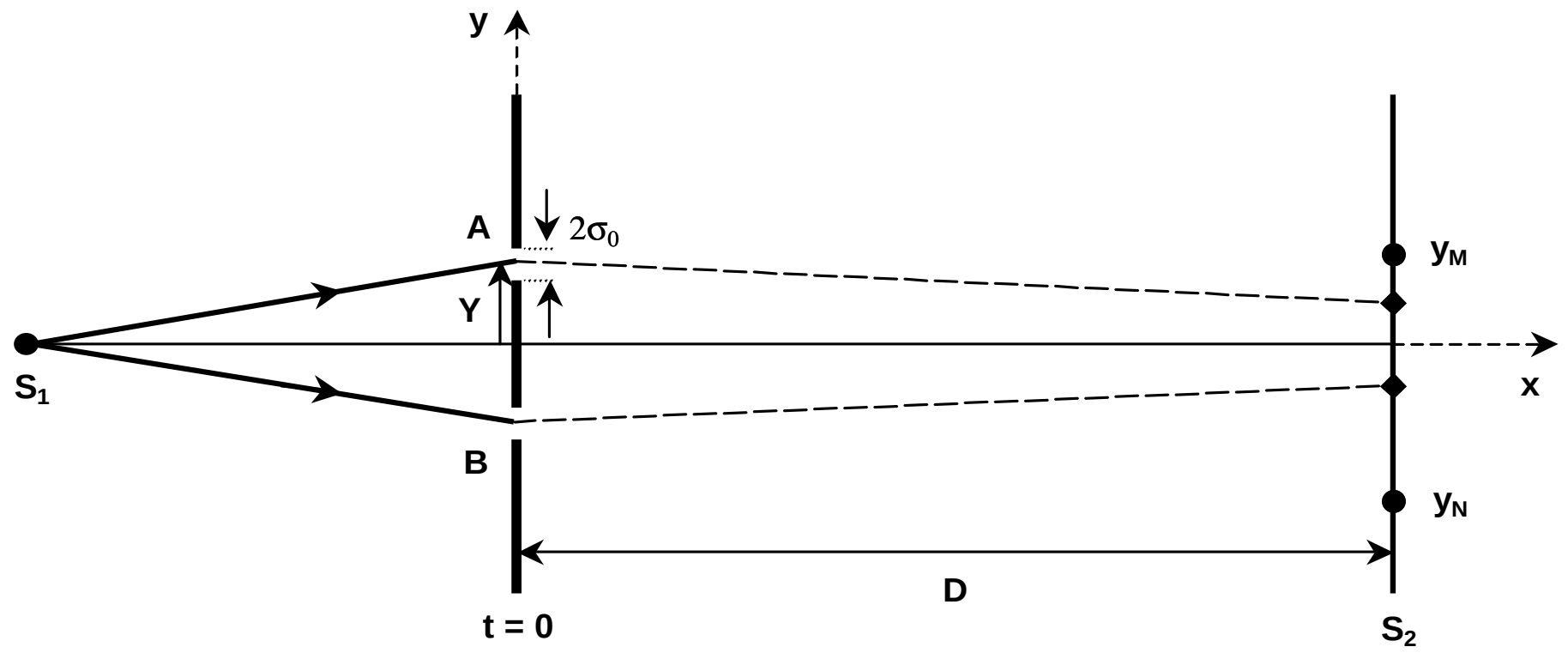

*Fig. 3.1:* A two-slit scheme in which two identical entangled particles are emitted from the source $S_1$. Then, they pass through the slits $A$ and $B$, and finally they are detected on the screen $S_2$, simultaneously. We have assumed that, $y(0) = 0$, or $\langle y(0)\rangle = 0$ under $\triangle y(0) \ll \sigma_0$ and $\hbar t/2m\sigma_0^2 \sim 1$ conditions. It is clear that dashed lines are not real trajectories.

$$= \frac{(\hbar/2m\sigma_0^2)^2}{1+(\hbar/2m\sigma_0^2)^2t^2}yt. \qquad (3.12)$$

Solving this differential equation, we get the path of the $y$-coordinate of the center of mass

$$y(t) = y(0)\sqrt{1+(\hbar/2m\sigma_0^2)^2t^2}. \qquad (3.13)$$

Using Eq. (3.13) and doing the same as what was done in chapter 1, one obtains the quantum potential for the center of mass motion

$$Q_{cm} = \frac{my^4(0)}{2y^2}(\frac{\hbar}{2m\sigma_0^2})^2 = \frac{1}{2}my^2(0)\frac{(\hbar/2m\sigma_0^2)^2}{1+(\hbar t/2m\sigma_0^2)^2}. \qquad (3.14)$$

If the center of mass of the two particles is exactly on the $x$-axis at $t = 0$ , then $y(0) = 0$, and the center of mass of the particles will always remain on the $x$-axis. In addition, the quantum potential for the center of mass of the two particles is zero at all times. Thus, we have $y_1(t) = -y_2(t)$ and the two particles, in both the bosonic and fermionic case, will be detected at points symmetric with respect to the $x$-axis, as is shown in Fig. 3.1. This differs from the prediction of SQM, as the probability relation (3.8) shows. SQM predicts that the probability of asymmetrical detection of the pair of particles can be different from zero in contrast to BQM's symmetrical prediction. Furthermore, according to SQM's prediction, the probability of finding two particles at one side of the $x$-axis can be non-zero while it is shown that BQM forbids such events, provided that $y(0) = 0$. Figure 3.1 shows one of the typical inconsistencies which can be considered at the individual level. Based on BQM, bosonic and fermionic particles have the same results, but, we know that if one bosonic particle passes through the upper (lower) slit, it must detected on the upper (lower) side on the $S_2$ screen, due to relations (3.11). Instead, there is no such constraint on fermionic particles. If $y(0) \neq 0$ and we have a distributed source with $\langle y(0) \rangle = 0$, then for the conditions $\hbar t/2m\sigma_0^2 \sim 1$ and $Y \sim \sigma_0$, it is easy to show that the constraint $\triangle y(0) \ll \sigma_0$ still yields reasonable symmetrical detection around the $x$-axis on the screen. But, now, in addition to the fermionic particles, the bosonic ones can cross the symmetry axis.

## 3.7 Bohmian predictions for the disentangled case

By considering the disentangled wave function (3.7) and the guidance condition (2.7), Bohmian velocities of particle 1 and 2 can be obtained as

$$\begin{aligned}
\dot{y}_1 = & \ \widetilde{N} \ \frac{\hbar}{m} Im\{\frac{1}{(\psi_{A_1}+\psi_{B_1})}([\frac{-2(y_1 - Y - \hbar k_y t/m)}{4\sigma_0\sigma_t} + ik_y]\psi_{A_1} \\
+ & \ [\frac{-2(y_1 + Y + \hbar k_y t/m)}{4\sigma_0\sigma_t} - ik_y]\psi_{B_1}\} \\
\dot{y}_2 = & \ \widetilde{N} \ \frac{\hbar}{m} Im\{\frac{1}{(\psi_{A_2}+\psi_{B_2})}([\frac{-2(y_2 - Y - \hbar k_y t/m)}{4\sigma_0\sigma_t} + ik_y]\psi_{A_2} \\
+ & \ [\frac{-2(y_2 + Y + \hbar k_y t/m)}{4\sigma_0\sigma_t} - ik_y]\psi_{B_2}\}.
\end{aligned} \tag{3.15}$$

Thus, as expected, the speed of each particle is independent of the other. Using relations (3.15) as well as Eq. (3.10), we obtain

$$\begin{aligned}
\dot{y}_1(x_1, y_1, t) = -\dot{y}_1(x_1, -y_1, t) \\
\dot{y}_2(x_2, y_2, t) = -\dot{y}_2(x_2, -y_2, t).
\end{aligned} \tag{3.16}$$

This implies that the $y$-component of the velocity of each particle would vanish on the $x$-axis. Although these relations are similar to the relations that were obtained for the entangled wave function, but here we have an advantage: none of the particles can cross the $x$-axis nor are tangent to it, independent of the other particle's position. This property can be used to show that BQM's predictions are incompatible with SQM's.

To see this incompatibility, we use a special detection process on the screen $S_2$ that we call it selective detection. In this selective detection, we register only those pair of particles which are detected on the two sides of the $x$-axis, simultaneously. That is, we eliminate the cases of detecting only one particle or detecting both particles of the pair on the upper or lower part of the $x$-axis on the screen. Again, it is useful to obtain the equation of motion of the center of mass in the $y$-direction. Using Eq. (3.15), one can show that,

$$\dot{y} = \frac{(\hbar/2m\sigma_0^2)^2 yt}{1+(\hbar/2m\sigma_0^2)^2t^2} + \widetilde{N}\frac{\hbar}{2m} Im\{\frac{1}{\psi}(\frac{Y+\hbar k_y t/m}{\sigma_0\sigma_t} + 2ik_y)(\psi_{A_1}\psi_{A_2} - \psi_{B_1}\psi_{B_2})\}. \tag{3.17}$$

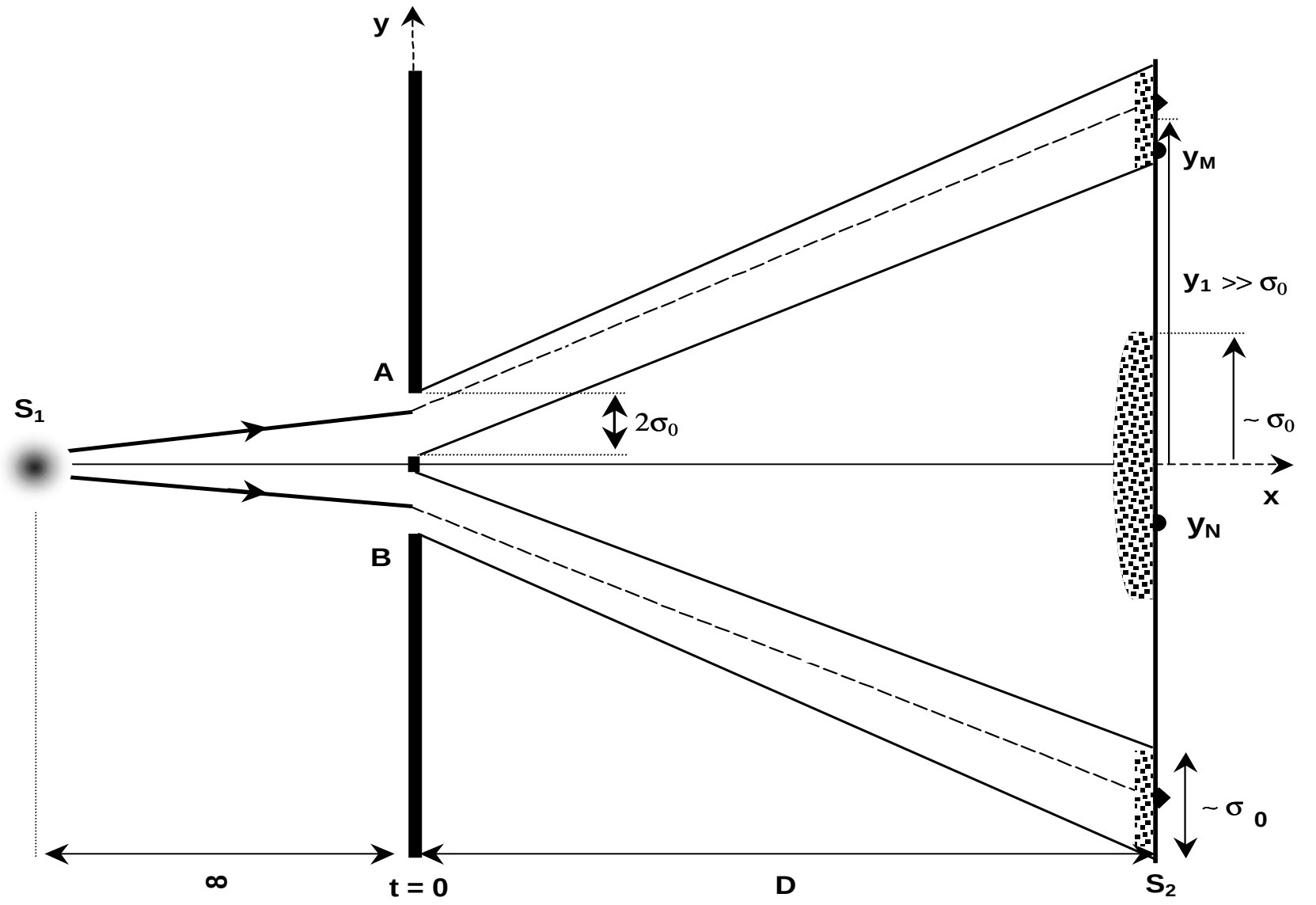

*Fig. 3.2:* Schematic drawing of a two-slit device in which two identical disentangled particles are simultaneously emitted by the source $S_1$. The symmetrical detection is not predicted at the central maximum, using both SQM and BQM. But, using BQM, we can have symmetrical detection at the other maxima (for example, at $y_1$ as the first acceptable maximum) under the condition $Y \ll 2\pi\sigma_0$.

We can assume that the distance between the source and the two-slit screen is so large that we have $k_y \simeq 0$. Then, using the special case $Y \ll \sigma_0$, the second term in Eq. (3.17) becomes negligible and the equation of motion for the $y$-coordinate of the center of mass is reduced to

$$\dot{y} \simeq \frac{(\hbar/2m\sigma_0^2)^2}{1+(\hbar/2m\sigma_0^2)^2t^2} yt \tag{3.18}$$

and similar to the entangled case, we have

$$y(t) \simeq y(0)\sqrt{1+(\hbar/2m\sigma_0^2)^2t^2}. \tag{3.19}$$

Since for this special source there was not any entanglement between the two particles, we must have $\triangle y(0) \sim \sigma_0$, according to quantum equilibrium hypothesis (QEH). Now, consider the case in which $\langle y(0)\rangle = 0$. To obtain symmetrical detection with reasonable approximation it is enough to assume that the center of mass variation is smaller than the distance between any two neighboring maxima on the screen. Then, according to Eq. (2.21), one can write

$$\triangle y(0) \ll \frac{\pi\hbar t}{Ym} \tag{3.20}$$

which yields

$$Y \ll 2\pi\sigma_0 \tag{3.21}$$

where in Eq. (3.20) we consider $\hbar t/2m\sigma_0^2 \sim 1$ so that $y(t) \sim y(0)$. Therefore, under these conditions, BQM's symmetrical prediction is incompatible with SQM's asymmetrical one. Figure 3.2 shows a schematic drawing of BQM's symmetrical detection occurred at a maximum for the conditions $Y \ll 2\pi\sigma_0$ and $\hbar t/2m\sigma_0^2 \sim 1$. It should be noted that, under these conditions, the two wave packets are overlapped on the screen in an interval of the order of $\sigma_0$. In this interval neither BQM nor SQM predict symmetrical detection around the $x$-axis. In fact, the symmetrical detection predicted by BQM happens far (relative to $\sigma_0$) from the $x$-axis on the screen. In other words, save the central peak, which does not show symmetry with respect to the $x$-axis, other less prominent maxima of the diffraction pattern appear at the locations

$$\begin{aligned} y_{n_+} &\simeq n_+\frac{\pi\hbar t}{2m\sigma_0} \pm \triangle y \\ y_{n_-} &\simeq -n_-\frac{\pi\hbar t}{2m\sigma_0} \pm \triangle y \end{aligned} \tag{3.22}$$

where $y_{n_\pm}$ refer to $y$-component of the maxima above or below the $x$-axis on the screen, respectively. In addition, $n_\pm$ represent positive integers. BQM's symmetrical prediction puts the following constraint:

$$n_+ = n_- \tag{3.23}$$

for great $n$'s. However, SQM's probabilistic prediction does not require this constraint.

Now, consider conditions in which $\langle y(0)\rangle \neq 0$, $\triangle y(0) \sim \sigma_0$, and $\hbar t/2m\sigma_0^2 \gg 1$. Then, the $x$-axis will not be an axis of symmetry and we have a new region on the $S_2$ screen around which all pairs of particles will be detected symmetrically. Thus, using Eqs. (3.16) and (3.19) as well as selective detection for the two particles, which requires registration of those two particles that are detected at

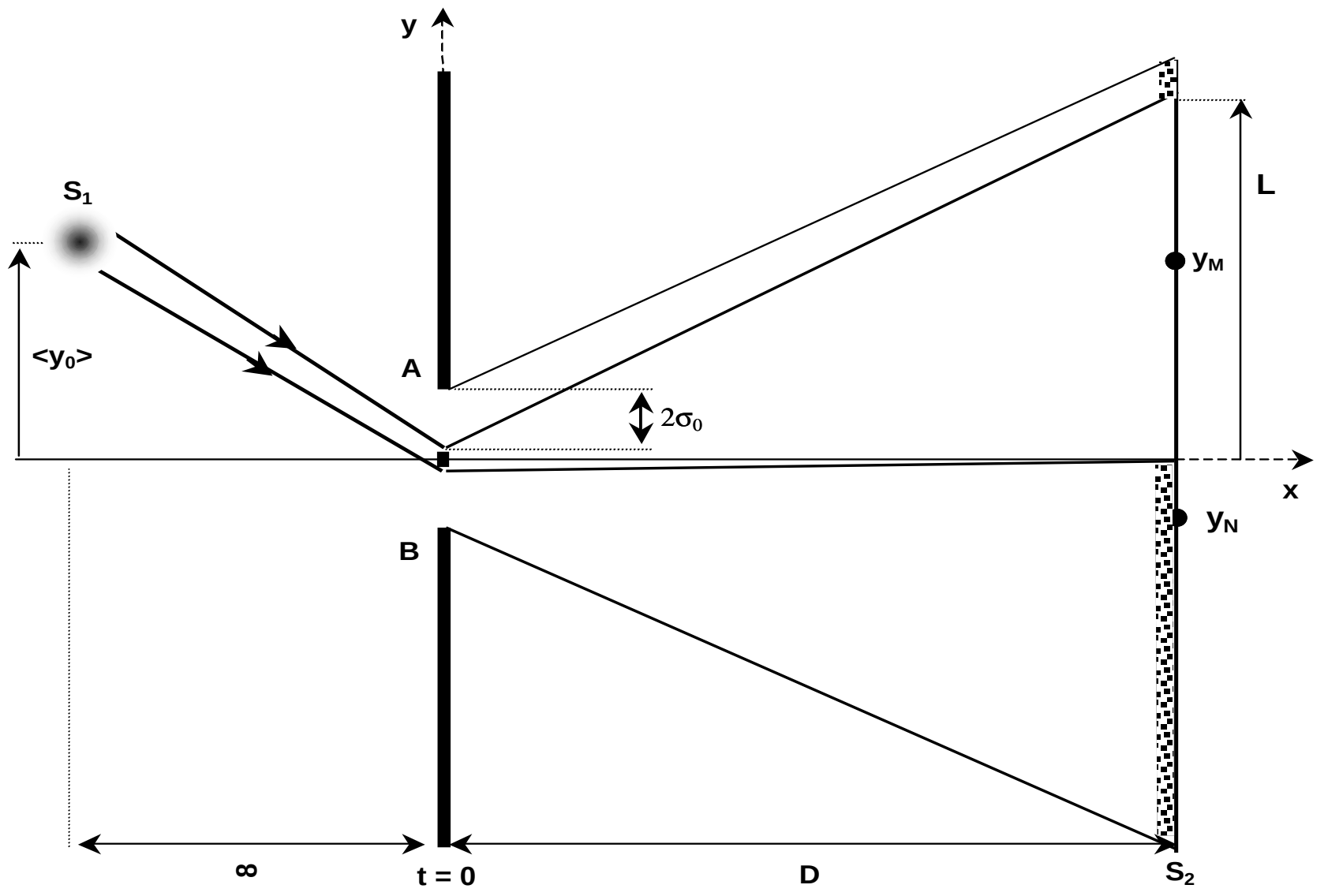

*Fig. 3.3:* Schematic drawing of a two-disentangled particle applied in a two-slit device in which the conditions $Y \ll \sigma_0 \ll \langle y(0)\rangle$ and $\hbar t/2m\sigma_0^2 \gg 1$ along with a special selective detection are considered. The length $L$ shows the low intensity interval in the final interference pattern.

the two sides of the $x$-axis simultaneously and omission of the others, BQM can predict a rather empty interval with low intensity of particles that has a length

$$L \simeq 2\langle y\rangle \simeq \frac{\hbar t}{m\sigma_0^2}\langle y(0)\rangle \tag{3.24}$$

if the constraint $\triangle y \ll L$ is satisfied. The last constraint at $\hbar t/2m\sigma_0^2 \gg 1$ condition, corresponds to $\triangle y(0) \ll \langle y(0)\rangle$. Figure 3.3 shows that according to BQM and under the conditions

$$\begin{aligned} &\frac{\hbar t}{2m\sigma_0^2} \gg 1 \\ &Y \ll \sigma_0 \ll \langle y(0)\rangle \end{aligned} \tag{3.25}$$

a considerable position change in the $y$-coordinate of the source produces a region with very low intensity on the screen which is not predicted by SQM. In fact, based on SQM, we have two alternatives:

*1.* The joint probability relation (3.8) is still valid and there is only a reduction in the intensity throughout the screen $S_2$, due to the selective detection.
*2.* SQM is silent about our selective detection.
In the first case, there is a disagreement between the predictions of SQM and BQM and in the second case, BQM has a better predictive power than SQM, even at the statistical level.

However, since the two particles in this scheme are emitted in a disentangled state, the results obtained may seem unbelievable. In this regard, Struyve [37] based on our aforementioned factorizable wave function (3.7), believes that the two independent particles of this experiment cannot produce different predictions for SQM and BQM. He argues that [37], the results of the experiment will not be altered if we emit the two particles simultaneously or emit only one particle at a time, because the two particles are totally independent. But, it should be noted that, we can use the selective detection only when the source emits two identical particles at the same time, and if the source emits just one particle at each time, then it is meaningless to utilize our selective detection. In the following, once again, we substantiate our previous arguments about this thought experiment, in some details.

At first, we examine the applied condition $Y \ll \sigma_0$, in a double-slit device with two disentangled particles. One may argue that this condition is meaningless, because based on the specifications of the set-up, Y represents the distance between the center of each slit to the $x$-axis, and therefore, the minimum value of $Y$ approaches $\frac{1}{2}\epsilon + \sigma_0$, where $\epsilon$ is considered to be very small and represents the length of the plane that separates the two slits. But, this objection can be answered by considering the overlapping of each particle's two Gaussian wave functions which are generated at the two near slits. The overlap causes the peak of each Gaussian wave to approach more and more to the $x$-axis. In addition, under this condition, the Gaussian wave functions lose their symmetrical form at each slit. Our argument becomes clearer when we consider $\epsilon = 0$ as a limiting case, i.e, we have only one slit. In this limiting case, it is clear that $Y = 0$. Therefore, when the two slits are very near together, the peak of Gaussian wave functions, i.e. $Y$, come very near to the $x$-axis and the condition $Y \ll \sigma_0$ is completely satisfied.

Another problem can arise when one thinks of the two independent particles. To handle this problem, let us reconsider the second term in Eq. (3.17), particularly the coefficient $(\psi_{A_1}\psi_{A_2} - \psi_{B_1}\psi_{B_2})$. Using the Gaussian wave (3.2) and the condition $k_y \simeq 0$, the latter coefficient can be written in the form

$$\begin{aligned} \psi_{A_1}\psi_{A_2} - \psi_{B_1}\psi_{B_2} &= (2\pi\sigma_t)^{-1/4} e^{2i(k_x x - Et/\hbar)} e^{-(y_1^2+y_2^2)/4\sigma_0\sigma_t} e^{-(Y+u_y t)^2/2\sigma_0\sigma_t} \\ &\times \left[ e^{(y_1+y_2)(Y+u_y t)/2\sigma_0\sigma_t} - e^{-(y_1+y_2)(Y+u_y t)/2\sigma_0\sigma_t} \right] \end{aligned} \tag{3.26}$$

where $u_y = \hbar k_y/m$. If we require that $y_1(t) + y_2(t) = 0$, i.e., the $y$-component of the two particles are entangled, then we would obtain the equation of motion (3.13), as expected. But our two particles in this scheme are initially disentangled, and it is not necessary to have $y_1(t) + y_2(t) = 0$. Instead, we can have another choice on the geometry of the two-slit set-up. In fact, if we apply the condition $Y \ll \sigma_0$, again the behavior of the equation of motion of the two particles in the $y$-direction is similar to the motion of the two entangled particles, while the two particles were disentangled. Hence, we can state that the classical interaction of the wave function of the two disentangled particles with the two-slit plane barrier for the condition $Y \ll \sigma_0$, results in a wave function which now guides the $y$-component of the center of mass of the two apparently disentangled particles in the same way as the case of two entangled particles with the initial condition $-\sigma_0 \leq (y_1 + y_2)_{t=0} \leq \sigma_0$, for those pairs of particles that pass through the two slits. Thus, we have shown that the results obtained in the two-slit device, using two synchronized identical particles along with the selective detection, are completely different from the ones obtained in a single-particle double-slit experiment. In fact, it seems that the motion of either particle is now dependent on its own location and the location of the other particle, although the apparent form of the wave function of the system can be efficiently represented by the use of the disentangled form in (3.7). Based on QEH, the factorizable wave function (3.7) results that both SQM and BQM must yield the same interference pattern for the whole particle arrived on the screen. But, one can see that this is not true for our specified conditions in which we have used a selective detection. If we study the interference pattern without using selective detection, we must obtain the same results for the two theories. However, using selective detection along with the guidance condition, it is clear that not only the two theories do not have the same statistical predictions, but also BQM clarifies and illuminates SQM, as Dürr *et al.* [38] said: *"by selectively forgetting results we can dramatically alter the statistics of those that we have not forgotten. This is a striking illustration of the way in which Bohmian mechanics does not merely agree with the quantum formalism, but, eliminating ambiguities, clarifies, and sharpens it."*. In our selective detection, we have forgotten detected single-particle and the two-particle contributions on the one side of the $x$-axis (of the screen $S_2$).

## 3.8 Conclusions

In this chapter, we have studied two nearly similar thought experiments which can give different results for the standard and Bohmian quantum mechanics in some particular conditions. The suggested experiments consist of a two-slit interferometer with a special source which emits two identical non-relativistic particles, simultaneously. We have shown that, according to the characteristic of the source, our two-particle system can be described by two kinds of wave functions: the entangled and the disentangled wave functions. For the entangled case, we have obtained some disagreement between SQM and BQM at the individual level while the two theories predict the same statistical results, as expected. For the disentangled case, the predictions of the two theories can be different at the individual level, too. Again, the results of the two theories were the same at the ensemble level. However, the use of selective detection can dramatically alter the interference pattern, so that not only the statistical results of BQM do not agree with those of SQM, but BQM may also increase our predictive power. Therefore, our suggested thought experiments provide examples for which the standard and Bohmian quantum mechanics may yield different predictions.

# Part II

# NEW PROPOSED EXPERIMENTS INVOLVING QUANTUM INFORMATION THEORY

# 4. INTRODUCTION-QUANTUM INFORMATION THEORY

## *4.1 Introduction*

In 1982 Feynman [39] observed that certain quantum mechanical effects cannot be simulated efficiently on a classical computer. This observation led to speculation that perhaps computation in general can be done more efficiently if it uses these quantum effects. But building quantum computers proved tricky, and as no one was sure how to use the quantum effects to speed up computation, the field developed slowly. It was not until 1994, when Shor [40, 41] surprised the world by describing a polynomial time quantum algorithm for factoring integers, that the field of quantum computing came into its own. This discovery prompted a flurry of activity, both among experimentalists trying to build quantum computers and theoreticians trying to find other quantum algorithms. Additional interest in the subject has been created by the invention of quantum dense coding [42], quantum teleportation [43] and quantum key distribution [44] as well as, more recently, popular press [45, 46] accounts of experimental successes in quantum teleportation and the demonstration of a three-bit quantum computer.

The concept of entanglement is the distinctive and responsible feature that allows quantum information to overcome some of the limitations posed by classical information, as exemplified by the new phenomena of teleportation, dense coding and key distribution which are explained in the following sections. In fact, entanglement leads to profound experimental consequences like non-local correlations: when two distantly apart parties Alice and Bob share, say, an EPR pair, the measurement by Alice on her state simultaneously determines the state on the Bob side. Apparently, this implies instant information transmission, in sharp contrast to Einstein's relativity. However, to reconcile both facts we must notice that the only way Bob has to know about his state (without measuring it) is by receiving a classical communication from Alice, which does propagate no faster than the speed of light.

## *4.2 Quantum dense coding*

Classical information can also be sent through quantum channels: to transmit the word 10011, it is enough that Alice prepares 5 qubits in the states $|1\rangle$, $|0\rangle$, $|0\rangle$, $|1\rangle$, $|1\rangle$, sends them to Bob through the quantum channel, and Bob measures each of them in the basis $|0\rangle$, $|1\rangle$. Each qubit carries a cbit, and this is the most it can do in isolation. But, as theoretically proposed by Bennett and Wiesner [42], if Alice and Bob share beforehand an entangled state, then 2 cbits of information can be sent from Alice to Bob with a single qubit. As a matter of fact, entanglement is a computing resource that allows more efficient ways of coding information. Assume, for instance, an entangled state of two photons. One of the photons goes to Alice, the other one to Bob. She performs one of the following operations on the polarization of her arriving photon: identity, flipping, change of $\pi$ in the relative phase, and the product of the last two. Once this is done, she sends back the photon to Bob, who measures in which of the four Bell states the photon pair is. Then, in this fashion we have been able to send 2 bits of information over one single particle with only 2 states, that is, by means of a qubit. It doubles what can be accomplished classically which results the name of quantum dense coding or super dense coding. Moreover, if Eve, as an eavesdropper, intercepts the qubit, she cannot get from it alone any information. Because, all the information lies in the entangled state, and Bob possesses half of the pair. Actually, Alice has sent Bob 2 qubits, but the first one long ago, as part of the initial entangled state. This fact has allowed them to communicate more efficiently, resorting to the entangled state they shared. Dense coding is kind of the inverse process to teleportation. In the latter the communication of two cbits allows us to reproduce a qubit state, while in the former the communication of a qubit carries along two cbits of information.

The following is a review of the dense coding protocol which is explained in [47]. Consider an EPR source which supplies Alice and Bob with a two-particle state like

$$|\psi\rangle = \frac{1}{\sqrt{2}}(|00\rangle + |11\rangle) \tag{4.1}$$

one of whose particles goes to Alice and the other one to Bob, who keep them. Alice is supplied with 2 cbits, which represent the numbers 0, 1, 2, 3 as 00, 01, 10, 11.

*Step 1.* Coding: According to the value of that number, Alice effects on her EPR half the unitary operation I,X,Z,Y, which brings the EPR state to 00+11, 10+01, 00-11, 10-01. Once this is done, she sends her half to Bob.

*Step 2.* Decoding: Upon reception, Bob effects on the EPR pair first a CNOT operation, such that the state becomes 00+10, 11+01, 00-10, 11-01. He then measures the second qubit; if Bob finds 0, he already knows that the message was 0 or 2, and if he finds 1, the message was 1 or 3. That is, he has gotten the second bit of the two-bit message. In order to know the first one, Bob next applies a Hadamard transformation on the first qubit, thereby the state becomes 00, 01, 10, -11, and after measuring the first bit, if he finds 0, he knows that the message was 0 or 1, and if he finds 1, the message was 2 or 3, that is, he has just gotten the second bit of the message.

An experiment of this nature has been performed in Innsbruck [48], by using the polarized-entangled photons of type II parametric down conversion that a non-linear crystal of $\beta$-barium borate produces: UV photons get disintegrated (though with low probability) in a pair of softer photons, with polarizations which in a certain geometric configuration are entangled. In that experiment, they managed to send $\log_2 3 = 1.58$ cbits per qubit. In a recent experiment [49], in which the qubits are the spins of $^1$H and $^{13}$C in a clorophorm molecule $^{13}CHCl_3$ marked with $^{13}$C, and RMN techniques are employed to initialize, manipulate and read out the spins, the authors claim to have reached the 2 cbits per qubit.

The initial preparation of the entangled pair and the posterior transmission of the information qubit may have opposite senses; for example, Bob sends to Alice one half of the entangled state, keeping the other half for himself, and then Alice uses her qubit to send to Bob the desired information. This may be of interest if the cost in the transmission in one way is higher than in the reverse way. On the other hand, intercepting the message from Alice to Bob does not provide any information to Eve, because the message is entangled with the part of the EPR pair possessed by Bob. Therefore it is automatically a secure emission of information (except if Eve intercepts both the original pair and the message and she replaces them).

## 4.3 Quantum teleportation

Copying classical states has never posed unsurmountable difficulties to experts. It suffices to thoroughfully observe the original as much as it may be required, avoiding to damage it, to retrieve the information needed to make a copy of it. This careful observation does not alter, in a noticeable way, its state. But if the original to be reproduced is a quantum system in an unknown state $\psi$, then any measurement (incompatible with $P_\psi$) made on the system to get information on $\psi$ will disturb, uncontrollably, the state destroying the original. Moreover, even in the case of having an unlimited number of copies of that state, infinitely many measurements will be necessary to determine that unknown state.

For example, assume that Alice has one spin-$\frac{1}{2}$) particle as a qubit in a pure state. Bob needs it, but Alice does not have any quantum channel to transmit it to him. If Alice knows the precise state of her qubit (for example, if she knows that her spin-$\frac{1}{2}$ is oriented in the direction $\mathbf{n}$), it is enough for her to give Bob in a letter (classical channel) that information (the components of $\mathbf{n}$) to enable him preparing a qubit exactly equal to Alice's. But if she happens not to know the state, she may choose to confess it to Bob, who would then be inevitably driven to prepare his qubit in a random way, obtaining a 50% fidelity on average. But Alice can also try to be more cooperative, making for example a measurement on her qubit of $\mathbf{n}' \cdot \sigma$, with $\mathbf{n}'$ arbitrarily chosen, and then transmitting to Bob both the components of $\mathbf{n}'$ and the result $\epsilon = \pm 1$ thus obtained. Armed with this information, Bob can prepare his qubit in the state

$$\frac{1}{2}(I + \epsilon \mathbf{n}' \cdot \sigma). \tag{4.2}$$

The average fidelity so obtained is larger than before: 2/3. However, it is not still enough.

If Alice and Bob share an EPR pair, there exists a protocol known as quantum teleportation, devised by Bennett *et al.* [43] in 1993, which resorts to the quantum entanglement of states, and the non-locality of quantum mechanics allows Bob to reproduce Alice's unknown quantum state with the assistance of only 2 cbits of information sent by Alice to Bob through a classical channel. This procedure necessarily destroys Alice's state (otherwise it would violate the quantum no-cloning theorem). The following is a description of the teleportation scheme as explained in [47].

Assume that Alice likes to teleport a qubit with the following quantum state

$$|\phi\rangle = \alpha|0\rangle + \beta|1\rangle \tag{4.3}$$

in which

$$\alpha = \cos\frac{1}{2}\theta, \beta = e^{i\phi}\sin\frac{1}{2}\theta. \tag{4.4}$$

In addition, consider

$$|\psi\rangle = \frac{1}{\sqrt{2}}(|00\rangle + |11\rangle) \tag{4.5}$$

be the EPR state shared between Alice and Bob, with Alice having the first of its qubits, and Bob the second. The initial state is thus $|\phi\rangle \otimes |\psi\rangle$, of which Alice can locally manipulate its two first bits and Bob the third one.
*Step 1.* Alice applies to the initial state the unitary operator

$$U = ((\mathrm{H} \otimes I)\mathrm{CNOT}) \otimes I \tag{4.6}$$

acting with the CNOT gate on the first two qubits and next with the Hadamard unitary gate H on the first one. The resulting state is

$$\frac{1}{2}(|00\rangle \otimes |\phi\rangle + |01\rangle \otimes X|\phi\rangle + |10\rangle \otimes Z|\phi\rangle + |11\rangle \otimes Y|\phi\rangle). \tag{4.7}$$

*Step 2.* Alice then measures the first two qubits, obtaining $|00\rangle, |01\rangle, |10\rangle$, or $|11\rangle$ equiprobably.[1]

Alice lets Bob know the result thus obtained, sending him two cbits: the pair of binary digits 00, 01, 10, 11 that characterizes it. As a byproduct of Alice's measurement, the first bit ceases to be in the original state $|\phi\rangle$, while the third qubit gets projected onto $|\phi\rangle, X|\phi\rangle, Z|\phi\rangle, Y|\phi\rangle$, respectively.

[1] Steps 1+2 amount to performing a Bell measurement on the initial state, thus correlating the Bell states $00 \pm 11, 01 \pm 10$ of Alice's two qubits with the states of Bob's qubit. It suffices to note that

$$|\phi\rangle|\psi\rangle = \frac{1}{\sqrt{2}}|\phi\rangle(|00\rangle + |11\rangle) = \frac{1}{2\sqrt{2}}((|00\rangle + |11\rangle)|\phi\rangle +$$
$$(|01\rangle + |10\rangle)X|\phi\rangle + (|00\rangle - |11\rangle)Z|\phi\rangle + (|01\rangle - |10\rangle)Y|\phi\rangle).$$

*Step 3.* Once Bob receives the classical information sent by Alice, he just needs to apply on his qubit the corresponding gate I, $X, Z, Y$, in order to drive it to the desired state $|\phi\rangle$.

Notice that this teleportation sends an unknown quantum state from one place (where it vanishes) to another place (where it shows up) without really traversing the intermediate space. It does not violates causality, though. In the first part of the process, quantum correlations get established between the Bell states obtained by Alice and the associated states of Bob's qubit. In the remaining part to conclude the teleportation, information is transmitted by classical means, in the standard non-superluminal fashion. Notice also that in this "noncorporeal" process, it is the information about the quantum state, the qubit, and not the physical state itself, what gets passed from Alice to Bob. There has been no transportation whatsoever of matter, energy or information at a speed larger than the speed of light.

It is nevertheless surprising in the quantum teleportation that all the information needed to reproduce the state $|\phi\rangle = (\cos\frac{1}{2}\theta)|0\rangle + e^{i\phi}(\sin\frac{1}{2}\theta)|1\rangle$ (information that is infinite for it requires to fix a point $(\theta, \phi)$ on the Bloch sphere with infinite precision, thus requiring infinitely many qubits), can be accomplished with only 2 cbits, provided an EPR state is shared. This state, by itself, only generates potentially an infinite number of random and correlated bit pairs.

Quantum teleportation was realized experimentally with photons for the first time in two laboratories [50, 51]. This is at least what these authors claim, although several criticisms have been raised [48-50] (see however [55, 56]). In the experiment by the Roma group [51], the initial state to be teleported from Alice to Bob was a photon polarization, but not an arbitrary one, because it coincided with that of the Alice's photon in the shared EPR photon pair. In the experiment by the Innsbruck group [50], however, the teleported state was arbitrary. Teleportation was reached with a high fidelity of $0.80 \pm 0.05$,[2] but with a reduced efficiency (a 25% of cases). Teleportation has also been realized for states which are parts of entangled states [57].

It is also worthwhile to mention that quantum teleportation of states of infinite dimensional systems, namely, the teleportation of coherent optical states leaning on pairs of EPR squeezed states, which is theoretically proposed by Braunstein ana Kimble [58], is realized experimentally by Furasawa *et al.* [59]. In this experiment, whose fidelity is $0.58 \pm 0.02$ (higher than the maximum $\frac{1}{2}$ expected without resorting to entanglement), a third party, the *verifier* Victor, supplies Alice with one state that is known to him, but not to her. After teleporting that state from Alice to Bob, Victor verifies on the output if Bob's state is similar to the one he provided to Alice. In this sense, this experiment is different from all the others, and led the authors to claim priority in the realization of teleportation.

It does not seem to be easy to implement the theoretical protocol with a 100% effectiveness. The Bell operator (which distinguishes among the four Bell states of 2 qubits) cannot be measured unless both qubits interact appreciably with each other (as it occurs with the CNOT gate used in the protocol explained above), something which is very hard to achieve with photons. However, with atoms in EM cavities the hopes are high.

[2] This fidelity overcomes the value $\frac{2}{3}$ corresponding to the case in which Alice measures her qubit and communicates the result to Bob classically.

Perhaps the most realistic application of quantum teleportation outside of pure physics research is in the field of quantum computation. In fact, quantum teleportation, which doubtlessly will be extended to entangled states from different kinds of systems (photons and atoms, ions and phonons, etc.), might have, in the future, remarkable applications for quantum computers and in computer networks (for example, combined with prior distillation of good EPR pairs), as well as in the production of quantum memory records by means of teleportation of information on systems such as photons to other systems as trapped and well-isolated ions in cavities [50, 60]. A quantum computer can work on superposition of many different inputs at once. For instance, it can run at algorithm simultaneously on one million inputs, using only as many qubits as a conventional computer would need bits to run the algorithm once on a single input. Theorists have proved that some algorithms running on quantum computers can solve certain problems faster (i.e., in fewer computational steps) than any known algorithm running on a classical computer [40, 41, 61]. The problems include, for example, factoring large numbers, which is of great interest for breaking secret codes, and searching for items in a database. So far only the most basic elements of quantum computers have been built: logic gates that can process one or two qubits [62, 63]. The realization of even a small scale quantum computer is still far away. A key problem is transferring quantum data reliably between different logic gates or processors, whether within a single quantum computer or across quantum networks. Quantum teleportation is one solution. In fact, Gottesman and Chuang [64] recently proved that a general quantum computer can be built out of three basic components: entangled particles, quantum teleporters, and gates that operate on a single qubit at a time. This result provides a systematic way to construct two-qubit gates. The trick of building a two-qubit gate from a teleporter is to teleport two qubits from the gate's input to its output, using carefully modified entangled pairs. The entangled pairs are modified in just such a way that the gate's output receives the appropriately processed qubits. Performing quantum logic on two unknown qubits is thus reduced to the tasks of preparing specific predefined entangled states and teleporting. Meantime, the complete Bell state measurement needed to teleport 100% efficiency is itself a type of two-qubit processing.

Teleportation of complicate objects can be also considered as an interesting and dreamy subject which one can think about [65]. Can we really hope to teleport complicated objects? It seems that, concerning this, there are many severe obstacles. For instance, the object has to be in a pure quantum state, and such states are very fragile. Photons do not interact with environment much, so our experiments can be done in the open space, but experiments with atoms and larger objects must be done in a isolated vacuum to avoid interactions with gas molecules and environments. In addition, the larger an object becomes, the easier it is to disturb its quantum state-even by thermal radiation from the walls of the apparatus. This is why we do not routinely see quantum effects in our everyday world. Quantum interference, an easier effect to produce than entanglement or teleportation, has been demonstrated with buckyballs, sphere made of 60 carbon atoms. Such work will proceed to larger objects, perhaps even small viruses, but we might not expect to continue this for much larger objects. Other problems is due to finding proper procedures for teleporting all desired kinds of objects. Foe example, it has not yet been proposed a realizable scheme to make possible teleportation of the spatial wave function of an object-even a particle. Also, the Bell state measurement is an obstacle. In addition to some experimental difficulties relating to this kind of measurement, what would it mean to do a Bell state measurement of, e.g., a virus consisting of about $10^7$ atoms? Clearly it is a serious problem to extract the $10^8$ bits of information that such a measurement would generate. Meantime, for an object of just a few grams the numbers become apparently impossible: $10^{24}$ bits of data. Teleportation of more complicated objects particularly a person with $10^{32}$ bits of information can be considered as an exciting but perhaps a fictitious subject. Being in the same quantum state does not seem necessary for being the same person. It seems that we change our quantum state all the time and remain the same people. Conversely, identical twins or biological clones are not the same people, because they have different memories. So may be Heisenberg uncertainty as well as quantum no-cloning theorem do not prohibit us from replicating a person precisely enough for him to think he was the same as the original. Therefore, it seems that the teleportation of a person does not require quantum accuracy, and provides a surprising procedure for long travel, while the time for the person is frozen.

In the part two of this dissertation, we have studied quantum dense coding in a spatial scheme resulting in more efficiency than some other well-known protocols, and also we have proposed a theoretical scheme which can provide position state teleportation of an object. To do this, once again we have considered an entangled source between two parties which works very similar to that is used in our two double-slit scheme.

# 5. QUANTUM DENSE CODING BY SPATIAL STATE ENTANGLEMENT

## *5.1 Introduction*

The quantum entanglement property is providing new methods of information transfer, in some cases much more powerful than their classical counterparts. In quantum information theory, entanglement, as a key concept is used for a wide range of applications, such as quantum dense coding [42], teleportation [43], secret sharing [66] and key distribution [44]. To find more about the mentioned topics and the efforts done on them, both theoretically and experimentally, one can refer to [45, 67] and references therein.

Quantum dense coding protocol was proposed originally by Bennett and Wiesner [42] in 1992. The protocol describes a way to transmit two bits of classical information through manipulation of only one of the entangled pair of spin-$\frac{1}{2}$ particles, while each of the pair individually could carry only one bit of classical information. The first experimental realization of dense coding has been reported by Mattle *et al.* [48] in 1996.

There have been attempts to generalize dense coding protocol to achieve higher channel capacity. The first proposition in this regard was due to the original reference of dense coding [42] by using a pair of $n$-state particles prepared in a completely entangled state (instead of an EPR spin pair in a singlet state) to encode $n^2$ values. In practice, there might be some limitations for finding $n$-state particles with high $n$'s and controlling them. Very recently, some progress has been made in this direction [68], but it is still worthy, both theoretically and experimentally, to examine other alternatives for achieving this goal. For example, Bose *et al.* [69] studied $\mathsf{N}$ pairwise entangled states in which each party gets one particle except Bob with $\mathsf{N}$ qubits. Also in a more efficient scheme, they considered $\mathsf{N}+1$ parties sharing maximally entangled qubits in a way that each party, including Bob, possesses one qubit. Moreover, some recent attempts on generalization of quantum dense coding can be found in [66-69]. In an elegant alternative, Vaidman [74] proposed a method for utilizing canonical continuous variables $x$ (position variable) and $p$ (linear momentum variable) to perform a quantum communication. In this regard, Braunstein and Kimble [75] presented a typical realization for continuous variable dense coding using the quadrature amplitudes of the electromagnetic fields in which the mean photon number in each channel should be considered very large. Recently, in another way, some effort has been also done on experimental realization of continuous variables dense coding [76].

Here, we have presented a theoretical extended version of dense coding protocol using entangled position state of two particles shared between two parties. In this regard, all necessary Bell states and their corresponding unitary operators are presented to encode and decode information [77]. This version, at large $N$'s, can be considered as a conceivable scheme for Vaidman's idea [74] except that we have considered just the position variable (not both canonical variables $x$ and $p$) for communication. Finally, the efficiency of our scheme is compared with some other known ones.

## 5.2 Description of the dense coding set-up

Consider an original EPR source [20] which emits a pair of identical (fermionic or bosonic) particles with vanishing total linear momentum, isotropically. Many examples similar to this process can be found in different branches of physics [78]. For instance, the molecule NO can decay from an excited state to a state of two free atoms N and O which propagate in opposite directions, and in another example, the $\Lambda$ hyperon decays into a proton and a negatively charged $\pi$ meson. In our scheme, the source $S$ is placed exactly in the middle of the two parties, "Alice" and "Bob", where each one has an array of receivers aligned on a vertical line. The receivers just *receive* the particles and do not perform any destructive measurement. Total number of the receivers of each party is considered to be $2N$. Since Alice and Bob are assumed to be very far from each other and the source is considered isotropic, there is an equal probability for every receiver to obtain one of the emitted particles. We label the receivers placed at the upper (lower) part of the $x$-axis with positive (negative) integers; 1, 2, ..., $N$ (-1, -2, ..., $-N$). Figure 5.1 shows an illustration of this scheme. Now, the position state of the *system* (the arrays + the source) can be written in the form

$$|\psi_{1,2}\rangle = \frac{1}{\sqrt{2N}} \sum_{n=1}^{N} [|n, -n\rangle \pm |-n, n\rangle] \tag{5.1}$$

where the subscripts 1 and 2 are related to $\pm$ signs, respectively, and $n$ refers to label of the receivers. In addition, the order in writing the state is according to the direct product of Alice's state and Bob's. The signs $\pm$ in $|\psi_{1,2}\rangle$ indicate the symmetry and anti-symmetry property of the position state with respect to the exchange of particles. In the following, without loss of generality, we assume bosonic property for our system, that is, $|\psi_1\rangle$ represents the initial state of the system.

## 5.3 A representation for Bell states

In general, the entangled state (5.1) can be considered as a member of a larger family, that is,

$$|\psi_{(1,j)}\rangle = \frac{1}{\sqrt{2N}} \sum_{n=1}^{N} [h_{j,2n-1}|n, -n\rangle + h_{j,2n}|-n, n\rangle]$$

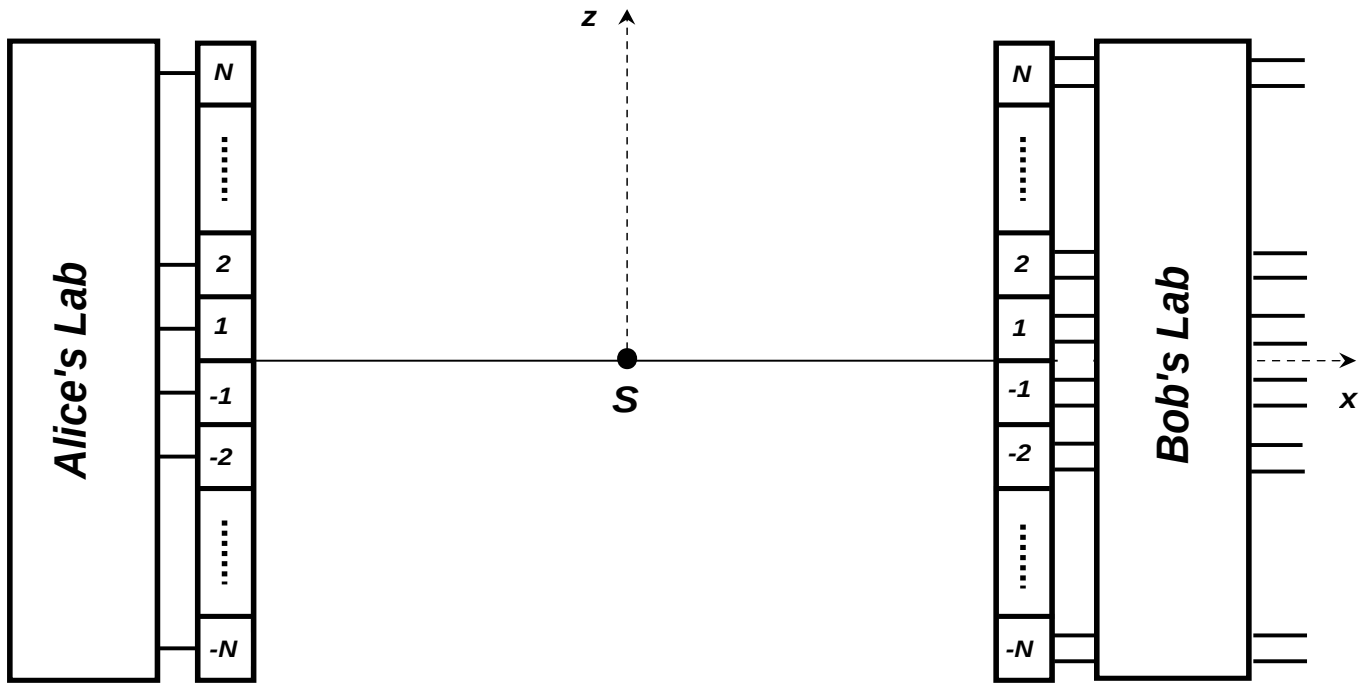

*Fig. 5.1:* Schematic drawing of the proposed dense coding scheme. An original EPR source is located between two array of receivers. The $2N$ receivers of each party are connected to its corresponding lab by quantum channels. The main quantum channels can be connected together in each lab using side quantum channels. Alice's lab is equipped with $O_{(k^{\pm},j)}$ encoder unitary operators, and Bob's lab contains a decoder unitary operator like $\mathcal{H}$. The Bell state measurement is completed by just a simple position measurement on Bob's outgoing channels.

$$1 \leq j \leq 2N \tag{5.2}$$

in which

$$h_{i,j} = \sqrt{2N}[\mathbf{H}]_{i,j} \tag{5.3}$$

where $\mathbf{H}$ is a $2N$-dimensional *normalized symmetric* Hadamard matrix which satisfies the property $\mathbf{H}^2 = I$. Furthermore, the above entangled states can be generalized to a more complete set of orthonormal and maximally entangled states, which can be defined as

$$|\psi_{(k^{\pm},j)}\rangle = \frac{1}{\sqrt{2N}} \sum_{n=1}^{N} [h_{j,2n-1}|n, f_{k^{\pm}}(n)\rangle + h_{j,2n}|-n, -f_{k^{\pm}}(n)\rangle]$$
$$1 \leq k \leq N, 1 \leq j \leq 2N \tag{5.4}$$

where $k^{\pm}$ and $j$ are called family and member indices, respectively, and

$$f_{k^{\pm}}(n) = \pm(n + k - 1)_{\mathrm{mod}(N)}. \tag{5.5}$$

If one uses the convention

$$(k^{\pm}, j) \longrightarrow 2N(k^{\pm} - 1) + j \tag{5.6}$$

in which

$$
\begin{aligned}
k^{+} &\equiv 2k \\
k^{-} &\equiv 2k-1
\end{aligned}
\tag{5.7}
$$

then the mentioned subscript $(k^{\pm}, j)$ will find a natural form, as we have shown in appendix B.

Now, the states (5.4) form *Bell bases* for our dense coding scheme. In fact, it is straightforward to check that these states have the orthonormal and maximally entanglement properties, as we have shown in the following

$$
\begin{aligned}
\langle\psi_{(k^{\pm},j)}|\psi_{(k'^{\pm},j')}\rangle &= \frac{1}{2N}\sum_{n,m=1}^{N}[h_{j,2n-1}h_{j',2m-1}\langle n, f_{k^{\pm}}(n)|m, f_{k'^{\pm}}(m)\rangle \\
&\quad +h_{j,2n-1}h_{j',2m}\langle n, f_{k^{\pm}}(n)|-m,-f_{k'^{\pm}}(m)\rangle \\
&\quad +h_{j,2n}h_{j',2m-1}\langle -n,-f_{k^{\pm}}(n)|m, f_{k'^{\pm}}(m)\rangle \\
&\quad +h_{j,2n}h_{j',2m}\langle -n,-f_{k^{\pm}}(n)|-m,-f_{k'^{\pm}}(m)\rangle] \\
&= \frac{1}{2N}\sum_{n=1}^{N}(h_{j,2n-1}h_{j',2n-1}+h_{j,2n}h_{j',2n})\delta_{k,k'} \\
&= \frac{1}{2N}\sum_{n=1}^{N}(h_{j,n}h_{j',n})\delta_{k,k'} \\
&= \delta_{j,j'}\delta_{k,k'}
\end{aligned}
\tag{5.8}
$$

which shows the orthonormal property of our Bell bases. Meantime, for the maximally entanglement property we can write

$$
\begin{aligned}
& \mathrm{tr}_2(|\psi_{(k^{\pm},j)}\rangle\langle\psi_{(k^{\pm},j)}|) \\
= \; & \mathrm{tr}_2(\sum_{n,m=1}^{N}\frac{1}{2N}[h_{j,2n-1}h_{j,2m-1}|n, f_{k^{\pm}}(n)\rangle\langle m, f_{k^{\pm}}(m)| \\
& +h_{j,2n-1}h_{j,2m}|n, f_{k^{\pm}}(n)\rangle\langle -m,-f_{k^{\pm}}(m)| \\
& +h_{j,2n}h_{j,2m-1}|-n,-f_{k^{\pm}}(n)\rangle\langle m, f_{k^{\pm}}(m)| \\
& +h_{j,2n}h_{j,2m}|-n,-f_{k^{\pm}}(n)\rangle\langle -m,-f_{k^{\pm}}(m)|]) \\
= \; & \frac{1}{2N}\sum_{n,m=1}^{N}[h_{j,2n-1}h_{j,2m-1}\delta_{n,m}|n\rangle\langle m| + h_{j,2n-1}h_{j,2m}\delta_{n,-m-2k+2}|n\rangle\langle -m| \\
& +h_{j,2n}h_{j,2m-1}\delta_{n,-m-2k+2}|-n\rangle\langle m| + h_{j,2n}h_{j,2m}\delta_{n,m}|-n\rangle\langle -m|] \\
= \; & \frac{1}{2N}\sum_{n,m=1}^{N}[h_{j,2n-1}h_{j,2m-1}|n\rangle\langle n| + h_{j,2n}h_{j,2n}|-n\rangle\langle -n|] \\
= \; & \frac{1}{2N}\sum_{n=-N(n\neq 0)}^{N}|n\rangle\langle n| = \frac{1}{2N}I_2.
\end{aligned}
\tag{5.9}
$$

Here it should be noted that, the order of a typical real Hadamard matrix can be 1, 2 and $4k$ where $k$ is a positive integer [79, 80]. Physically, it means that in other cases, i.e. for odd and half odd $N$ cases, one cannot make enough necessary entangled orthonormal states to perform an efficient dense coding. However, it is not a serious limitation in our proposed scheme.

## 5.4 Alice's encoding process

For the allowed $N$'s, Alice can find unitary encoding operators which transform the state describing the system in Eq. (5.1), $|\psi_1\rangle$, as a Bell state with $j=1$ in the first family into the others given in Eq. (5.4). Then, the representation of $4N^2$ suitable unitary operators for this task can be considered, for example, as

$$O_{(k^\pm,j)} = \sum_{n=1}^{N} [h_{j,2n-1}|n\rangle\langle f_{k^\pm}(n)| + h_{j,2n}|-n\rangle\langle -f_{k^\pm}(n)|] \quad (5.10)$$

which operate on Bell bases in Alice's lab as follows

$$O_{(k^\pm,j)}|\psi_{(k'^\pm,j')}\rangle = |\psi_{(k''^\pm,j'')}\rangle \quad (5.11)$$

in which

$$h_{j'',i} = h_{j,i}h_{j',i} \quad (5.12)$$

$$(5.13)$$

when $j'=1$, and

$$k''^\pm = (k+k'-1)^{ss'}_{\text{mod}(N)} \quad (5.14)$$

where $s$ and $s'$ are signs of superscripts of $k$ and $k'$, respectively. After the completion of Alice's encoding process, she sends her particle to Bob in a parallel line to the $x$-axis.

It is useful to see the act of $O_{(k^\pm,j)}$ operator on the Bell states in details. Thus, we have

$$\begin{aligned} O_{(k^\pm,j)}|\psi_{(k'^\pm,j')}\rangle \quad = \quad & \frac{1}{\sqrt{2N}} \sum_{n=1}^{N}\sum_{m=1}^{N} [h_{j,2n-1}h_{j',2m-1}\langle f_{k^\pm}(n)|m\rangle|n, f_{k'^\pm}(m)\rangle \\ & + h_{j,2n-1}h_{j',2m}\langle f_{k^\pm}(n)|-m\rangle|n, -f_{k'^\pm}(m)\rangle \\ & + h_{j,2n}h_{j',2m-1}\langle -f_{k^\pm}(n)|m\rangle|-n, f_{k'^\pm}(m)\rangle \\ & + h_{j,2n}h_{j',2m}\langle -f_{k^\pm}(n)|-m\rangle|-n, -f_{k'^\pm}(m)\rangle] \end{aligned} \quad (5.15)$$

which by using orthonormality condition of the states leads to

$$\begin{aligned} O_{(k^\pm,j)}|\psi_{(k'^\pm,j')}\rangle \quad = \quad & \frac{1}{\sqrt{2N}} \sum_{n=1}^{N}\sum_{m=1}^{N} [h_{j,2n-1}h_{j',2m-1}\delta_{\pm(n+k-1),m}|n, f_{k'^\pm}(m)\rangle \\ & + h_{j,2n-1}h_{j',2m}\delta_{\pm(n+k-1),-m}|n, -f_{k'^\pm}(m)\rangle \\ & + h_{j,2n}h_{j',2m-1}\delta_{\mp(n+k-1),m}|-n, f_{k'^\pm}(m)\rangle \\ & + h_{j,2n}h_{j',2m}\delta_{\mp(n+k-1),-m}|-n, -f_{k'^\pm}(m)\rangle]. \end{aligned} \quad (5.16)$$

Now, there are two $k^+$ and $k^-$ cases which should be considered. For the $k^+$ case, we obtain

$$O_{(k^+,j)}|\psi_{(k'^\pm,j')}\rangle = \\ h_{j,2n-1}h_{j',2m-1}|n, f_{k''^\pm}(n)\rangle + h_{j,2n}h_{j',2m}|-n, -f_{k''^\pm}(n)\rangle \quad (5.17)$$

in which $m = n + k - 1$ and $k'' \equiv k + k' - 1$. In a similar way, for the $k^-$ case we have

$$O_{(k^-,j)}|\psi_{(k'^\pm,j')}\rangle = \\ h_{j,2n-1}h_{j',2m}|n, f_{k''^\mp}(n)\rangle + h_{j,2n}h_{j',2m-1}|-n, -f_{k''^\mp}(n)\rangle \quad (5.18)$$

where again $m = n + k - 1$ and Eq. (5.14) is satisfied. Since Alice's encoding process is done on the initial state of the system with $j' = 1$, and also in a normalized symmetric Hadamard matrix $h_{1,i} = +1$, therefore, the act of the $O_{(k^\pm,j)}$ on the $|\psi_{(k'^\pm,j')}\rangle$ results in the Bell states introduced in Eq. (5.4) with $h_{j'',i} = h_{j,i}$ as well as Eq. (5.14).

## 5.5 Introducing basic gates and their realizability

It is a relevant question to ask about how to implement $O_{(k^\pm,j)}$ operators in practice. So here, we propose one way of performing this task by using some basic and conceivable operators. For example, similar to Pauli's operators in the spin-$\frac{1}{2}$ space, suitable basic unitary operators in the position space can be considered to be

$$N_n|\pm n\rangle = \pm|\pm n\rangle \quad (5.19)$$

$$P_n|\pm n\rangle = |\mp n\rangle \quad (5.20)$$

where the subscript $n$ means that the operator acts as a local gate on the $\pm n$-th channels. The latter operator, $P_n$, is defined to relate the two groups of receivers in the upper and lower halves of the $x$-axis. We also need to define a ladder operator $L_+$ to relate the receivers located on each half of the $x$-axis, so that

$$L_+|\pm n\rangle = |\pm(n+1)\rangle_{\mathrm{mod}(N)}. \quad (5.21)$$

Now, it is worthy to consider theoretical realization of the above mentioned operators in this protocol. The operation of $N_n$ gate can be easily conceived, for instance, by a kind of phase shifter, whose action can be represented as

$$N_n = e^{i\theta(\mp n)\pi} \quad (5.22)$$

where $\theta(n)$ is the conventional unit step function. To conceive this operator, one can consider a typical switch gate acting on the $-n$-th channel and its side channel, as is shown in Fig. 5.2. If $N_n$ turned ON (OFF), then the direct (side) channel is closed and the side (direct) channel is opened. It is clear that this set-up is required only for the channels having negative label.

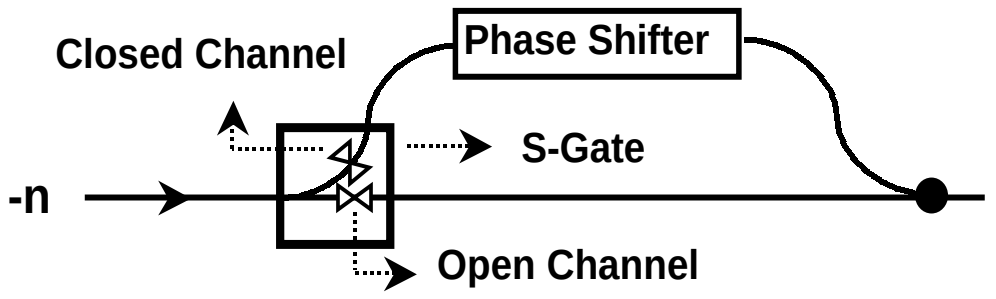

*Fig. 5.2:* Quantum circuit diagram of $N_n$ gate in an OFF state.

On the other hand, $P_n$ operator can be considered as a kind of swap gate between the channels $n$ and $-n$. So, to design this operation, one can use two local switch gates acting on these two channels. If $P_n$ turned ON (OFF) it means that direct (side) channels are closed and side (direct) channels are open (closed). Figure 5.3 schematically shows a $P_n$ gate which is constructed with two quantum switch gates and side channels.

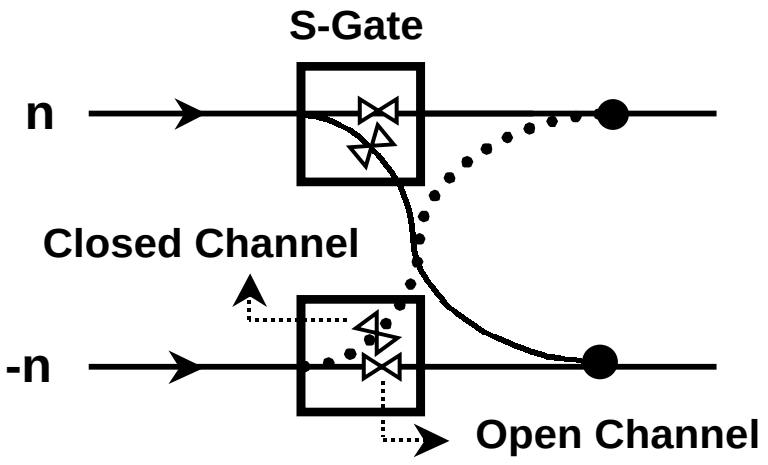

*Fig. 5.3:* Circuit diagram of $P_n$ gates in an OFF state.

In a similar manner, the action of the $L_+$ operations can be conceived using $2N(N-1)$ numbers of the switch gates and side quantum channels, as shown in Fig. 5.4. Therefore, our basic operators, i.e. $N_n$, $P_n$ and $L_+$, can be constructed using some quantum switch gates and side channels.

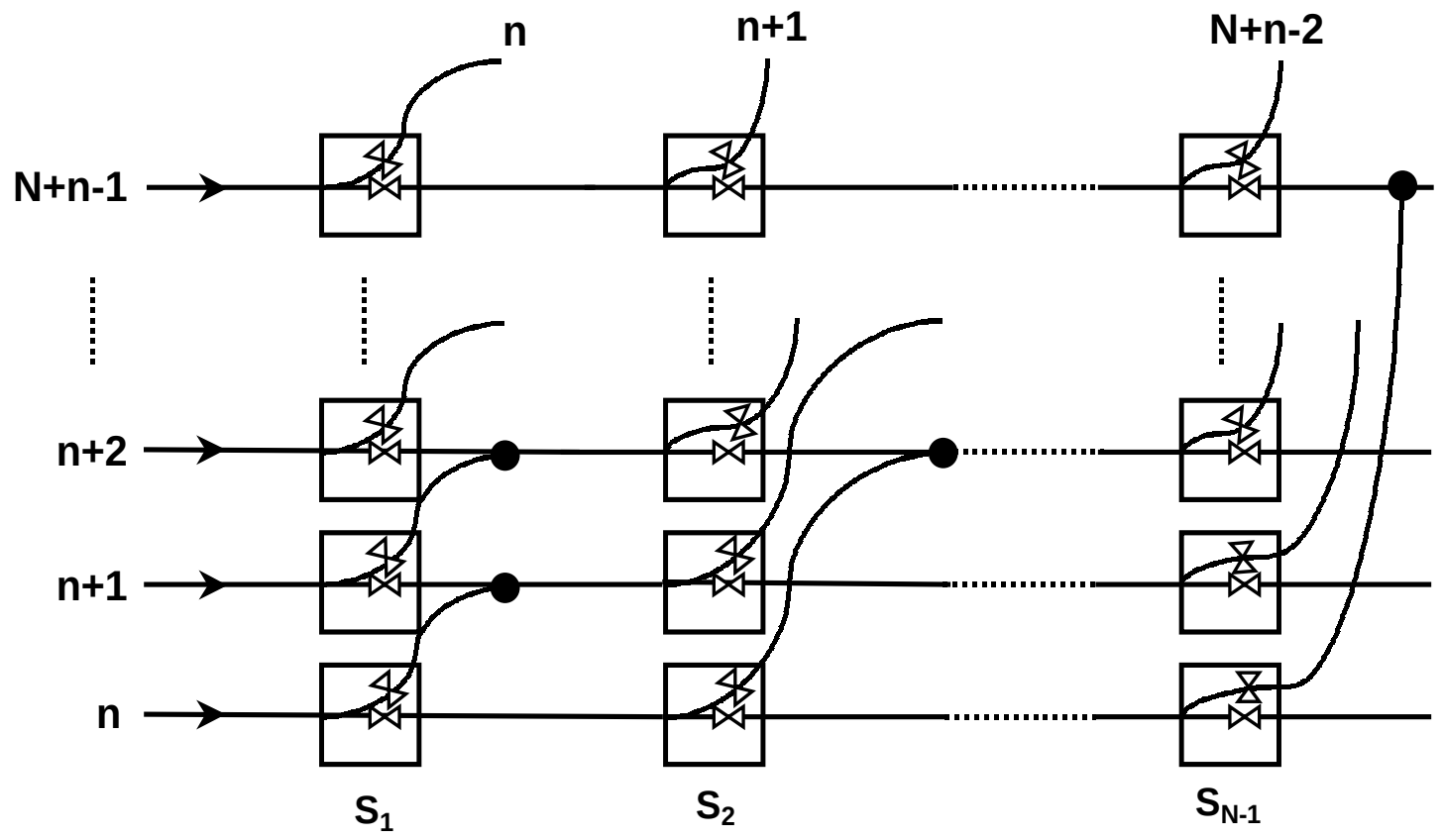

*Fig. 5.4:* Schematic application of $L_+$ operators, by using $2N(N-1)$ switch gates and side quantum channels, for $n > 0$. If $L_+^k$ is required, then the all $S_k$ gates must be turned ON.

As a proposal, if our introduced channels in the scheme are considered, for instance, superconductor wires containing an entangled current according to (5.1), then these basic operators can be realized by using switching process in the superconductor circuits. In fact, when a basic gate is OFF, its main channel(s) is (are) superconducting and the side channel(s) is (are) not superconducting, and conversely, if the gate is ON then the main channel(s) is (are) not superconducting and the suitable and corresponding side channel(s) is (are) superconducting. However, full realization of this proposition is left as an experimental challenge.

Now using the above basic gates, it is possible to show that one can construct operators which transform any member of a family to any other member of the other ones, that is, $O_{(k^\pm,j)}$. Here, our construction mechanism is to find a set of operators, for instance $O_j$, to transform the $j$ label of the Bell states of any given family. Next, we should introduce another set of operators, $F_{k^\pm}$, to transform $k$ label of the Bell states of any given member. Thus, our desired total operators have the form

$$O_{(k^\pm,j)} = F_{k^\pm} O_j. \tag{5.23}$$

The explicit form of the $O_j$ operators based on the basic gates can be constructed using a normalized symmetric Hadamard matrix as follows

$$O_j = \prod_{i=1}^{N} (P_i N_i P_i)^{(h_{a,2i-1}-h_{j,2i})/2} N_i^{(h_{a,2i}-h_{j,2i})/2}$$
$$1 \le j, a \le 2N \tag{5.24}$$

where $a$ is an arbitrary fixed positive integer. One can easily check that, the operator $O_j$ has the property

$$O_j|\psi_{(k^\pm,j')}\rangle = |\psi_{(k^\pm,j'')}\rangle \tag{5.25}$$

where $h_{j'',i}$ obeys the rule in Eq. (5.12) for every $j'$. In fact, in a detailed form we can write

$$\begin{aligned} O_j|\psi_{(k^\pm,j')}\rangle &= \frac{1}{\sqrt{2N}}\sum_{n=1}^{N}[h_{j',2n-1}(-1)^{(h_{a,2n-1}-h_{j,2n-1})/2}|n, f_{k^\pm}(n)\rangle \\ &\qquad +h_{j',2n}(-1)^{(h_{a,2n}-h_{j,2n})/2}|-n,-f_{k^\pm}(n)\rangle] \\ &= \frac{1}{\sqrt{2N}}\sum_{n=1}^{N}[h_{j',2n-1}h_{j,2n-1}|n, f_{k^\pm}(n)\rangle + h_{j',2n}h_{j,2n}|-n,-f_{k^\pm}(n)\rangle] \\ &= \frac{1}{\sqrt{2N}}\sum_{n=1}^{N}[h_{j'',2n-1}|n, f_{k^\pm}(n)\rangle + h_{j'',2n}|-n,-f_{k^\pm}(n)\rangle]. \end{aligned} \tag{5.26}$$

Furthermore, the explicit form for $F_{k^\pm}$ can be considered as

$$F_{k^\pm} = L_+^{(N-k+1)}\prod_{i=1}^{N}P_i^{(2k-k^\pm-1)} \tag{5.27}$$

in which $k^+ \equiv 2k$ and $k^- \equiv 2k-1$. It can be easily seen that they transform families to each other according to

$$F_{k^\pm}|\psi_{(k'^\pm,j')}\rangle = |\psi_{(k''^\pm,j')}\rangle \tag{5.28}$$

where $k''^\pm$ satisfies the rule in Eq. (5.14). Therefore, the obtained explicit form for $O_{(k^\pm,j)}$ operators is equivalent to their representation in Eq. (5.10) and (5.11). In appendix B, one can find examples concerning explicit forms of $O_{(k^\pm,j)}$ operators for some initial cases.

## 5.6 Bob's decoding process

It was seen that after completing the encoding process, Alice sends her particle to Bob. Thus, to each of Bob's receivers, two outgoing quantum channels should be considered for which we use the labels "control" and "target" channel (in brief; C and T-channel, respectively). Now we assume that, at first Bob opens all T-channels and closes all C-channels. Thus, he receives his own particle, which was devoted to him by the source, at one of the T-channels. Then, he sends it to a quantum storage media and closes all T-channels as well as opens all C-channels and waits to receive the processed particle which is sent to him by Alice. After synchronizing the two particles, Bob is ready to perform a Bell state measurement (BSM) on them. To do this, Bob needs to apply a *grand* unitary and Hermitian operator on all the channels in his lab which should be followed by a position state measurement on his outgoing channels. The representation of the grand operator can be considered as

$$\mathcal{H} = \sum_{j=1}^{2N} \sum_{k=1}^{N} |j, f'_{k^\pm}(j)\rangle\langle\psi'_{(k^\pm,j)}| \tag{5.29}$$

where

$$f'_{k^\pm}(j) = \pm(j+k-1)_{mod(2N)} \tag{5.30}$$

and

$$|\psi'_{(k^\pm,j)}\rangle = \frac{1}{\sqrt{2N}} \sum_{n=1}^{2N} h_{j,n}|j, f'_{k^\pm}(j)\rangle \tag{5.31}$$

which is a compact form of Eq. (5.4), and can be transformed to it by applying some local unitary operator. It is easy to check that the properties

$$\mathcal{H}|\psi'_{(k^\pm,j)}\rangle = |j, f'_{k^\pm}(j)\rangle \tag{5.32}$$

and also $\mathcal{H}^2 = I$ are satisfied. Here, to make the former property consistent with our channel labelling it suffices, just as an example, to use the following convention: if in a ket like $|i\rangle$ we have $1 \leq |i| \leq N$, then we should change $i$ to $-i$, and in the case of $N+1 \leq |i| \leq 2N$, to $|i| - N$. Moreover, in the following, we have shown that the grand unitary operator is also a reversible one.

$$\begin{aligned}
\mathcal{H}^2 \quad &= \sum_{j,j'=1}^{2N} \sum_{k,k'=1}^{N} |j, f'_{k^\pm}(j)\rangle\langle\psi'_{(k^\pm,j)}|j', f'_{k'^\pm}(j')\rangle\langle\psi'_{(k'^\pm,j')}| \\
&= \sum_{j,j'=1}^{2N} \sum_{k,k'=1}^{N} \sum_{n,n'=1}^{2N} h_{j,n}h_{j',n'}|j, f'_{k^\pm}(j)\rangle\langle n', f'_{k'^\pm}(n')|\delta_{j',n}\delta_{k,k'} \\
&= \sum_{j,k,n,n'} h_{j,n}h_{n,n'}|j, f'_{k^\pm}(j)\rangle\langle n', f'_{k^\pm}(n')| \\
&= \sum_{j,k} |j, f'_{k^\pm}(j)\rangle\langle j, f'_{k^\pm}(j)| = 1.
\end{aligned} \tag{5.33}$$

It can be argued that any general quantum physical process is required to operate by "finite means", i.e., it is equipped only with the possibility of applying any operation of some finite fixed set of basic unitary operations [81]. Also, it is shown that various quite small collections of unitary operators (so called "universal sets" of operations) suffice to approximate any unitary operation on any number of qubits to arbitrary accuracy [82, 84]. Furthermore, in a recent work, Bremner *et al.* [85] showed that any unitary operation can be constructed using finite numbers of an arbitrary conditional operator. Therefore, the grand unitary operator $\mathcal{H}$ is constructible using some finite basic operators. In the following, as an example, a realizable theoretical scheme for the BSM will be presented, using our introduced basic gates.

## 5.7 A conceivable scheme for Bell state measurement

In this section, we want to find an explicit form for $\mathcal{H}$ operator based on the introduced basic operators. At first, Bob needs one kind of logic gates to entangle two particles. The first presentation and demonstration of these kinds of gates, called CNOT, was performed in 1995 by Barenco *et al.* [62] and Monroe *et al.* [63], respectively. Here, we have considered another similar gate as a non-local operator which acts conditionally on the $|l,m\rangle$ state as

$$\mathrm{PCS}|l,m\rangle = \theta(-l)(I\otimes P_m)|l,m\rangle + \theta(l)|l,m\rangle \tag{5.34}$$

where PCS stands for position controlled swap operator and $\theta(l)$ is the conventional unit step function.

To understand the action of this operator there is a way using four typical spin-$\frac{1}{2}$ CNOT gates. Figure 5.5 shows a theoretical sketch of this construction. The $\overline{\mathrm{CNOT}}$ is somehow a complement of the usual CNOT in the sense that

$$\begin{aligned}\mathrm{CNOT}|x,y\rangle &= |x,x+y\rangle_{\mathrm{mod}(2)}\\ \overline{\mathrm{CNOT}}|x,y\rangle &= |x,\overline{x}+y\rangle_{\mathrm{mod}(2).}\end{aligned} \tag{5.35}$$

If a spin-$\frac{1}{2}$ particle passes through the control channel then the two first CNOT and $\overline{\mathrm{CNOT}}$ definitely change spin of an intermediate spin-$\frac{1}{2}$ particle, named spin target, which is preset to a fixed state , e.g., $|0\rangle$. It is assumed that any change in the state describing the spin target means switching the swap gate for the channels $m$ and $-m$. The two latter CNOT and $\overline{\mathrm{CNOT}}$ are just to reset the spin target. To show this PCS gate in our quantum circuits, we adopt a simplified representation for it, as depicted in Fig. 5.6. Our PCS gate can be realized using, for example, superconductor circuits. It is obvious that any change in the spin target results into a change in the magnetic flux passing through a superconductor circuit. For switching the swap gate the induced current can be amplified by a superconducting LC circuit in order to produce at least a critical magnetic field to set the direct (side) channel not superconducting (superconducting).

On the other hand, the same as the spin-$\frac{1}{2}$ case, Bob can use Hadamard operators in the position space with the form

$$H_{x_n} = \frac{1}{\sqrt{2}}(P_n + N_n) \tag{5.36}$$

where $H_{x_n}$ is a local operator acting on the $\pm n$-th channels. Although the position Hadamard operator has not a multiplicative form, but its design is relatively straightforward in our scheme. Figure 5.7 shows a schematic circuit for the Hadamard gate acting on the $n$-th channel in which we have assumed that the probability of passing a particle through each basic gate is the same. In fact, in our scheme as an advantage, it is not necessary to find a multiplicative form to design such operators using the basic ones.

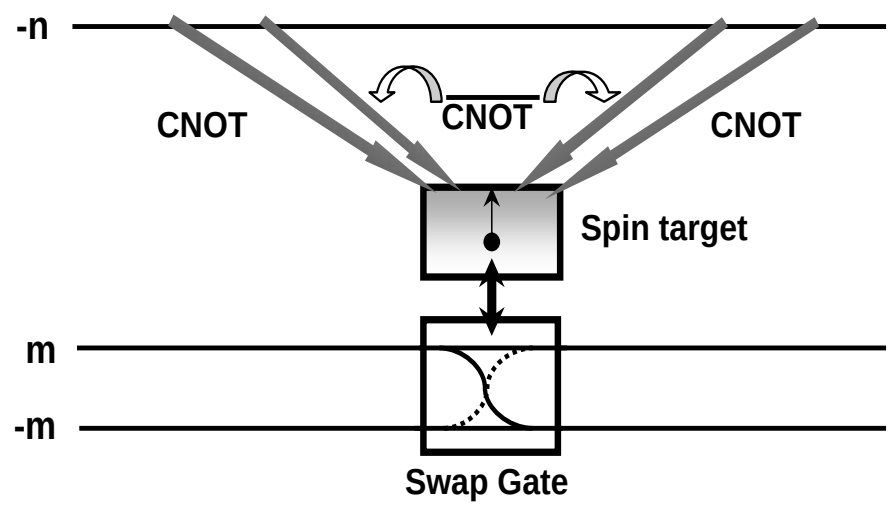

*Fig. 5.5:* Schematic diagram of a position controlled-swap gate.

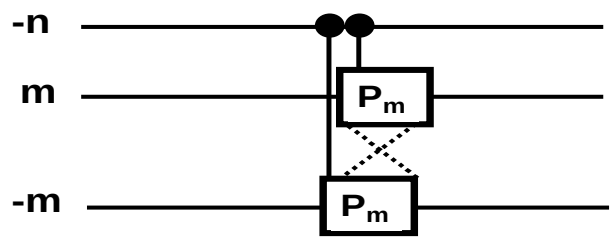

*Fig. 5.6:* Schematic representation of a PCS gate. For simplicity, we omit dotted cross lines in our quantum circuits.

For $N \geq 2$ cases, however, applying just PCS and $H_{x_n}$ operators does not produce pure position states in a disentangled and measurable form. So, we should introduce a non-local unitary operator for this mean which acts like

$$U_{(N)}|\pm l, \pm m\rangle = \frac{1}{\sqrt{N}} \sum_{n=1}^{N} h_{m,m+n-1} |f_{l^{\pm}}(n), f_{m^{\pm}}(n)\rangle_{\mathrm{mod}(N)} \tag{5.37}$$

where $l$ and $m$ are positive integers. Moreover, here, $h_{m,n}$ is an element of an $N$-dimensional normalized symmetric Hadamard matrix. It is straightforward to check that $U^2_{(N)} = I$, as we have shown in the following

$$U_{(N)}U_{(N)}|\pm l, \pm m\rangle =$$

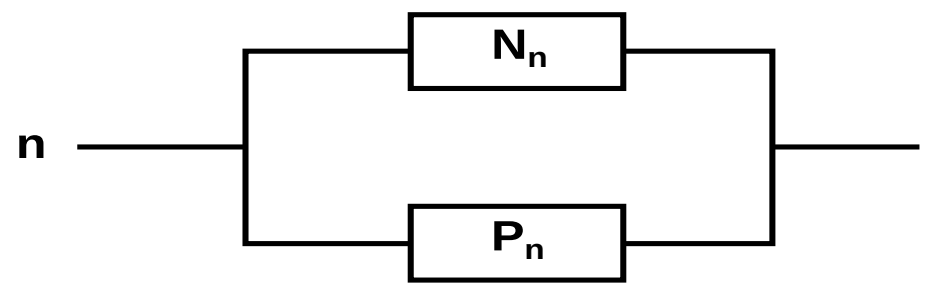

*Fig. 5.7:* Quantum circuit diagram of $H_{x_n}$ gate.

$$\frac{1}{N}\sum_{n=1}^{N}\sum_{k=1}^{N} h_{m,m+n-1}h_{m+n-1,m+n+k-2}|f_{l^\pm}(n+k-1), f_{m^\pm}(n+k-1)\rangle_{\mathrm{mod}(N)}$$
$$= \frac{1}{N}\sum_{a}\sum_{k} h_{m,a-k+1}h_{a-k+1,a}|\pm(l+a-m), \pm a\rangle$$
$$= \sum_{a}\delta_{m,a}|\pm(l+a-m), \pm a\rangle = |\pm l, \pm m\rangle. \quad (5.38)$$

The same as $O_{(k^\pm,j)}$ operator, it is also possible to find an explicit form for $U_{(N)}$ operator based on the basic gates. Here, for example, we have considered it as

$$U_{(N)} = \frac{1}{\sqrt{N}}\sum_{n=0}^{N/2-1}\{L_+^{2n}\otimes[\prod_{i=1}^{N}P_i^{(h_{b,i}-h_{2n+i,i})/2}N_iP_i^{(h_{b,i}-h_{N-2n+i,i})/2}]L_+^{2n}$$
$$+L_+^{2n+1}\otimes[\prod_{i=1}^{N}(P_iN_iP_iN_i)^{(h_{b,i}-h_{N-2n-1+i,i})/2}]L_+^{2n+1}\} \quad (5.39)$$

where $h_{i,j}$ is an element of an $N$-dimensional real symmetric Hadamard matrix. In addition, the subscript $b$ is an arbitrary constant positive integer. This relatively complex form results from the generalization procedure introduced in appendix B. If we represent $U_{(N)}$ operator in the form

$$U_{(N)} = \frac{1}{\sqrt{N}}\sum_{n=0}^{N-1}(L_+^n\otimes A_{(n+1)}L_+^n) \quad (5.40)$$

then, in a similar manner used to design $H_{x_n}$ operator, it will be possible to conceive $U_{(N)}$ operator using the local operator $A_{(n+1)}$ which can be designed by the basic gates, and a nonlocal operator working like PCS gate, as is shown in Fig. 5.7. Here, we have assumed that each of Bob's channels is divided into $N$ identical branches, so that the probability of entering the particle into each branch is the same. Finally, each branch is joined to its correspondence channels in Bob's lab. It should be noted that, the mechanism of the required conditional gate in this quantum circuit is the same as PCS gate, but it should turn on the $S_k$ gates (instead of $P_n$ gate) to operate $L_+^k$ on Alice's particle.

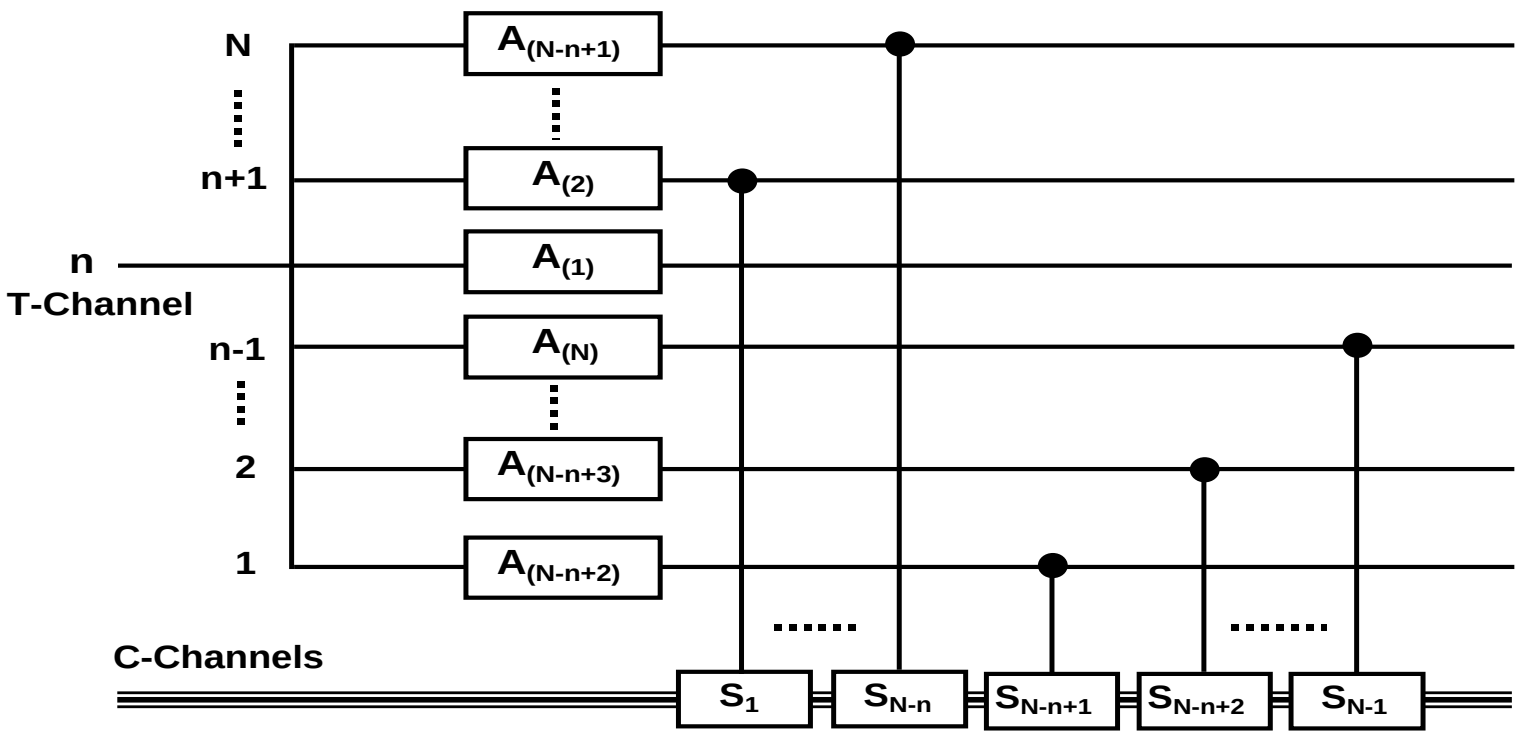

*Fig. 5.8:* Quantum circuit diagram of $U_{(N)}$ operator for the $n$-th T-channel. Here, each branch of the T-channel acts as a control channel for the $S_k$ gates applied on the C-channels.

Now, using the above mentioned operators, Bob can perform his BSM in this way

$$U_{(N)}(H_{x_1}H_{x_2}\dots H_{x_n}\otimes I)\mathrm{PCS}|\psi_{(k^\pm,j)}\rangle = |m,n\rangle \tag{5.41}$$

which should be followed by a position measurement on the outgoing channels. Figure 5.9 shows a schematic quantum circuit of Bob's lab. In Eq. (5.41), $m$ and $n$ are some unique functions of $k^\pm$ and $j$. In addition, action of the set of operators in this equation is equivalent to the action of the grand operator (5.29).

## 5.8 The rate of classical information gain

We have seen that, there are $4N^2$ Bell basis and the same number of different unitary operators $O_{(k^\pm,j)}$ for encoding in our dense coding scheme. This obviously corresponds to encoding $4N^2$ different messages by Alice. Thus, she can send $2\ \log_2(2N)$ bits of classical information per particle to Bob. Now, he needs $N$ PCS, $2N$ Hadamard and one $U_{(N)}$ gates to read out the sent classical information during the BSM. Since Bob performs one BSM on just two particles, it is possible to consider that all PCS and then all Hadamard gates operate in a parallel form, i.e. *concurrently.* If the operation times for the PCS, Hadamard and $U_{(N)}$ gates are $t_p$, $t_h$ and $t_u$, respectively, the rate of classical information gain R, defined as sent classical bits of information per unit time and sent particle, is

$$R_x = \frac{2\ \log_2(2N)}{t_p + t_h + t_u}. \tag{5.42}$$

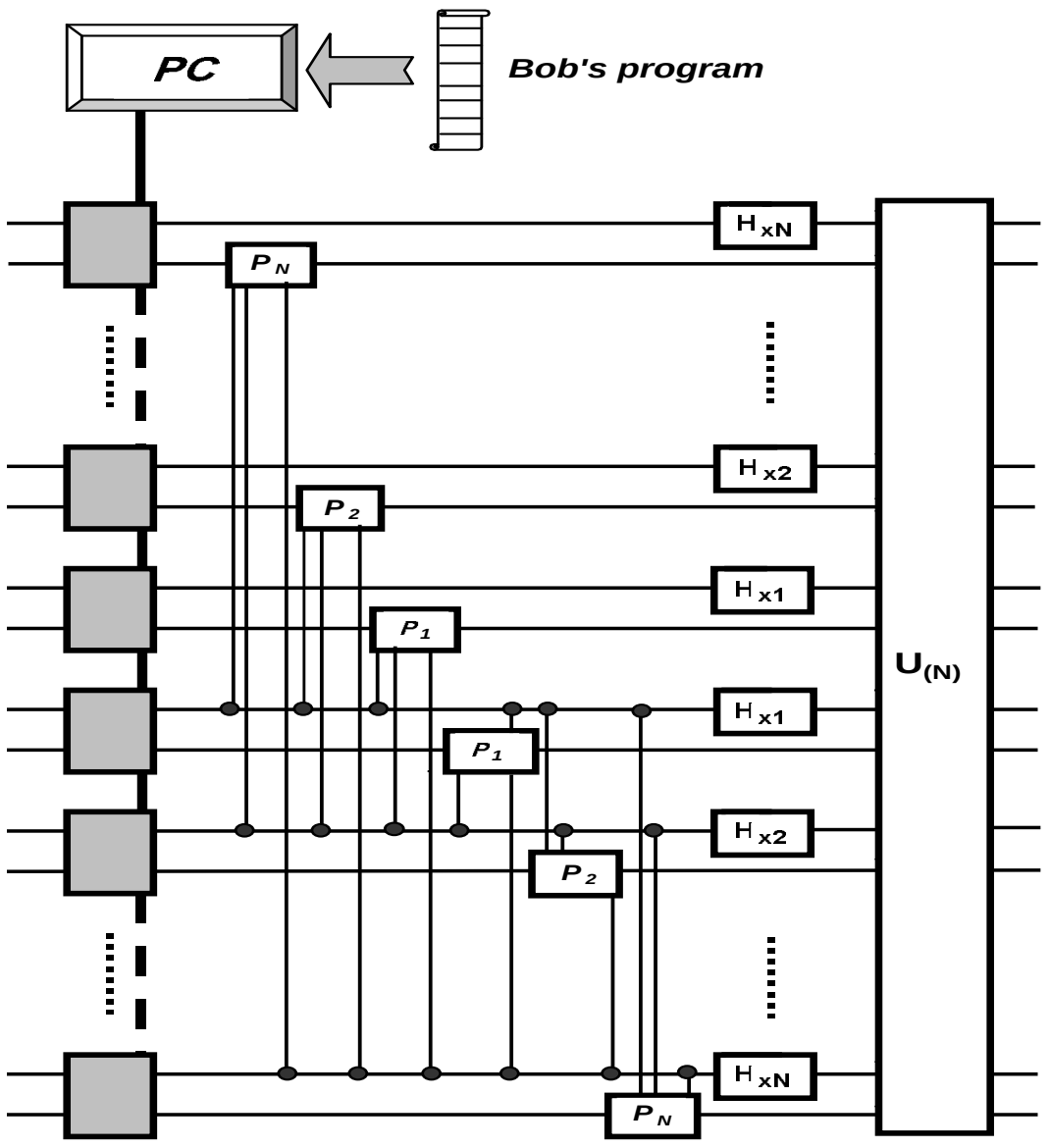

*Fig. 5.9:* Some details of Bob's laboratory. Bold lines, between gray boxes and PC show classical channels (i.e., ordinary wires), and the others are quantum ones. Gray boxes represent quantum memories which our programmed PC controls their whole performance, including opening and closing C and T-channels on time and the synchronization of the two particles. The connection points between quantum channels are shown by the bold points.

In a similar way, one can calculate R for a dense coding protocol which works using N pairwise entangled qubits and/or N maximally entangled qubits shared between two parties [69, 86]. In the pairwise entangled case, $2\mathsf{N}$ classical bits of information are transferred from Alice to Bob. Meantime, in this case, $\mathsf{N}$ separate CNOT and Hadamard gates are required by Bob to decode the Bell states. Thus, its R in terms of bits per unit time per sent particle is

$$R_p = \frac{2\mathsf{N}}{\mathsf{N}^2(t_c + t_h)} \tag{5.43}$$

where $t_c$ is the operation time of a CNOT gate. Recently, however, Lee *et al.* [87] assert that the rate of information gain for this case is as high as $2^N/N(t_c + t_h)$ bits per unit time. Concerning this, we have shown in appendix C that their claims are unfounded. On the other hand, for the maximally entangled case, the number of sent classical bits is $\mathsf{N}$. In this case, Bob needs $(\mathsf{N}-1)$ successive CNOT and one Hadamard gates so that he gains

$$R_m = \frac{\mathsf{N}}{(\mathsf{N}-1)[(\mathsf{N}-1)t_c + t_h]} \tag{5.44}$$

bits per time and particle. Now, if we assume that both $N$ and $\mathsf{N}$ are very large and all basic gates operate in an equal time interval, i.e. $t_c \sim t_h \sim t_p/4 \sim t_u/N \equiv t$, then $R_x = 2\log_2(2N)/Nt$ and $R_p = R_m = 1/\mathsf{N}t$. Therefore, it can be easily seen that at large $N$'s, if $N = \mathsf{N}$ is considered, i.e. dimensions of Hilbert spaces are identical, then our protocol is more efficient than both the pairwise and the maximally entangled cases with a *logarithmic* factor.

Furthermore, other degrees of freedom such as spin, polarization and squeezation can be added to our protocol in order to obtain a more powerful quantum dense coding. Here, for instance, we consider that each particle can also have spin $S$. To introduce spin, it is sufficient to assume that the source emits entangled pair of particles not only with vanishing total momentum but also with zero total spin. Therefore, the quantum state of the system would be

$$\begin{aligned} |\psi_1\rangle_{xs} &= \frac{1}{\sqrt{2N(2S+1)}}\sum_{n=1}^{N}[|n,-n\rangle + |-n,n\rangle] \\ &\quad\times\sum_{s=0}^{2S}(\pm 1)^s|(S-s),-(S-s)\rangle \end{aligned} \tag{5.45}$$

which is simply a tensor product of position and spin states of the system. Therefore, now Alice is capable of sending $2\ \log_2[2N(2S+1)]$ bits per particle.

## 5.9 Conclusions

We have proposed a theoretical extended version of dense coding protocol by using entangled spatial states of the two particles shared between two parties. Our construction is based on using the well known real symmetric Hadamard matrix for representing orthogonal states, required encoding and decoding operators, and hence is subject to its intrinsic characteristics. Furthermore, we have given a typical proposition for initial realization of the scheme. In this regard, some basic gates as well as a position conditional gate have been introduced and it is shown that whole of the scheme can be established based upon them. By comparing our scheme with some of the previously proposed multi-qubit protocols, it is shown that the rate of classical information gain in our case is better than them by a logarithmic factor. Also we have shown that capability of considering internal degrees of freedom, like spin, strengthens the scheme in a straightforward manner.

# 6. A SCHEME TOWARDS COMPLETE STATE TELEPORTATION

## 6.1 Introduction

Teleportation scheme proposed by Bennett *et al.* [43] is a protocol for disembodied transmission of an *unknown* quantum state of a spin-$\frac{1}{2}$ particle by a sender, Alice, to a receiver, Bob, by conveying two classical bits of information. In addition, the successful experimental realizations of this protocol, particularly for polarization states of photons [50, 51], stimulated studying of teleportation protocol for more complex systems. For instance, extensions to $N$-dimensional Hilbert spaces recently have attracted much attentions. In [83-86] an $N$-level source and in [92, 93] some two level EPR sources have been utilized to accomplish $N$-level state teleportation. Using the above entangled versions, it is theoretically possible to teleport any system with discretizable Hilbert space.

But, what about teleportation of position states of a quantum *object*, $\psi(\mathbf{x})$? The theoretical answer to this question was first provided by Vaidman [74, 94], utilizing perfect entanglement between position and momentum which results in a scheme for teleportation of continuous variables, or the wave function $\psi(\mathbf{x})$. Interestingly, reliable teleportation of continuous variables is shown to be implementable in experiments. In fact, Braunstein and Kimble [58] made a *realistic* proposal for teleportation of quantum state of a single mode of the electromagnetic field. In other words, their scheme was a typical implementation of Vaidman's method. Then, experimental realization of Braunstein-Kimble method was performed by Furasawa *et al.* [59]. But, it is believed that this type of experiment cannot be expected to improve, due to difficulties in establishing highly squeezed light fields [53], and even in ideal experimental conditions, details of Bob's reconstruction process results in an *almost* perfect replica of light fields [45]. Also, as Vaidman pointed out, Braunstein-Kimble method is not applicable directly for teleporting $\psi(\mathbf{x})$, where $\mathbf{x}$ is the spatial position of a quantum object [94]. As a solution, Vaidman proposed a way to overcome this disability by introducing a quantum-quantum interaction which can convert the continuous variable of a real position to the electromagnetic field amplitude variable. However, other methods or alternatives might be possible to solve this problem.

In this chapter, we have presented an alternate theoretical scheme to achieve realization of teleporting wave function $\psi_S(\mathbf{x})$ of an object having spin. To do this, we have offered a scheme, including a special EPR source and some arrays of receivers to teleport a multi-level *entangled position-spin* state in our three-dimensional space [95]. Moreover, some theoretical details on Bell state measurement (BSM) and the reconstruction process have been illustrated in our novel spatial scheme.

## 6.2 Description of the teleportation set-up

Consider a theoretical set-up including two parties, Alice and Bob, who have labs which contain $2N$-receiver arrays, as shown in Fig. 6.1. An EPR source is placed in the center of the symmetry axis ($z$-axis) of the two receiver lines, which emits, *isotropically*, two identical *maximally entangled* particles with vanishing total momentum. The receivers are labelled by $\pm 1$, $\pm 2, \ldots$, $\pm N$. We assume that they do not perform any destructive measurements on the received particles. All lines that connect each receiver to the related lab are representatives of quantum channels. Each quantum object with an unknown state, as an initial import to Alice's lab, is sent, after quantum scanning, to her lab to be combined with her own particle. Then, she performs a BSM on her two particles. Based upon the result of this measurement, she communicates with Bob through a classical channel, sending related binary codes. Bob, after receipt of Alice's message, reconstructs the initial quantum state using his lab.

## 6.3 A representation for Bell bases

After this illustration, it is possible to explain more details about the scheme for teleportation of quantum position state of an object. At the first step, we study the teleportation of one-dimensional position state of a spinless particle. In this case, the total state describing the *system* (EPR source+the arrays) is

$$|\psi_{1(2)}\rangle_x = \frac{1}{\sqrt{2N}} \sum_{n=1}^{N} [|n, -n\rangle \pm |-n, n\rangle] \tag{6.1}$$

where $n$ refers to the label of the receivers. Additionally, we adopt the convention that in the tensor products the left (right) states belong to Alice (Bob). The signs $\pm$ in $|\psi_{1(2)}\rangle_x$, in which 1(2) indicates +(-), stand for symmetry and antisymmetry of the state with respect to the two-party exchange which simply means bosonic or fermionic property of our system. Here, without any loss of generality, we admit that our system is bosonic. In general, the entangled state (6.1) can be considered as a member of a larger family, that is,

$$|\psi_{(1,j)}\rangle_x = \frac{1}{\sqrt{2N}} \sum_{n=1}^{N} [h_{j,2n-1}|n, -n\rangle + h_{j,2n}|-n, n\rangle] \qquad 1 \leq j \leq 2N \tag{6.2}$$

in which $h_{i,j} = \sqrt{2N}[\mathbf{H}]_{i,j}$, and $\mathbf{H}$ is a *normalized symmetric* $2N \times 2N$ Hadamard matrix which satisfies the property $\mathbf{H}^2 = I$. Furthermore, the above states can

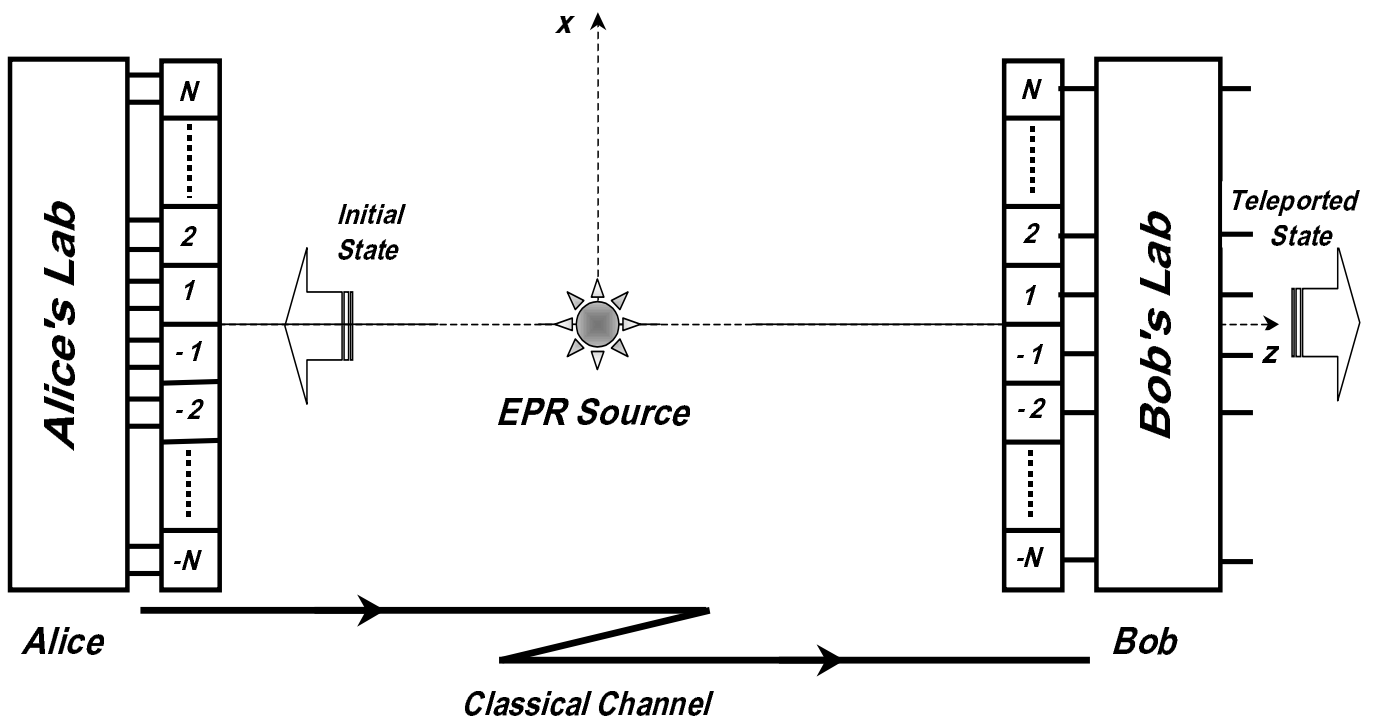

*Fig. 6.1:* A primary scheme for teleportation of a position state.

be generalized to a more complete set of orthonormal and maximally entangled states, which are defined as

$$|\psi_{(k^\pm,j)}\rangle_x = \frac{1}{\sqrt{2N}}\sum_{n=1}^{N}[h_{j,2n-1}|n, f_{k^\pm}(n)\rangle + h_{j,2n}|-n, -f_{k^\pm}(n)\rangle]$$
$$1 \le k \le N, 1 \le j \le 2N \tag{6.3}$$

where $j$ and $k^\pm$ are member and family indices, respectively, and we have also adopted the convention $f_{k^\pm}(n) = \pm(n+k-1)_{\mathrm{mod}(N)}$. By the way, the states (6.3) form *Bell bases* for our teleportation scheme. In fact, it can be simply checked that these states have the properties

$$\langle\psi_{(k^\pm,j)}|\psi_{(k'^\pm,j')}\rangle = \delta_{j,j'}\delta_{k,k'}$$
$$\mathrm{tr}_{1(2)}(|\psi_{(k^\pm,j)}\rangle\langle\psi_{(k^\pm,j)}|) = \tfrac{1}{2N}I_{1(2)}. \tag{6.4}$$

In addition, as known, the order of a real Hadamard matrix is 1, 2 and $4k$ with $k$ as a positive integer [79, 80]. Thus, in other conditions, there does not exist any related teleportation scheme. In other words, we cannot make all necessary entangled orthonormal states to perform teleportation for odd and half odd $N$ cases.

## 6.4 Unitary transformation of Bell bases

For the permitted $N$'s, one can find unitary operators which transform the initial state of the system into the other Bell states given in Eq. (6.3). The representation of suitable operator for this mean is

$$O'_{(k^\pm,j)} = \sum_{n=1}^{N}[h_{j,2n-1}|n\rangle\langle -f_{k^\pm}(n)| + h_{j,2n}|-n\rangle\langle f_{k^\pm}(n)|] \tag{6.5}$$

which acts on the Bell bases as

$$O'_{(k^{\pm},j)}|\psi_{(k'^{\pm},j')}\rangle_x = |\psi_{(k''^{\pm},j'')}\rangle_x \tag{6.6}$$

where

$$h_{j'',i} = h_{j,i}h_{j',i} \tag{6.7}$$

at $j' = 1$, and

$$k''^{\pm} = (k + k' - 1)^{-ss'}_{\mathrm{mod}(N)} \tag{6.8}$$

in which $s$ and $s'$ are signs of superscripts of $k$ and $k'$, respectively. In the following, it will be seen that the operator $O'_{(k^{\pm},j)}$ are applied by Bob to reconstruct the unknown wave functions which are teleported to him by Alice. Moreover, an example for realization of these operators can be considered the same as $O_{(k^{\pm},j)}$ mentioned for the dense coding scheme in the previous chapter.

Here, it is useful to have a review on the details of action of this unitary operator on the Bell bases. Thus, using orthonormality of the position bases, we have

$$\begin{aligned} O'_{(k^{\pm},j)}|\psi_{(k'^{\pm},j')}\rangle_x &= \frac{1}{\sqrt{2N}}\sum_{n,m=1}^{N}[h_{j,2n-1}h_{j',2m-1}\delta_{\mp(n+k-1),m}|n, f_{k'^{\pm}}(m)\rangle \\ &+h_{j,2n-1}h_{j',2m}\delta_{\mp(n+k-1),-m}|n, -f_{k'^{\pm}}(m)\rangle \\ &+h_{j,2n}h_{j',2m-1}\delta_{\pm(n+k-1),m}|-n, f_{k'^{\pm}}(m)\rangle \\ &+h_{j,2n}h_{j',2m}\delta_{\pm(n+k-1),-m}|-n, -f_{k'^{\pm}}(m)\rangle]. \end{aligned} \tag{6.9}$$

Now, there are two $k^+$ and $k^-$ cases which can be examined separately. For the $k^+$ case, we have

$$\begin{aligned} O'_{(k^{+},j)}|\psi_{(k'^{\pm},j')}\rangle_x &= \frac{1}{\sqrt{2N}}\sum_{n=1}^{N}[h_{j,2n-1}h_{j',2(n+k-1)}|n, f_{k''^{\mp}}(n)\rangle \\ &+h_{j,2n}h_{j',2(n+k-1)-1}|-n, -f_{k''^{\mp}}(n)\rangle] \end{aligned} \tag{6.10}$$

where $k'' = k + k' - 1$. Similarly, for the $k^-$ case we can write

$$\begin{aligned} O'_{(k^{-},j)}|\psi_{(k'^{\pm},j')}\rangle_x &= \frac{1}{\sqrt{2N}}\sum_{n=1}^{N}[h_{j,2n-1}h_{j',2(n+k-1)-1}|n, f_{k''^{\pm}}(n)\rangle \\ &+h_{j,2n}h_{j',2(n+k-1)}|-n, -f_{k''^{\pm}}(n)\rangle] \end{aligned} \tag{6.11}$$

where once again $k'' = k + k' - 1$. Therefore, it is clear now that for $j' = 1$ Eqs. (6.7) and (6.8) are satisfied.

## 6.5 General procedure for teleporting an object

In the one-dimensional scheme, the most general position states which Alice is able to teleport is

$$|\phi\rangle_x = \sum_{n=1}^{N}[a_n|n\rangle + a_{-n}|-n\rangle]. \tag{6.12}$$

Concisely, the mapping of the object's wave function to this state can be conceived by a typical scanning, for instance, using two approaching arrays of receivers sweeping the object. After this scanning, Alice has to combine her own particle with the given one. By rewriting the total state of the three particles based on our proposed Bell bases, she must perform a BSM on her two possessed particles and then send the classical obtained results, which encode the retrieving process, to Bob as the following

$$\begin{aligned} |\phi\rangle_x|\psi_1\rangle_x \quad &= \frac{1}{\sqrt{2N}}\sum_{n=-N(n\neq 0)}^{N}\sum_{m=1}^{N} a_n[|n,m,-m\rangle + |n,-m,m\rangle] \\ &= \frac{1}{2N}\sum_{k=1}^{N}\sum_{j=1}^{2N}|\psi_{(k^\pm,j)}\rangle_x O'^{\dagger}_{(k^\pm,j)}|\phi\rangle_x \\ &\overset{\text{Alice's}}{\underset{\text{BSM}}{\longrightarrow}} \; |\psi_{(k^\pm,j)}\rangle_x O'^{\dagger}_{(k^\pm,j)}|\phi\rangle_x \end{aligned} \tag{6.13}$$

where

$$O'^{\dagger}_{(k^\pm,j)} = \sum_{n=1}^{N}[h_{j,2n-1}|-f_{k^\pm}(n)\rangle\langle n| + h_{j,2n}|f_{k^\pm}(n)\rangle\langle -n|]. \tag{6.14}$$

Now, it suffices for Bob to know the classical information identifying the corresponding operator, i.e. $(k^\pm, j)$, to reconstruct the initial state $|\phi\rangle_x$ using a suitable $O'_{(k^\pm,j)}$ operator. Because this operator is a unitary operator, as we can see in the following

$$\begin{aligned} O'_{(k^\pm,j)}O'^{\dagger}_{(k^\pm,j)} \quad &= \quad \sum_{n,m=1}^{N}[h_{j,2n-1}h_{j,2m-1}\delta_{n,m}|n\rangle\langle m| \\ &\quad +h_{j,2n-1}h_{j,2m}\delta_{n,-m-2k+2}|n\rangle\langle -m| \\ &\quad +h_{j,2n}h_{j,2m-1}\delta_{n,-m-2k+2}|-n\rangle\langle m| \\ &\quad +h_{j,2n}h_{j,2m}\delta_{n,m}|-n\rangle\langle -m|] \\ &= \quad \sum_{n=1}^{N}|n\rangle\langle n| + |-n\rangle\langle -n| = I. \end{aligned} \tag{6.15}$$

Here, it should be pointed out that in the original scheme for the continuous variables teleportation [74, 94], Alice needs to apply two separate kinds of measurements on the position and momentum to accomplish a BSM for teleportation of $\psi(x)$. But, in our proposed scheme, Alice needs to perform just one kind of BSM on the position variable.

It is straightforward to check the rewriting done in the second line of Eq. (6.13). To do this, we first consider

$$\begin{aligned} O'^{\dagger}_{(k^{\pm},j)}|\phi\rangle_x &= \sum_{n,m=1}^{N}[h_{j,2n-1}a_m\delta_{n,m}|-f_{k^{\pm}}(n)\rangle + h_{j,2n-1}a_{-m}\delta_{n,-m}|-f_{k^{\pm}}(n)\rangle \\ &\quad + h_{j,2n}a_m\delta_{-n,m}|f_{k^{\pm}}(n)\rangle + h_{j,2n}a_{-m}\delta_{n,m}|f_{k^{\pm}}(n)\rangle] \\ &= \sum_{m=1}^{N}[h_{j,2m-1}a_m|-f_{k^{\pm}}(m)\rangle + h_{j,2m}a_{-m}|f_{k^{\pm}}(m)\rangle]. \end{aligned} \tag{6.16}$$

Then, by substituting this result in the second line of Eq. (6.13), we obtain

$$\begin{aligned} &\frac{1}{2N}\sum_{k=1}^{N}\sum_{j=1}^{2N}|\psi_{(k^{\pm},j)}\rangle_x O'^{\dagger}_{(k^{\pm},j)}|\phi\rangle_x \\ &= \frac{1}{2N}\sum_{k=1}^{N}\sum_{j=1}^{2N}(\frac{1}{\sqrt{2N}}\sum_{l=1}^{N}[h_{j,2l-1}|l,f_{k^{\pm}}(l)\rangle + h_{j,2l}|-l,-f_{k^{\pm}}(l)\rangle]) \\ &\qquad \times(\sum_{m=1}^{N}[h_{j,2m-1}a_m|-f_{k^{\pm}}(m)\rangle + h_{j,2m}a_{-m}|f_{k^{\pm}}(m)\rangle]) \\ &= \frac{1}{2N\sqrt{2N}}\sum_{k=1}^{N}\sum_{j=1}^{2N}\sum_{l,m=1}^{N}[h_{j,2l-1}h_{j,2m-1}a_m|l,f_{k^{\pm}}(l),-f_{k^{\pm}}(m)\rangle \\ &\qquad + h_{j,2l-1}h_{j,2m}a_{-m}|l,f_{k^{\pm}}(l),f_{k^{\pm}}(m)\rangle \\ &\qquad + h_{j,2l}h_{j,2m-1}a_m|-l,-f_{k^{\pm}}(l),-f_{k^{\pm}}(m)\rangle \\ &\qquad + h_{j,2l}h_{j,2m}a_{-m}|-l,-f_{k^{\pm}}(l),f_{k^{\pm}}(m)\rangle] \\ &= \frac{1}{\sqrt{2N}}\sum_{k,m=1}^{N}[a_m|m,f_{k^{\pm}}(m),-f_{k^{\pm}}(m)\rangle + a_{-m}|-m,-f_{k^{\pm}}(m),f_{k^{\pm}}(m)\rangle] \\ &= \frac{1}{\sqrt{2N}}\sum_{n=-N,(n\neq 0)}^{N}\sum_{m=1}^{N}[a_m|m,n,-n\rangle + a_{-m}|-m,-n,n\rangle] \\ &= |\phi\rangle_x|\psi_1\rangle_x. \end{aligned} \tag{6.17}$$

In the last three lines of this proof, we have used the orthogonal property of rows or columns of the Hadamard matrix, that is,

$$\sum_j h_{j,m}h_{j,n} = 2N\delta_{m,n}. \tag{6.18}$$

## 6.6 Alice's Bell state measurement

To do a BSM, Alice needs to apply a *grand* unitary and Hermitian operator on all the channels followed by a position state measurement on Alice's outgoing

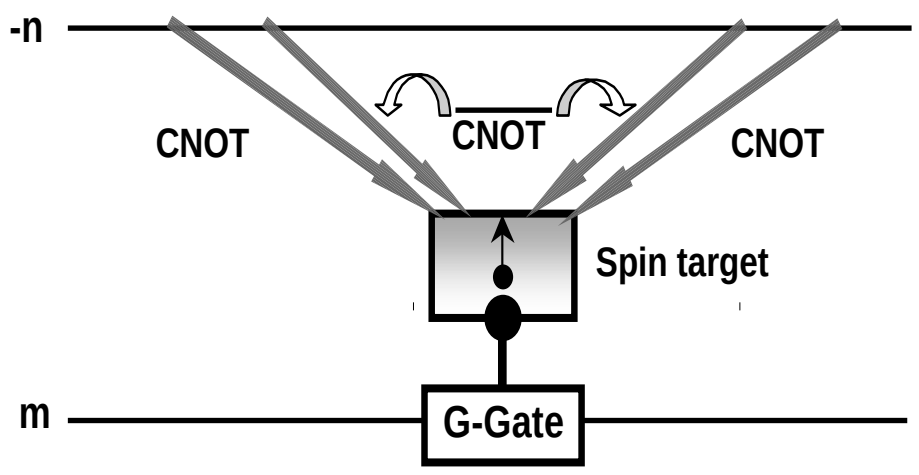

*Fig. 6.2:* Schematic diagram of a position controlled-gate.

receivers. Here, for instance, we can consider the representation of the grand operator as

$$\mathcal{H}_x = \sum_{j=1}^{2N} \sum_{k=1}^{N} |j, f'_{k^\pm}(j)\rangle \langle \psi'_{(k^\pm,j)}| \tag{6.19}$$

where $f'_{k^\pm}(j) = \pm(j+k-1)_{\mathrm{mod}(2N)}$ and $|\psi'_{(k^\pm,j)}\rangle_x = \frac{1}{\sqrt{2N}} \sum_{n=1}^{2N} h_{j,n} |j, f'_{k^\pm}(j)\rangle_x$. It is easy to check that $\mathcal{H}_x |\psi'_{(k^\pm,j)}\rangle_x = |j, f'_{k^\pm}(j)\rangle_x$, and $\mathcal{H}_x^2 = I$. In addition, to make the former property consistent with our channel labelling, it suffices to apply a proper unitary operator. It is seen that the required grand operator can be considered as a conditional gate which acts in a $2N$-dimensional (channel) Hilbert space. Furthermore, based on the works of Deutsch [81] and Barenco *et al.* [82] as well as Bremner *et al.* [85], we can construct the grand operator using, for example, a set of controlled gates. Thus, we introduce our position controlled gate (PCG) which acts like

$$\mathrm{PCG}|n, m\rangle = \theta(-n)(I \otimes G)|n, m\rangle + \theta(n)|n, m\rangle \tag{6.20}$$

where $\theta(n)$ is the unit step function, and $G$ is a general form for $N_m$, $P_m$ and $L_+$ basic gates. To understand the action of this operator there is a way using four well-known spin-$\frac{1}{2}$ CNOT gates. Figure 6.2 shows a sketch of this theoretical construction. The $\overline{\mathrm{CNOT}}$ is somehow a complement to the usual CNOT in the sense that $\mathrm{CNOT}|x, y\rangle = |x, x+y\rangle_{\mathrm{mod}(2)}$ and $\overline{\mathrm{CNOT}}|x, y\rangle = |x, \overline{x}+y\rangle_{\mathrm{mod}(2)}$. If, for instance, a spin-$\frac{1}{2}$ particle passes through the control channel, then the two first CNOT and $\overline{\mathrm{CNOT}}$ definitely change spin of an intermediate spin-$\frac{1}{2}$ particle, named spin target, which is preset to a fixed state , e.g. $|0\rangle$. It is assumed that any change in the state describing the spin target means switching the gate for acting on the channel $m$. The two latter CNOT and $\overline{\mathrm{CNOT}}$ are just to reset the spin target. This conditional gate can be realized using, for example, superconductor circuits. It is obvious that any change in the spin target results into a change in the magnetic flux passing through a superconductor circuit. Thus, for switching the gate, the induced current can be amplified by a superconducting LC circuit in order to produce at least a critical magnetic field to set the direct (side) channel not superconducting (superconducting).

## 6.7 Teleportation of an object having spin

At this stage, one can think of an entangled position-spin state teleportation of a particle having spin $S$. To accomplish this, it is only sufficient to assume that the source emits entangled pair of particles not only with vanishing total momentum but also with zero total spin. Therefore, the quantum state of the system would be

$$|\psi_1\rangle_{xs} = \frac{1}{\sqrt{2N(2S+1)}} \sum_{n=1}^{N} \sum_{s=0}^{2S} (\pm 1)^s [|n, -n\rangle + |-n, n\rangle] |(S-s), -(S-s)\rangle \tag{6.21}$$

which is simply a tensor product of position and spin states of the source. So, as an advantage, it is not necessary to have any entanglement between these two degrees of freedom. By this system, Alice would be able to send unknown quantum states which have the following general form

$$|\phi\rangle_{xs} = \sum_{n=-N(n\neq 0)}^{N} \sum_{s=0}^{2S} a_{ns} |n, S-s\rangle. \tag{6.22}$$

As is seen, to teleport spin state along with position state, it does not need to change our formalism drastically, but it simply needs the map $N \to N(2S+1)$ in our mathematical stages. So, just by adapting our protocol to work for $4N^2(2S+1)^2$ dimensions of Hilbert space, we can simultaneously accomplish spin state as well as position state teleportation for the entangled state in the form $|\phi\rangle_{xs}$. This procedure can be generalized to teleportation of any other degrees of freedom of an object, provided that its necessary entanglement is supplied in the source.

## 6.8 Teleportation of a 2-dimensional object using a planar quantum scanner

To extend the scheme to the teleportation of a two-dimensional position state of an object, we assume that our arrays are planar so that the state of our system would be

$$|\psi_{1(2)}\rangle_{xy} = \frac{1}{\sqrt{4N_x N_y}} \sum_{n_x=1}^{N_x} \sum_{n_y=1}^{N_y}$$

$$[|(n_x, n_y), (-n_x, -n_y)\rangle \pm |(-n_x, -n_y), (n_x, n_y)\rangle] +$$
$$[|(n_x, -n_y), (-n_x, n_y)\rangle \pm |(-n_x, n_y), (n_x, -n_y)\rangle]. \tag{6.23}$$

By this simple assumption, Alice can now teleport the states in general form

$$|\phi\rangle_{xy} = \sum_{n_x=-N_x(n_x\neq 0)}^{N_x} \sum_{n_y=-N_y(n_y\neq 0)}^{N_y} a_{n_x n_y} |(n_x, n_y)\rangle \tag{6.24}$$

to Bob, which are representatives of states that result in two-dimensional wave functions $\psi(x, y)$ at large $N$'s. Now, by insight obtained from one-dimensional case, it is clear that two-dimensional position state teleportation is mathematically equivalent to one-dimensional case just with a larger (channel) Hilbert space dimension, that is, $N = N_x N_y$.

## 6.9 Teleportation of the 3rd dimension using momentum basis

There is an intricacy in extension of the protocol for teleportation of three-dimensional wave functions $\psi(x, y, z)$. It does not seem trivial how to achieve teleportation of three-dimensional objects, utilizing our *planar* receivers. To overcome this problem, we use the fact that any quantum state $|\phi\rangle_z$ can be expanded in terms of the momentum basis as well as the position ones, that is

$$|\phi\rangle_z = \sum_z c_z |z\rangle = \sum_p b_p |p\rangle. \tag{6.25}$$

Thus, to teleport a state represented in the momentum basis we add another property to our source, without any change in the form of our planar receivers. Suppose that the emitted entangled particles from the source, can also get a definite momentum whose value is $p_1, p_2, \ldots, p_M$ with the same probability in the $z$-direction. Fulfilling these properties for the source might be an experimental challenge (for example, as a primary proposal in this regard, see [97]). Anyway, the state of the system in this direction can be written as

$$|\varphi_1\rangle_z = \frac{1}{\sqrt{2M}} \sum_{m=1}^{M} [|p_m, -p_m\rangle + |-p_m, p_m\rangle]. \tag{6.26}$$

Utilizing this source, Alice can send the most general states in the form

$$|\phi\rangle_z = \sum_{m=1}^{M} [b_m |p_m\rangle + b_{-m}|-p_m\rangle] \equiv \sum_p b_p |p\rangle. \tag{6.27}$$

All teleportation processes of $|\phi\rangle_z$ once again can be performed very similar to the one-dimensional position state teleportation. In fact, it is sufficient to consider the conversion $|\pm n\rangle \longleftrightarrow |\pm p_n\rangle$ in all relations which we have obtained

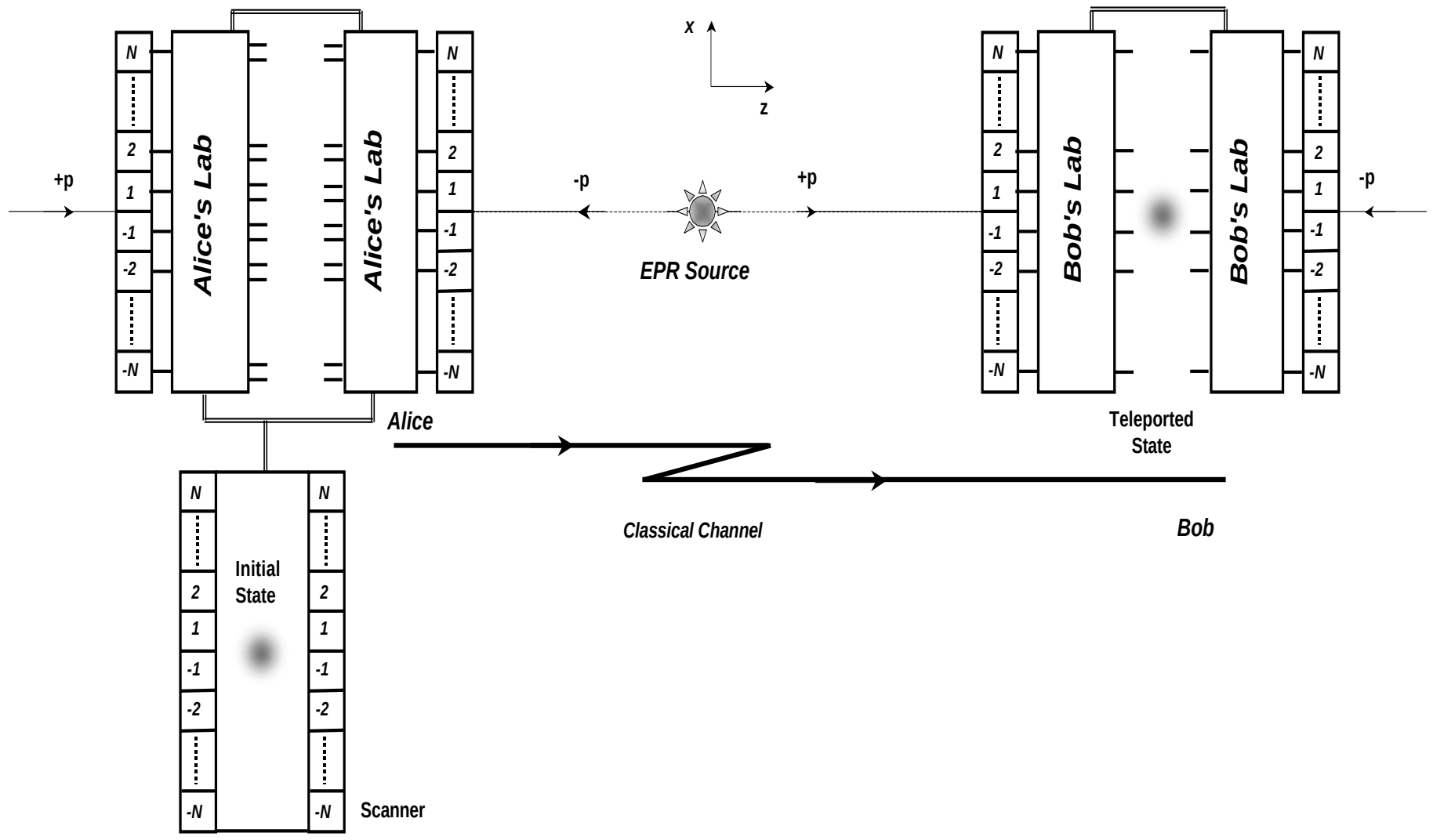

*Fig. 6.3:* Schematic drawing of the proposed theoretical set-up for teleporting the quantum state of an object in our three-dimensional space. The initial state is scanned by two approaching arrays of receivers which are properly connected to Alice's lab.

in the one-dimensional case. For example, similar to Eq. (6.3), the momentum Bell bases can be represented as

$$|\varphi_{(q^{\pm},r)}\rangle_z = \frac{1}{\sqrt{2M}} \sum_{m=1}^{M} [h_{r,2m-1}|p_m, \pm p_{f_q(m)}\rangle + h_{r,2m}| - p_m, \mp p_{f_q(m)}\rangle]$$
$$1 \leq q \leq M, 1 \leq r \leq 2M \tag{6.28}$$

where

$$f_q(m) = (m + q - 1)_{\mathrm{mod}(M)}. \tag{6.29}$$

## 6.10 Towards complete teleportation of a 3-dimensional object

Equipped with this novel kind of source, Alice is now able to teleport three-dimensional states. Regarding the facts of the previous sections, it is sufficient to consider $x$ (as a representative of $xy$) and $z$ components to explain function of the protocol for three-dimensional case. Therefore, we consider general states

like

$$\begin{aligned} |\phi\rangle_{xz} &= \sum_{x,z} c_{xz}|\phi\rangle_x|\phi\rangle_z \\ &= \sum_{x,z} \sum_{n,m=-N,-M(n,m\neq 0)}^{N,M} c_{xz}a_{nx}b_{mz}|n\rangle_x|p_m\rangle_z \end{aligned} \tag{6.30}$$

which Alice wishes to teleport. Here, the proposed teleportation set-up in Fig. 6.1 should be changed a little so that the system (arrays+source) shown in Fig. 6.3 can now describe the three-dimensional quantum state $|\psi_1\rangle_x|\varphi_1\rangle_z$. Again, Alice combines her own particle with the scanned particle corresponding to the state $|\phi\rangle_{xz}$. This combination can be represented as

$$\begin{aligned} |\phi\rangle_{xz}|\psi_1\rangle_x|\varphi_1\rangle_z &= \frac{1}{4NM}\sum_{x,z}\sum_{n=-N(n\neq 0)}^{N}\sum_{m=-M(m\neq 0)}^{M} c_{xz}a_{nx}b_{mz} \\ &\times \sum_{k,j=1}^{N,2N}|\psi_{(k^\pm,j)}\rangle_x O'^\dagger_{(k^\pm,j)}|n\rangle_x \sum_{q,r=1}^{M,2M}|\varphi_{(q^\pm,r)}\rangle_z T^\dagger_{(q^\pm,r)}|p_m\rangle_z. \end{aligned} \tag{6.31}$$

in which $|\varphi_{(q^\pm,r)}\rangle_z$ states are momentum analog of Bell bases $|\psi_{(k^\pm,j)}\rangle_x$, and also $T_{(q^\pm,r)}$ operators work like momentum analog of $O'_{(k^\pm,j)}$ operators generated by the replacement of $|\pm n\rangle \longleftrightarrow |\pm p_n\rangle$ in Eq. (6.5). After performing two separate BSM's on the position and momentum bases of the two particles in the $x$ and $z$ directions, respectively, the protocol is completed and one can obtain

$$\begin{aligned} &|\psi_{(k^\pm,j)}\rangle_x|\varphi_{(q^\pm,r)}\rangle_z\sum_{x,z,n,m} c_{xz}a_{nx}b_{mz}O'^\dagger_{(k^\pm,j)}T^\dagger_{(q^\pm,r)}|n\rangle_x|p_m\rangle_z \\ &= |\psi_{(k^\pm,j)}\rangle_x|\varphi_{(q^\pm,r)}\rangle_z O'^\dagger_{(k^\pm,j)}T^\dagger_{(q^\pm,r)}|\phi\rangle_{xz}. \end{aligned} \tag{6.32}$$

Now Bob needs to know the classical information $(k^\pm, j)$ and $(q^\pm, r)$ via a classical channel in order to reconstruct the initial state $|\phi\rangle_{xz}$.

## 6.11 Examination on the realizability of the momentum gates

### 6.11.1 Momentum basic gates

Similar to the position basic gates introduced in the previous chapter, here, it is also possible to consider some basic gates for the momentum space. Concerning this, we have introduced the following operators

$$\begin{aligned} &\mathcal{P}_m|\pm p_m\rangle = \pm|\pm p_m\rangle \\ &\mathcal{R}_m|\pm p_m\rangle = |\mp p_m\rangle \\ &D_+|\pm p_m\rangle = |\pm p_{(m+1)_{\mathrm{mod}(M)}}\rangle \end{aligned} \tag{6.33}$$

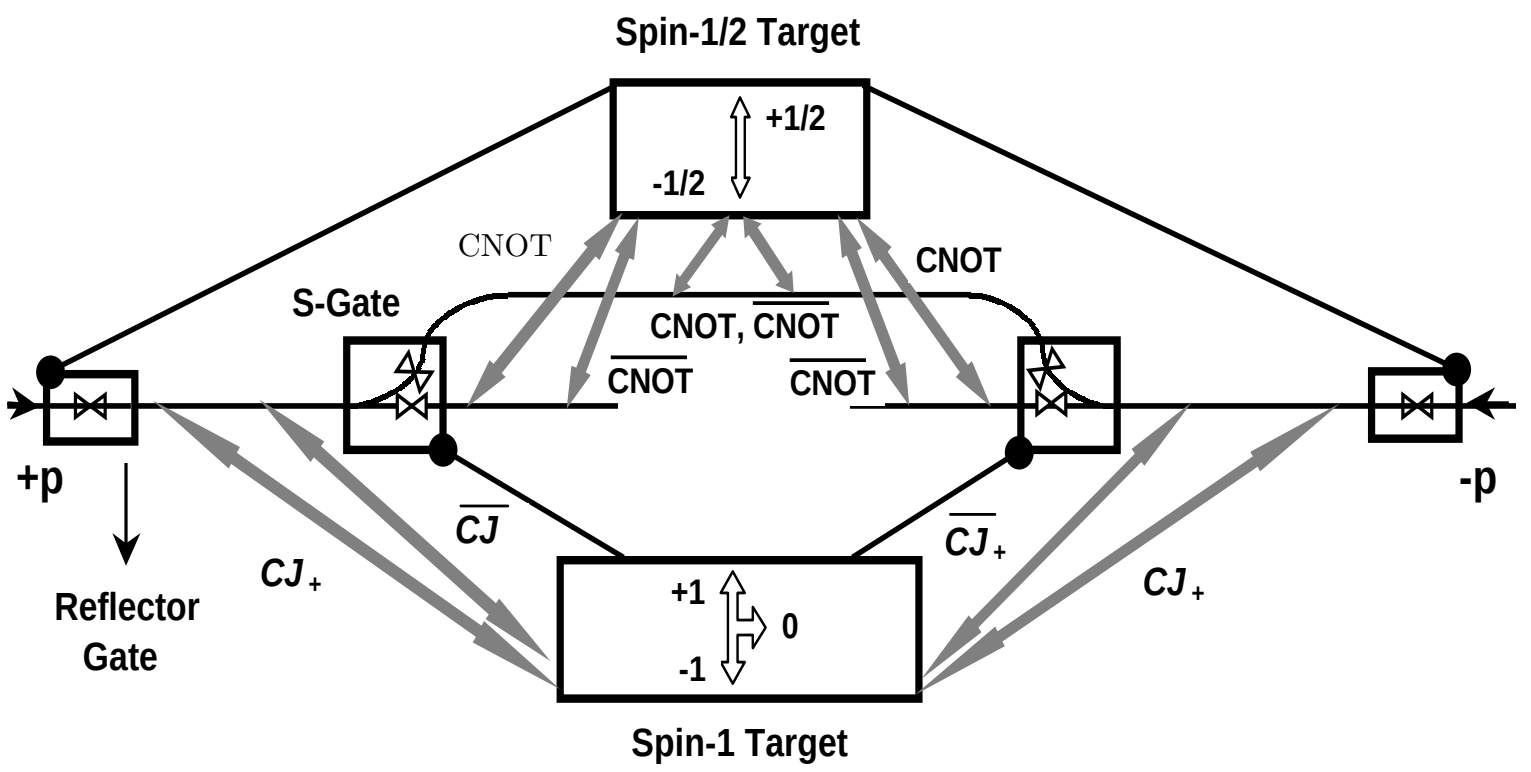

*Fig. 6.4:* Schematic diagram of a reversal operator ($\mathcal{R}$) in the momentum space.

in the momentum space which are called normalized momentum, reversal and drift operators corresponding to $N_n$, $P_m$ and $L_+$ operators in the position space, respectively. This correspondence helps one to use all the relations obtained up to now for the position case, just by the following replacements:

$$\begin{aligned} |\pm n\rangle &\longrightarrow |\pm p_m\rangle \\ N_n &\longrightarrow \mathcal{P}_m \\ P_n &\longrightarrow \mathcal{R}_m \\ L_+ &\longrightarrow D_+. \end{aligned} \tag{6.34}$$

Anyway, again it is relevant to see question how one can conceive these operators in practice. For $\mathcal{P}_m$ operator, it is clear that it acts like a phase factor as follows:

$$\mathcal{P}_m = e^{i\theta(\mp p_m)\pi} \tag{6.35}$$

which works the same as $N_n$ operator in the position space. However, the design of the other operators is relatively different. For instance, $\mathcal{R}$ operator can be considered to operate as Fig. 6.4 shows. In this scheme, in addition to a spin-$\frac{1}{2}$ target which controls the two reflector gates, we have introduced a spin-1 target for controlling the two switch gates. It is assumed that, the spin-1 target is initially adjusted in a fixed state, e.g. $|0\rangle \equiv |s_z = -1\rangle$. Each quantum interaction between the particle passing through the channel and the spin-1 particle results in a quantum rotation for the target's particle in the spin

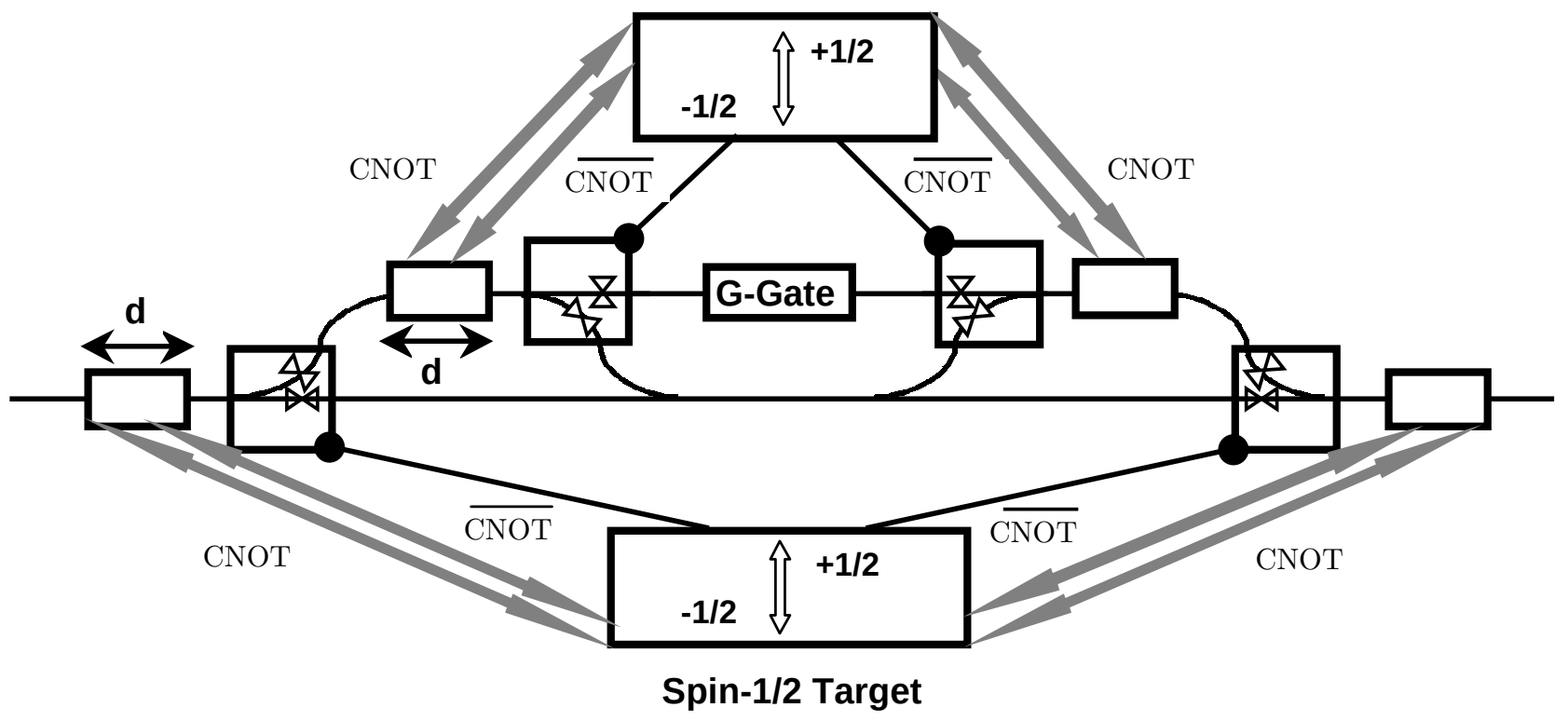

*Fig. 6.5:* Schematic drawing of a filtering quantum circuit. If in the first quantum interaction box $t \geq t_m$, where $t$ is the time of interaction and $t_m = d/v_m$, then the initial state of the lower spin-$\frac{1}{2}$ target will change and the two related $S$-gate will be turned ON. Moreover, if in the second quantum interaction box $t \geq t_{m+1}$, then the initial state of the upper spin-$\frac{1}{2}$ target will change and its related $S$-gates will be turned ON. The G-gate can be considered any desired local gate such as a phase shifter or the $\mathcal{R}$ operator.

space, which can be represented as

$$J_+^n|j\rangle = |j+n\rangle_{\text{mod}(3)}; \qquad j = 0,1,2 \tag{6.36}$$

where $J_+$ is a typical ladder operator in the spin space, $|j\rangle$ is the target's spin state and $n$ shows number of the quantum interactions. Furthermore, $CJ_+$ operators work in the sense that

$$\begin{aligned} CJ_+|x,y\rangle &= |x, x+y\rangle_{\text{mod}(3)} \\ \overline{CJ_+}|x,y\rangle &= |x, \overline{x}+y\rangle_{\text{mod}(3)} \end{aligned} \tag{6.37}$$

where $x = 0,1$ and $y = 0,1,2$. Here, we have assumed that, if $j = 0(1,2)$, then the two switch gates are turned OFF (ON), i.e., the main channels are open (closed) and the side channels are closed (open). For the spin-$\frac{1}{2}$ target, it is also assumed that any change in the target's initial spin state corresponds to turning ON the reflector gate. If this gate is turned OFF, it permits particles

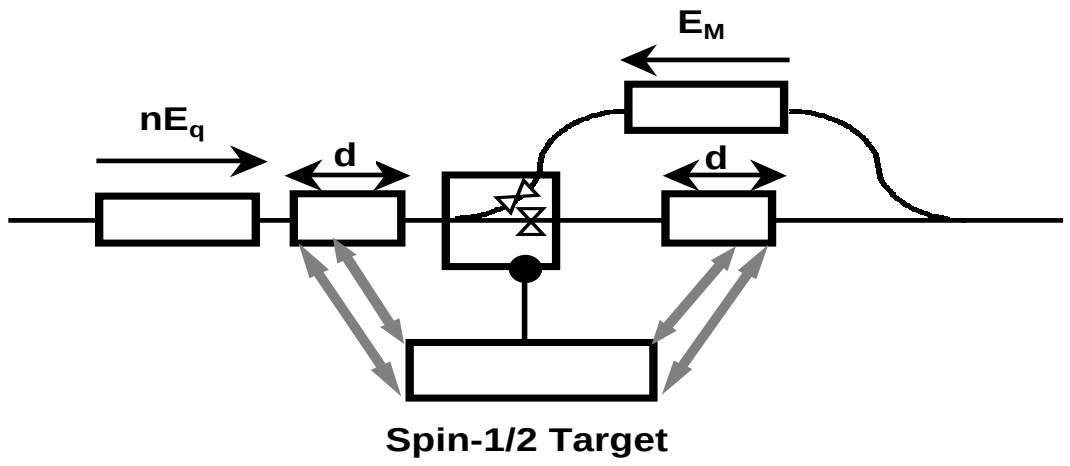

*Fig. 6.6:* A typical scheme fo $D_+^n$ operator. In the first box from the right hand side, the entering charged particle obtains some energy corresponding to an electric field $nE_q$. We have assumed that applying $E_q$ changes $p_m$ to $p_{m+1}$. Now, if in the quantum interaction box $t \geq t_M$, where $t_M = d/v_M$ and $t$ is the time of interaction between the particle and the spin-$\frac{1}{2}$ target, then the initial state of target will change, and consequently, the $S$-gate will be turned OFF. Otherwise, the $S$-gate is turned ON and the opposite field $E_M$ acts as the mod($M$) function.

to pass, and if it is turned ON, then the particles are elastically reflected from it.

So far, we have introduced just $\mathcal{P}$ and $\mathcal{R}$ operators. But in our scheme, particularly in the reconstruction process, we need $\mathcal{P}_m$ and $\mathcal{R}_m$ operators which should act just on the particles having the momentum $p_m$. Thus, we should design a quantum filtering circuit which leads only the particles with the momentum $p_m$ into the considered gate. Figure 6.5 shows such helpful circuit. In this scheme, we have assumed that each quantum interaction between the particle and the spin-$\frac{1}{2}$ targets changes the target's state only if the time of interactions are taken long enough.

For the $D_+$ operator, it is reasonable to consider that, for example, an electric field can easily control the momentum of charged particles. Meantime, it should be noted that $D_+$ operator has a cyclic behavior as shown in Eq. (6.33). Therefore, a conceivable scheme for this operator can be considered as represented in Fig. 6.6.

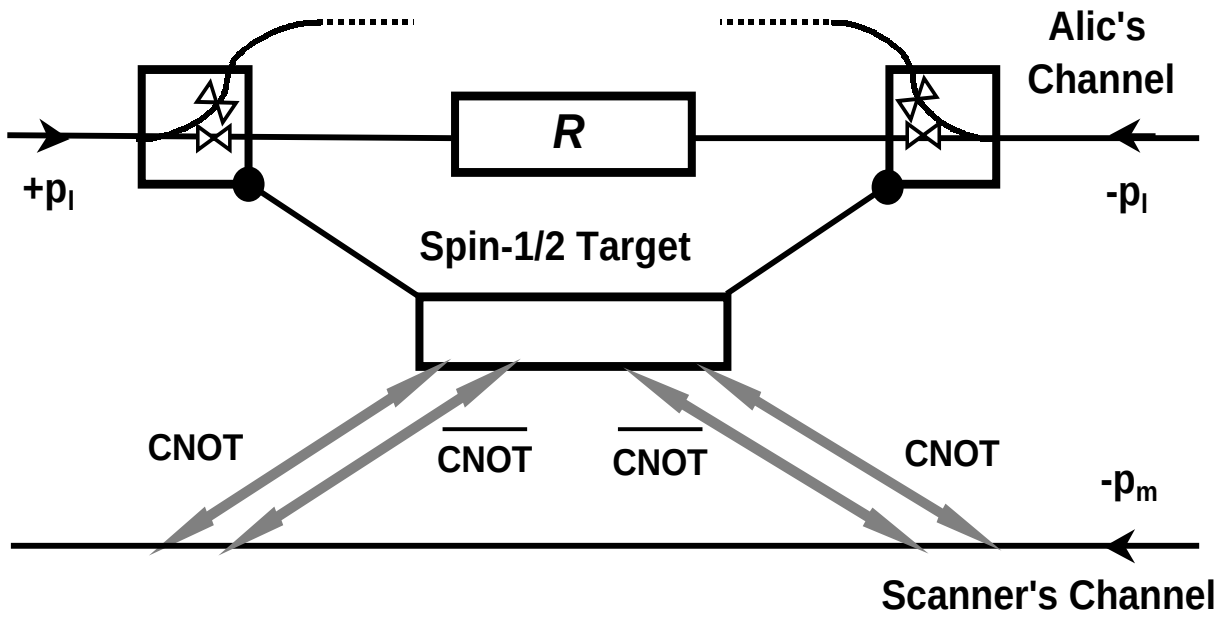

*Fig. 6.7:* Schematic diagram of a momentum controlled gate (MCR). If the scanned particle has a momentum in the opposite of the $z$-direction, then it interacts with the spin-$\frac{1}{2}$ target and changes its initial state. This change results turning OFF of the S-gates.

### 6.11.2 Momentum Bell state measurement

In the momentum space, we should also entangled the momentum of the two particles and perform a BSM on them in the $z$-direction. The same as the grand unitary operator $\mathcal{H}_x$ represented in Eq. (6.19), one can consider another grand unitary operator in the momentum space to perform a BSM as follows:

$$\mathcal{H}_p = \sum_{r=1}^{2M} \sum_{q=1}^{M} |p_r, \pm p_{f'_q(r)}\rangle\langle\varphi'_{(q^\pm,r)}| \tag{6.38}$$

where

$$|\varphi'_{(q^\pm,r)}\rangle_z = \frac{1}{\sqrt{2M}} \sum_{m=1}^{2M} h_{r,m}|p_r, \pm p_{f'_q(r)}\rangle \tag{6.39}$$

and

$$f'_q(r) = (r+q-1)_{\mathrm{mod}(2M)}. \tag{6.40}$$

We have seen that these grand unitary operators can be theoretically con-structed using our proposed conditional gates. Thus, here, we need to introduce a general momentum controlled gate (MCR) which acts like

$$\text{MCR}|p_l, p_m\rangle = \theta(-p_l)(I \otimes \mathcal{R})|p_l, p_m\rangle + \theta(p_l)|p_l, p_m\rangle. \tag{6.41}$$

Fortunately, for this operator we do not need any filtering process, because the only important parameter is the direction of particle's momentum. So, the scheme of this gate, based on our introduced basic gates and side channels, can be easily conceived as shown in Fig. 6.7.

From the previous chapter remember that, in the BSM we have used a position Hadamard operator represented in Eq. (5.36). Here, we introduce the same operator but in the momentum space, that is,

$$H_{p_m} = \frac{1}{\sqrt{2}}(\mathcal{R}_m + \mathcal{P}_m). \tag{6.42}$$

Since, similar to Eq. (5.41), these kinds of operators should be applied on all $p_m$'s, thus it is not necessary to consider any filtering process at this stage. So, the circuit diagram of $H_{p_1} H_{p_2} \ldots H_{p_M}$ operator can be shown as in Fig. 6.8.

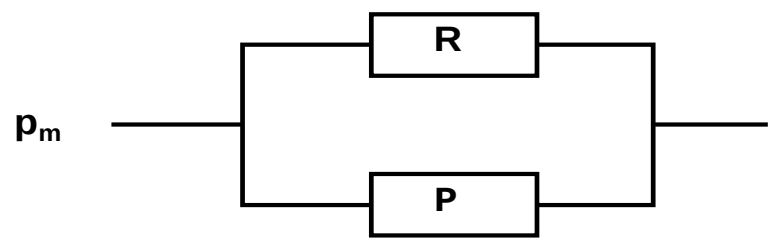

*Fig. 6.8:* Schematic diagram of a momentum Hadamard operator using two $\mathcal{R}$ and $\mathcal{P}$ basic gates. Here, it is assumed that the probability of passing the particle through the upper or lower channel is the same.

At the next stage, we need to design a quantum circuit for $V_{(M)}$ operator which acts like $U_{(N)}$ operator introduced in Eq. (5.39), but in the momentum space. Thus, we can represent it as

$$V_{(M)} = \frac{1}{\sqrt{M}} \sum_{m=0}^{M/2-1} \{D_+^{2m} \otimes [\prod_{i=1}^{M} \mathcal{R}_i^{(h_{b,i}-h_{2n+i,i})/2} \mathcal{P}_i \mathcal{R}_i^{(h_{b,i}-h_{M-2m+i,i})/2}] D_+^{2m}$$
$$+ D_+^{2m+1} \otimes [\prod_{i=1}^{M} (\mathcal{R}_i \mathcal{P}_i \mathcal{R}_i \mathcal{P}_i)^{(h_{b,i}-h_{M-2m-1+i,i})/2}] D_+^{2m+1}\}. \tag{6.43}$$

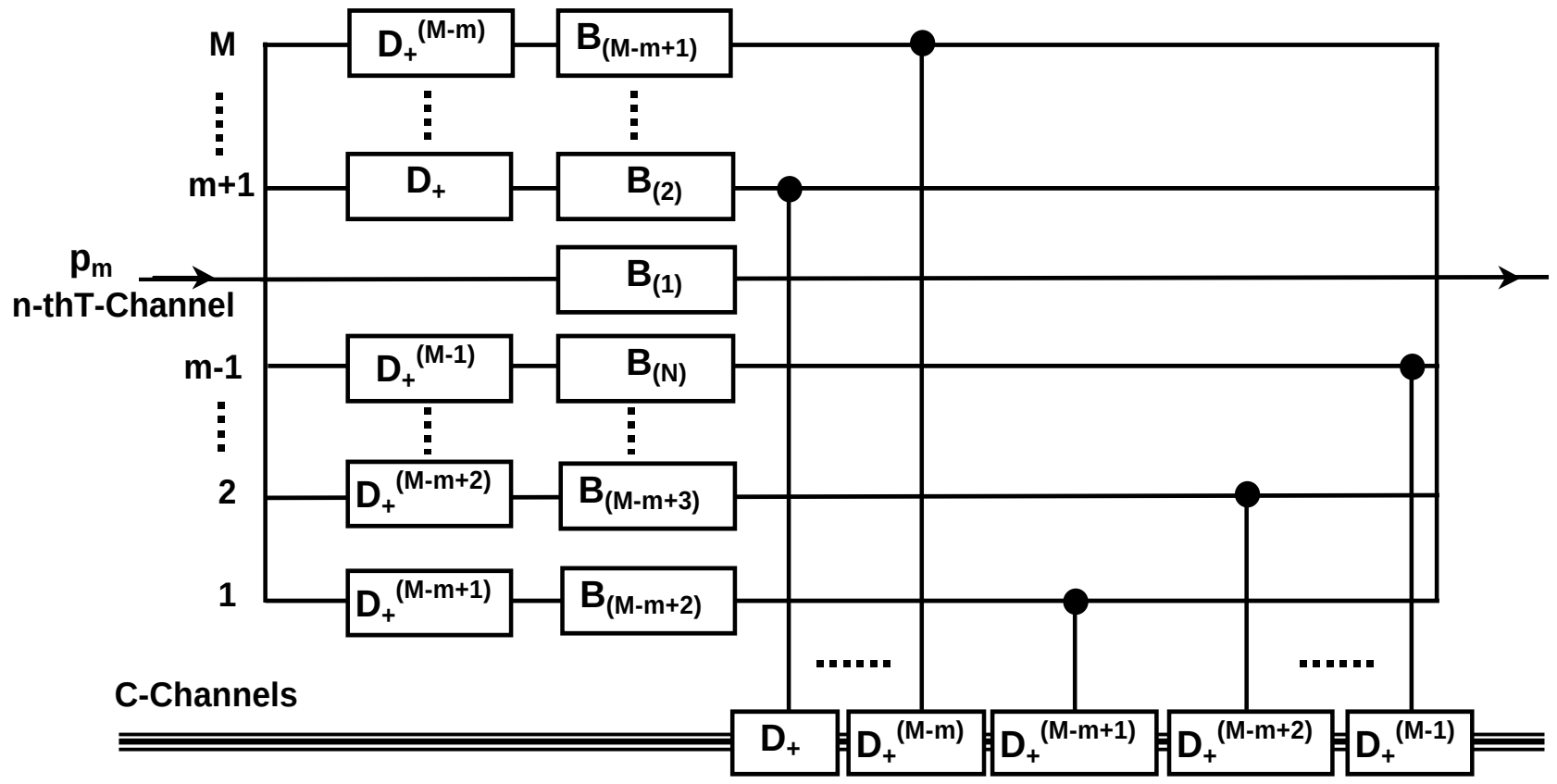

*Fig. 6.9:* Quantum circuit diagram of $V_{(M)}$ operator fot the $n$-th T-channel. Here, each branch of the T-channel acts as a control channel for the $D_+^n$ operators applied on the C-channels.

Now, if we rewrite this operator as

$$V_{(M)} = \frac{1}{\sqrt{M}} \sum_{m=0}^{M-1} (D_+^m \otimes B_{(m+1)} D_+^m) \tag{6.44}$$

then it would be possible to design theoretically a quantum circuit for this operator, very similar to the one performed for $U_{(N)}$ operator in the previous chapter. Figure 6.9 shows a conceivable quantum circuit of the $V_{(M)}$ operator for just a typical T-channel, using the introduced momentum basic gates.

## 6.12 Conclusions

In conclusion, we have proposed a theoretical scheme which introduces $N$-level position state teleportation of a particle having other degrees of freedom in three-dimensional space. We have given representations for Bell states and necessary unitary operators, using just symmetric normalized Hadamard matrices. This scheme uses just one special EPR pair of particles and a planar quantum scanner. Therefore, if realized, this scheme can be used to provide a complete wave function teleportation of a particle including spin, $\psi_S(\mathbf{x})$, at large $N$'s, for the first time. We have shown that an entangled position-spin state can be teleported using an EPR pair entangled in position and spin variables separately, without the need to assume any kind of entanglement between the two variables in the source. By the way, as an advantage, the scheme needs only one kind of mechanism for BSM in each spatial dimension. So our scheme can be considered economically better than the original one, which needed two kinds of measurement on position and momentum spaces per spatial dimension. Furthermore, one can now think of moving towards understanding teleportation of multi-particle objects, using EPR sources which only emit the constituent particles of the object. Today, the technology for teleporting states of individual atoms is at hand. For instance, the group led by Haroche [98] in Paris has demonstrated entanglement of atoms. The entanglement of molecules and then their teleportation may reasonably be expected within the next decades. However, what happens beyond that, is anybody's guess.

# APPENDIX

# A. A CLARIFICATION ON THE DEFINITION OF CENTER OF MASS COORDINATE OF THE EPR PAIR

Here we want to show that the two different definitions for the center of mass position of the two entangled particles, that is, $y_1 - y_2$ in Eq. (2.1) and $y_1 + y_2$ in Eq. (2.13), are two consistent representations for the center of mass coordinate.

At first, Concerning the center of mass coordinate $y_1 - y_2$ represented in Eq. (2.1), we can consider that

$$\begin{aligned} \mathbf{k_1}.\mathbf{y_1} &= |\mathbf{k_1}||\mathbf{y_1}|cos(\widehat{\mathbf{k_1},\mathbf{y_1}}) = \pm|\mathbf{k_1}||\mathbf{y_1}| \\ \mathbf{k_1}.\mathbf{y_2} &= |\mathbf{k_1}||\mathbf{y_2}|cos(\widehat{\mathbf{k_1},\mathbf{y_2}}) = \pm|\mathbf{k_1}||\mathbf{y_2}|. \end{aligned} \tag{A.1}$$

In addition, we should assume that $cos(\widehat{\mathbf{k_1},\mathbf{y_1}}) = cos(\widehat{\mathbf{k_1},\mathbf{y_2}})$. Because this guarantees that one of the entangled particles moves into the positive and the other one into the negative direction, which is a necessary constraint in the EPR experiment.[1] Then, the $y$ component of Eq. (2.1) can be rewritten in a clearer form, that is

$$\begin{aligned} \psi(y_1, y_2) &= \int e^{i\mathbf{k_1}.\mathbf{y_1}} e^{i\mathbf{k_2}.(\mathbf{y_2}+\mathbf{y_0})} \delta(\mathbf{k}_1 + \mathbf{k}_2) d\mathbf{k}_1 d\mathbf{k}_2 \\ &= \int e^{i\mathbf{k_1}.\mathbf{y_1}} e^{-i\mathbf{k}_1.(\mathbf{y_2}+\mathbf{y_0})} d\mathbf{k}_1 \\ &= \int_{-\infty}^{+\infty} e^{ik_1(|\mathbf{y_1}|-|\mathbf{y_2}|-|\mathbf{y_0}|)} dk_1 = 2\pi\delta(|\mathbf{y_1}| - |\mathbf{y_2}| - |\mathbf{y_0}|) \end{aligned} \tag{A.2}$$

where $|\mathbf{y_0}| = |\mathbf{y_1}| - |\mathbf{y_2}|$ obviously represents the center of mass position of the two particles (with a 1/2 factor) if either $y_1 > 0$ and $y_2 < 0$, or vice versa. This means that it is possible to have simultaneous eigenfunctions for the total momentum and the center of mass coordinate for the two entangled particles, as was previously shown by Einstein *et al.* [20]. Now, It is proper to select $\mathrm{y}_0 = 0$. Therefore, in Eq. (2.1), $y_1 - y_2$ really represents $|\mathbf{y_1}| - |\mathbf{y_2}|$ which is clearly the center of mass coordinate. These arguments are also completely consistent with the common realization of the EPR experiment.

[1] It is well known that a plane wave with representation $e^{ikx}$ goes out from the origin of coordinate while $e^{-ikx}$ goes into the origin. When the two particles of an EPR pair are considered at the one side of the origin on the $x$-axis, one of them goes away from the origin and another approaches to it. On the other hand, if the origin is considered between the two entangled particles, then the two particles either go away from the origin or go into it.

On the other hand, in Eq. (2.5) and consequently in Eqs. (2.13) and (2.30), the $y$ parameters are variables or vectors in one dimension, not absolute values. Thus, in these equations, $\mathbf{y_1} + \mathbf{y_2} \equiv y_1(t) + y_2(t)$ naturally represents the center of mass coordinate of the two particles. Again, one can show that having simultaneous eigenfunction for the total momentum $\mathbf{p_{y_1}} + \mathbf{p_{y_2}}$ and the center of mass position $\mathbf{y_1} + \mathbf{y_2}$ for the two entangled particles is feasible. To see this, consider the following properties of our considered particles

$$\begin{aligned} \mathbf{p_{y_1}} &= p_{y_1}\widehat{j}, \qquad & \mathbf{p_{y_2}} &= p_{y_2}\widehat{j} = -p_{y_1}\widehat{j} \\ \mathbf{y}_1 &= y_1\widehat{j}, \qquad & \mathbf{y}_2 &= -y_2\widehat{j} \end{aligned} \tag{A.3}$$

where similar to the EPR's wave function the center of mass position is represented by $\mathbf{y}_1 + \mathbf{y}_2 = (y_1 - y_2)\widehat{j}$, in which the values of $y_1$ and $y_2$ have the same signs. Now, it is easy to show that

$$[\mathbf{p_{y_1}} + \mathbf{p_{y_2}}, \mathbf{y_1} + \mathbf{y_2}] = [p_{y_1} + p_{y_2}, y_1 - y_2] = 0 \tag{A.4}$$

which means it is possible to determine the total momentum and the center of mass coordinate, simultaneously, for the two entangled particles.

# B. DETAILS ON PREPARING AND MEASURING PROCESSES FOR SOME INITIAL CASES

In this appendix, we have presented some details on preparing (encoding) and measuring (decoding) processes in the dense coding scheme introduced in Chapter 3 for $N = 1$, 2, 4, 8 and 12 special cases in order to clarify the procedure of obtaining some relatively complex relations. The explicit forms of the encoding and decoding operators represented in the text using the basic gates, are principally generalizations of the following calculations.

For the $N = 1$ case, there are four mutually entangled orthogonal states as

$$|\psi_{1,2}^{(1)}\rangle = \frac{1}{\sqrt{2}}[|1,-1\rangle \pm |-1,1\rangle] \tag{B.1}$$

$$|\psi_{3,4}^{(1)}\rangle = \frac{1}{\sqrt{2}}[|1,1\rangle \pm |-1,-1\rangle] \tag{B.2}$$

where all required processes are clearly very similar to what can be done for the well known system of two spin-$\frac{1}{2}$ particles. The preparation and measuring processes for this case are summarized in Tables B.1 and B.2, respectively,

*Tab. B.1:* Alice's preparation process for $N = 1$.

| Transformation | New state |
|---|---|
| $(I \otimes I)\|\psi_1^{(1)}\rangle$ | $\|\psi_1^{(1)}\rangle$ |
| $(N_1 I_s \otimes I)\|\psi_1^{(1)}\rangle$ | $\|\psi_2^{(1)}\rangle$ |
| $(P_1 I_s \otimes I)\|\psi_1^{(1)}\rangle$ | $\|\psi_3^{(1)}\rangle$ |
| $(N_1 P_1 I_s \otimes I)\|\psi_1^{(1)}\rangle$ | $\|\psi_4^{(1)}\rangle$ |

*Tab. B.2:* Bob's measurement process for $N = 1$.

| Initial state | PCS | $H_{x_1} \otimes I$ | Renamed |
|---|---|---|---|
| $\|\psi_1^{(1)}\rangle$ | $\frac{1}{2}(\|1\rangle + \|-1\rangle)\|-1\rangle$ | $\|1,-1\rangle$ | 01 |
| $\|\psi_2^{(1)}\rangle$ | $\frac{1}{2}(\|1\rangle - \|-1\rangle)\|-1\rangle$ | $\|-1,-1\rangle$ | 11 |
| $\|\psi_3^{(1)}\rangle$ | $\frac{1}{2}(\|1\rangle + \|-1\rangle)\|1\rangle$ | $\|1,1\rangle$ | 00 |
| $\|\psi_4^{(1)}\rangle$ | $\frac{1}{2}(\|1\rangle - \|-1\rangle)\|1\rangle$ | $\|-1,1\rangle$ | 10 |

where we have applied the renaming convention

$$|n\rangle \equiv 0, \ |-n\rangle \equiv 1 \tag{B.3}$$

in the last column of Tab. B.2.

For the $N = 2$ case, we have 16 entangled orthonormal spatial states as follows: which form our Bell basis in the $N = 2$ case. Now, it is proper to

*Tab. B.3:* entangled orthonormal position states for $N = 2$.

| |
|---|
| $\lvert\psi^{(2)}_{1,2}\rangle = \frac{1}{2}[(\lvert 1,-1\rangle \pm \lvert -1,1\rangle) + (\lvert 2,-2\rangle \pm \lvert -2,2\rangle)]$ |
| $\lvert\psi^{(2)}_{3,4}\rangle = \frac{1}{2}[(\lvert 1,-1\rangle \pm \lvert -1,1\rangle) - (\lvert 2,-2\rangle \pm \lvert -2,2\rangle)]$ |
| $\lvert\psi^{(2)}_{5,6}\rangle = \frac{1}{2}[(\lvert 1,1\rangle \pm \lvert -1,-1\rangle) + (\lvert 2,2\rangle \pm \lvert -2,-2\rangle)]$ |
| $\lvert\psi^{(2)}_{7,8}\rangle = \frac{1}{2}[(\lvert 1,1\rangle \pm \lvert -1,-1\rangle) - (\lvert 2,2\rangle \pm \lvert -2,-2\rangle)]$ |
| $\lvert\psi^{(2)}_{9,10}\rangle = \frac{1}{2}[(\lvert 2,-1\rangle \pm \lvert -2,1\rangle) + (\lvert 1,-2\rangle \pm \lvert -1,2\rangle)]$ |
| $\lvert\psi^{(2)}_{11,12}\rangle = \frac{1}{2}[(\lvert 2,-1\rangle \pm \lvert -2,1\rangle) - (\lvert 1,-2\rangle \pm \lvert -1,2\rangle)]$ |
| $\lvert\psi^{(2)}_{13,14}\rangle = \frac{1}{2}[(\lvert 2,1\rangle \pm \lvert -2,-1\rangle) + (\lvert 1,2\rangle \pm \lvert -1,-2\rangle)]$ |
| $\lvert\psi^{(2)}_{15,16}\rangle = \frac{1}{2}[(\lvert 2,1\rangle \pm \lvert -2,-1\rangle) - (\lvert 1,2\rangle \pm \lvert -1,-2\rangle)]$ |

introduce the notion of family of states. By a family we mean sets of states which share the same kets in their structure and at most differ in their signs. So in Tab. B.3, four families can be distinguished. Tables B.4 and B.5 illustrate all preparation processes that can be selected and done by Alice, and also Bob's measurement process, for each family, respectively. In Tab. B.5 it is seen what

*Tab. B.4:* Alice's preparation process for $N = 2$.

| Transformation | New state |
|---|---|
| $(I \otimes I)\lvert\psi^{(2)}_1\rangle$ | $\lvert\psi^{(2)}_1\rangle$ |
| $(N_1N_2 \otimes I)\lvert\psi^{(2)}_1\rangle$ | $\lvert\psi^{(2)}_2\rangle$ |
| $(P_2N_2P_2N_2 \otimes I)\lvert\psi^{(2)}_1\rangle$ | $\lvert\psi^{(2)}_3\rangle$ |
| $(P_2N_1N_2P_2 \otimes I)\lvert\psi^{(2)}_1\rangle$ | $\lvert\psi^{(2)}_4\rangle$ |
| $(P_1P_2 \otimes I)\lvert\psi^{(2)}_1\rangle$ | $\lvert\psi^{(2)}_5\rangle$ |
| $(N_1N_2P_1P_2 \otimes I)\lvert\psi^{(2)}_1\rangle$ | $\lvert\psi^{(2)}_6\rangle$ |
| $(N_2P_1P_2N_2 \otimes I)\lvert\psi^{(2)}_1\rangle$ | $\lvert\psi^{(2)}_7\rangle$ |
| $(N_1P_1P_2N_2 \otimes I)\lvert\psi^{(2)}_1\rangle$ | $\lvert\psi^{(2)}_8\rangle$ |
| $(L_+ \otimes I)\lvert\psi^{(2)}_1\rangle$ | $\lvert\psi^{(2)}_9\rangle$ |
| $(N_1N_2L_+ \otimes I)\lvert\psi^{(2)}_1\rangle$ | $\lvert\psi^{(2)}_{10}\rangle$ |
| $(P_2N_2P_2N_2L_+ \otimes I)\lvert\psi^{(2)}_1\rangle$ | $\lvert\psi^{(2)}_{11}\rangle$ |
| $(P_2N_1N_2P_2L_+ \otimes I)\lvert\psi^{(2)}_1\rangle$ | $\lvert\psi^{(2)}_{12}\rangle$ |
| $(P_1P_2L_+ \otimes I)\lvert\psi^{(2)}_1\rangle$ | $\lvert\psi^{(2)}_{13}\rangle$ |
| $(N_1N_2P_1P_2L_+ \otimes I)\lvert\psi^{(2)}_1\rangle$ | $\lvert\psi^{(2)}_{14}\rangle$ |
| $(N_2P_1P_2N_2L_+ \otimes I)\lvert\psi^{(2)}_1\rangle$ | $\lvert\psi^{(2)}_{15}\rangle$ |
| $(N_1P_1P_2N_2L_+ \otimes I)\lvert\psi^{(2)}_1\rangle$ | $\lvert\psi^{(2)}_{16}\rangle$ |

differs from the $N = 1$ case is that applying just $H_{x_1}H_{x_2}$ does not produce Bob's desired orthonormal spatial states. Therefore, we have applied a new unitary

*Tab. B.5:* Bob's measurement process for $N = 2$

| Initial state | PCS | $H_{x_1} H_{x_2} \otimes I$ | $U_{(2)}$ | Renamed |
|---|---|---|---|---|
| $\vert\psi_1^{(2)}\rangle$ | $\frac{1}{2}[(\vert 1\rangle + \vert -1\rangle)\vert -1\rangle + (\vert 2\rangle + \vert -2\rangle)\vert -2\rangle]$ | $\frac{1}{\sqrt{2}}[\vert 1,-1\rangle + \vert 2,-2\rangle]$ | $\vert 1,-1\rangle$ | $0\sqcup 1\sqcup$ |
| $\vert\psi_2^{(2)}\rangle$ | $\frac{1}{2}[(\vert 1\rangle - \vert -1\rangle)\vert -1\rangle + (\vert 2\rangle - \vert -2\rangle)\vert -2\rangle]$ | $\frac{1}{\sqrt{2}}[\vert -1,-1\rangle + \vert -2,-2\rangle]$ | $\vert -1,-1\rangle$ | $1\sqcup 1\sqcup$ |
| $\vert\psi_3^{(2)}\rangle$ | $\frac{1}{2}[(\vert 1\rangle + \vert -1\rangle)\vert -1\rangle - (\vert 2\rangle + \vert -2\rangle)\vert -2\rangle]$ | $\frac{1}{\sqrt{2}}[\vert 1,-1\rangle - \vert 2,-2\rangle]$ | $\vert 2,-2\rangle$ | $\sqcup 0\sqcup 1$ |
| $\vert\psi_4^{(2)}\rangle$ | $\frac{1}{2}[(\vert 1\rangle - \vert -1\rangle)\vert -1\rangle - (\vert 2\rangle - \vert -2\rangle)\vert -2\rangle]$ | $\frac{1}{\sqrt{2}}[\vert -1,-1\rangle - \vert -2,-2\rangle]$ | $\vert -2,-2\rangle$ | $\sqcup 1\sqcup 1$ |
| $\vert\psi_5^{(2)}\rangle$ | $\frac{1}{2}[(\vert 1\rangle + \vert -1\rangle)\vert 1\rangle + (\vert 2\rangle + \vert -2\rangle)\vert 2\rangle]$ | $\frac{1}{\sqrt{2}}[\vert 1,1\rangle + \vert 2,2\rangle]$ | $\vert 2,2\rangle$ | $\sqcup 0\sqcup 0$ |
| $\vert\psi_6^{(2)}\rangle$ | $\frac{1}{2}[(\vert 1\rangle - \vert -1\rangle)\vert 1\rangle + (\vert 2\rangle - \vert -2\rangle)\vert 2\rangle]$ | $\frac{1}{\sqrt{2}}[\vert -1,1\rangle + \vert -2,2\rangle]$ | $\vert -2,2\rangle$ | $\sqcup 1\sqcup 0$ |
| $\vert\psi_7^{(2)}\rangle$ | $\frac{1}{2}[(\vert 1\rangle + \vert -1\rangle)\vert 1\rangle - (\vert 2\rangle + \vert -2\rangle)\vert 2\rangle]$ | $\frac{1}{\sqrt{2}}[\vert 1,1\rangle - \vert 2,2\rangle]$ | $\vert 1,1\rangle$ | $0\sqcup 0\sqcup$ |
| $\vert\psi_8^{(2)}\rangle$ | $\frac{1}{2}[(\vert 1\rangle - \vert -1\rangle)\vert 1\rangle - (\vert 2\rangle - \vert -2\rangle)\vert 2\rangle]$ | $\frac{1}{\sqrt{2}}[\vert -1,1\rangle - \vert -2,2\rangle]$ | $\vert -1,1\rangle$ | $1\sqcup 0\sqcup$ |
| $\vert\psi_9^{(2)}\rangle$ | $\frac{1}{2}[(\vert 1\rangle + \vert -1\rangle)\vert -2\rangle + (\vert 2\rangle + \vert -2\rangle)\vert -1\rangle]$ | $\frac{1}{\sqrt{2}}[\vert 2,-1\rangle + \vert 1,-2\rangle]$ | $\vert 2,-1\rangle$ | $\sqcup 01\sqcup$ |
| $\vert\psi_{10}^{(2)}\rangle$ | $\frac{1}{2}[(\vert 1\rangle - \vert -1\rangle)\vert -2\rangle + (\vert 2\rangle - \vert -2\rangle)\vert -1\rangle]$ | $\frac{1}{\sqrt{2}}[\vert -2,-1\rangle + \vert -1,-2\rangle]$ | $\vert -2,-1\rangle$ | $\sqcup 11\sqcup$ |
| $\vert\psi_{11}^{(2)}\rangle$ | $\frac{1}{2}[(\vert 1\rangle + \vert -1\rangle)\vert -2\rangle - (\vert 2\rangle + \vert -2\rangle)\vert -1\rangle]$ | $\frac{1}{\sqrt{2}}[\vert 2,-1\rangle - \vert 1,-2\rangle]$ | $\vert 1,-2\rangle$ | $0\sqcup\sqcup 1$ |
| $\vert\psi_{12}^{(2)}\rangle$ | $\frac{1}{2}[(\vert 1\rangle - \vert -1\rangle)\vert -2\rangle - (\vert 2\rangle - \vert -2\rangle)\vert -1\rangle]$ | $\frac{1}{\sqrt{2}}[\vert -2,-1\rangle - \vert -1,-2\rangle]$ | $\vert -1,-2\rangle$ | $1\sqcup\sqcup 1$ |
| $\vert\psi_{13}^{(2)}\rangle$ | $\frac{1}{2}[(\vert 1\rangle + \vert -1\rangle)\vert 2\rangle + (\vert 2\rangle + \vert -2\rangle)\vert 1\rangle]$ | $\frac{1}{\sqrt{2}}[\vert 2,1\rangle + \vert 1,2\rangle]$ | $\vert 1,2\rangle$ | $0\sqcup\sqcup 0$ |
| $\vert\psi_{14}^{(2)}\rangle$ | $\frac{1}{2}[(\vert 1\rangle - \vert -1\rangle)\vert 2\rangle + (\vert 2\rangle - \vert -2\rangle)\vert 1\rangle]$ | $\frac{1}{\sqrt{2}}[\vert -2,1\rangle + \vert -1,2\rangle]$ | $\vert -1,2\rangle$ | $1\sqcup\sqcup 0$ |
| $\vert\psi_{15}^{(2)}\rangle$ | $\frac{1}{2}[(\vert 1\rangle + \vert -1\rangle)\vert 2\rangle - (\vert 2\rangle + \vert -2\rangle)\vert 1\rangle]$ | $\frac{1}{\sqrt{2}}[\vert 2,1\rangle - \vert 1,2\rangle]$ | $\vert 2,1\rangle$ | $\sqcup 00\sqcup$ |
| $\vert\psi_{16}^{(2)}\rangle$ | $\frac{1}{2}[(\vert 1\rangle - \vert -1\rangle)\vert 2\rangle - (\vert 2\rangle - \vert -2\rangle)\vert 1\rangle]$ | $\frac{1}{\sqrt{2}}[\vert -2,1\rangle - \vert -1,2\rangle]$ | $\vert -2,1\rangle$ | $\sqcup 10\sqcup$ |

operator in the form

$$U_{(2)} = \frac{1}{\sqrt{2}}[(I \otimes P_1 N_1 N_2 P_1) + (L_+ \otimes L_+)] \tag{B.4}$$

which transforms the processed orthonormal states to the spatial bases which are manageable by Bob. Furthermore, in the last column of Tab. B.5, the following renaming convention is added to the previous one in Eq. (B.3). At first, according to Fig. B.1, Bob prepares a sequence of blank spaces the number of which is equal to his receivers, that is, $2N$. Now, the first half of these spaces belong to the particle received from Alice and the second half belong to his own particle. Then, he observes the outcomes of all C-channels. If one of them, e.g. $i$-th C-channel, is ON ($i \in (-N, \ldots, N)$), he fills the $|i|$−th space with 1 or 0, according to Eq. (B.3), otherwise all the remaining spaces are left blank; $\sqcup$. Then, he scans all T-channels and in the same manner he fills one of the blank spaces. Finally, Bob obtains a $2N$ character strip which obtains 2 digits including 0 and/or 1 as well as $(2N - 2)$ $\sqcup$'s.

At the next stage, it is natural to consider the $N = 3$ case. Here, one may expect that the foregoing mechanism works for this case, too. As we have seen, the first step is to find all entangled orthonormal states, like Eq. (B.3). At the first glance, one may consider

$$|\psi_{1,2}^{(3)}\rangle = \frac{1}{\sqrt{6}}[(|1,-1\rangle \pm |-1,1\rangle) + (|2,-2\rangle \pm |-2,2\rangle) + (|3,-3\rangle \pm |-3,3\rangle)] \tag{B.5}$$

as examples of two entangled orthonormal states of a family. But, the number of sets of parenthesis, i.e. ( ), in them is odd, so that constructing more entangled orthonormal states for this family, by permuting $\pm$ signs, is impossible. So, it is seen that there is not any other member for it. This means that the size of this family is smaller than the case of $N = 2$, in Tab. B.3. Thus, we do not have sufficient entangled orthonormal states to perform, say, dense coding process in this case. In other words, we only have 12 entangled orthonormal position states. A similar argumentation also works for the case of odd $N$.

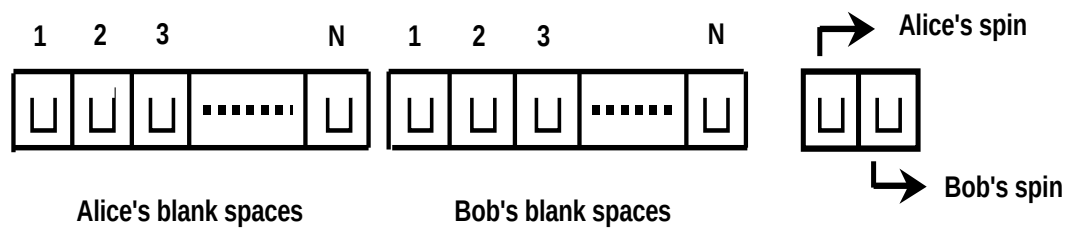

*Fig. B.1:* Initial (OFF) state of Bob's detector.

For the next case, consider $N = 4$. All of the 64 entangled orthonormal position states, classified in 8 families, are listed in Tab. B.6. Now, Alice can apply proper combinations of the basic gates on her particle to produce other members of the Bell bases. Table B.7 shows all preparation that can be selected and performed by Alice for each family. After receiving Alice's processed particle by Bob, he perform a BSM on the two particle, as is concisely shown in Tab. B.8. In this case, the last unitary operator for BSM can be considered like

$$\begin{aligned} U_{(4)} = \; & \frac{1}{\sqrt{4}} \; [I \otimes P_1 P_4 N_1 N_2 N_3 N_4 P_1 P_4 + L_+ \otimes P_4 N_4 P_4 N_4 L_+ \\ + \; & L_+^2 \otimes P_1 P_3 N_1 N_2 N_3 N_4 P_1 P_3 L_+^2 + L_+^3 \otimes P_3 N_3 P_3 N_3 L_+^3]. \end{aligned} \tag{B.6}$$

Now, consider the $N = 6$ case. Again, at first glance, it seems that a similar procedure works for this case. Thus, imagine that one has succeeded to perform the protocol and produce a set of final entangled orthonormal states from a complete set of initial entangled orthonormal states (Bell bases). Among them, there should be a state like

$$\frac{1}{\sqrt{6}}(|1,1\rangle + |2,2\rangle + |3,3\rangle + |4,4\rangle + |5,5\rangle + |6,6\rangle). \tag{B.7}$$

The first obvious entangled state which is orthogonal to this state can be considered to be

$$\frac{1}{\sqrt{6}}(|1,1\rangle - |2,2\rangle + |3,3\rangle - |4,4\rangle + |5,5\rangle - |6,6\rangle). \tag{B.8}$$

But, it can be tested that no other states of this family, formed by changing kets' signs, may be found which are orthogonal to both of them, simultaneously. So,

*Tab. B.6:* Entangled orthogonal position states for $N = 4$.

| |
|---|
| $\|\psi^{(4)}_{1,2}\rangle = \frac{1}{\sqrt{8}}[(\|1,-1\rangle \pm \|-1,1\rangle) + (\|2,-2\rangle \pm \|-2,2\rangle) + (\|3,-3\rangle \pm \|-3,3\rangle) + (\|4,-4\rangle \pm \|-4,4\rangle)]$ |
| $\|\psi^{(4)}_{3,4}\rangle = \frac{1}{\sqrt{8}}[(\|1,-1\rangle \pm \|-1,1\rangle) - (\|2,-2\rangle \pm \|-2,2\rangle) - (\|3,-3\rangle \pm \|-3,3\rangle) + (\|4,-4\rangle \pm \|-4,4\rangle)]$ |
| $\|\psi^{(4)}_{5,6}\rangle = \frac{1}{\sqrt{8}}[(\|1,-1\rangle \pm \|-1,1\rangle) - (\|2,-2\rangle \pm \|-2,2\rangle) + (\|3,-3\rangle \pm \|-3,3\rangle) - (\|4,-4\rangle \pm \|-4,4\rangle)]$ |
| $\|\psi^{(4)}_{7,8}\rangle = \frac{1}{\sqrt{8}}[(\|1,-1\rangle \pm \|-1,1\rangle) + (\|2,-2\rangle \pm \|-2,2\rangle) - (\|3,-3\rangle \pm \|-3,3\rangle) - (\|4,-4\rangle \pm \|-4,4\rangle)]$ |
| $\|\psi^{(4)}_{9,10}\rangle = \frac{1}{\sqrt{8}}[(\|1,1\rangle \pm \|-1,-1\rangle) + (\|2,2\rangle \pm \|-2,-2\rangle) + (\|3,3\rangle \pm \|-3,-3\rangle) + (\|4,4\rangle \pm \|-4,-4\rangle)]$ |
| $\|\psi^{(4)}_{11,12}\rangle = \frac{1}{\sqrt{8}}[(\|1,1\rangle \pm \|-1,-1\rangle) - (\|2,2\rangle \pm \|-2,-2\rangle) - (\|3,3\rangle \pm \|-3,-3\rangle) + (\|4,4\rangle \pm \|-4,-4\rangle)]$ |
| $\|\psi^{(4)}_{13,14}\rangle = \frac{1}{\sqrt{8}}[(\|1,1\rangle \pm \|-1,-1\rangle) - (\|2,2\rangle \pm \|-2,-2\rangle) + (\|3,3\rangle \pm \|-3,-3\rangle) - (\|4,4\rangle \pm \|-4,-4\rangle)]$ |
| $\|\psi^{(4)}_{15,16}\rangle = \frac{1}{\sqrt{8}}[(\|1,1\rangle \pm \|-1,-1\rangle) + (\|2,2\rangle \pm \|-2,-2\rangle) - (\|3,3\rangle \pm \|-3,-3\rangle) - (\|4,4\rangle \pm \|-4,-4\rangle)]$ |
| $\|\psi^{(4)}_{17,18}\rangle = \frac{1}{\sqrt{8}}[(\|1,-2\rangle \pm \|-1,2\rangle) + (\|2,-3\rangle \pm \|-2,3\rangle) + (\|3,-4\rangle \pm \|-3,4\rangle) + (\|4,-1\rangle \pm \|-4,1\rangle)]$ |
| $\|\psi^{(4)}_{19,20}\rangle = \frac{1}{\sqrt{8}}[(\|1,-2\rangle \pm \|-1,2\rangle) - (\|2,-3\rangle \pm \|-2,3\rangle) - (\|3,-4\rangle \pm \|-3,4\rangle) + (\|4,-1\rangle \pm \|-4,1\rangle)]$ |
| $\|\psi^{(4)}_{21,22}\rangle = \frac{1}{\sqrt{8}}[(\|1,-2\rangle \pm \|-1,2\rangle) - (\|2,-3\rangle \pm \|-2,3\rangle) + (\|3,-4\rangle \pm \|-3,4\rangle) - (\|4,-1\rangle \pm \|-4,1\rangle)]$ |
| $\|\psi^{(4)}_{23,24}\rangle = \frac{1}{\sqrt{8}}[(\|1,-2\rangle \pm \|-1,2\rangle) + (\|2,-3\rangle \pm \|-2,3\rangle) - (\|3,-4\rangle \pm \|-3,4\rangle) - (\|4,-1\rangle \pm \|-4,1\rangle)]$ |
| $\|\psi^{(4)}_{25,26}\rangle = \frac{1}{\sqrt{8}}[(\|1,2\rangle \pm \|-1,-2\rangle) + (\|2,3\rangle \pm \|-2,-3\rangle) + (\|3,4\rangle \pm \|-3,-4\rangle) + (\|4,1\rangle \pm \|-4,-1\rangle)]$ |
| $\|\psi^{(4)}_{27,28}\rangle = \frac{1}{\sqrt{8}}[(\|1,2\rangle \pm \|-1,-2\rangle) - (\|2,3\rangle \pm \|-2,-3\rangle) - (\|3,4\rangle \pm \|-3,-4\rangle) + (\|4,1\rangle \pm \|-4,-1\rangle)]$ |
| $\|\psi^{(4)}_{29,30}\rangle = \frac{1}{\sqrt{8}}[(\|1,2\rangle \pm \|-1,-2\rangle) - (\|2,3\rangle \pm \|-2,-3\rangle) + (\|3,4\rangle \pm \|-3,-4\rangle) - (\|4,1\rangle \pm \|-4,-1\rangle)]$ |
| $\|\psi^{(4)}_{31,32}\rangle = \frac{1}{\sqrt{8}}[(\|1,2\rangle \pm \|-1,-2\rangle) + (\|2,3\rangle \pm \|-2,-3\rangle) - (\|3,4\rangle \pm \|-3,-4\rangle) - (\|4,1\rangle \pm \|-4,-1\rangle)]$ |
| $\|\psi^{(4)}_{33,34}\rangle = \frac{1}{\sqrt{8}}[(\|1,-3\rangle \pm \|-1,3\rangle) + (\|2,-4\rangle \pm \|-2,4\rangle) + (\|3,-1\rangle \pm \|-3,1\rangle) + (\|4,-2\rangle \pm \|-4,2\rangle)]$ |
| $\|\psi^{(4)}_{35,36}\rangle = \frac{1}{\sqrt{8}}[(\|1,-3\rangle \pm \|-1,3\rangle) - (\|2,-4\rangle \pm \|-2,4\rangle) - (\|3,-1\rangle \pm \|-3,1\rangle) + (\|4,-2\rangle \pm \|-4,2\rangle)]$ |
| $\|\psi^{(4)}_{37,38}\rangle = \frac{1}{\sqrt{8}}[(\|1,-3\rangle \pm \|-1,3\rangle) - (\|2,-4\rangle \pm \|-2,4\rangle) + (\|3,-1\rangle \pm \|-3,1\rangle) - (\|4,-2\rangle \pm \|-4,2\rangle)]$ |
| $\|\psi^{(4)}_{39,40}\rangle = \frac{1}{\sqrt{8}}[(\|1,-3\rangle \pm \|-1,3\rangle) + (\|2,-4\rangle \pm \|-2,4\rangle) - (\|3,-1\rangle \pm \|-3,1\rangle) - (\|4,-2\rangle \pm \|-4,2\rangle)]$ |
| $\|\psi^{(4)}_{41,42}\rangle = \frac{1}{\sqrt{8}}[(\|1,3\rangle \pm \|-1,-3\rangle) + (\|2,4\rangle \pm \|-2,-4\rangle) + (\|3,1\rangle \pm \|-3,-1\rangle) + (\|4,2\rangle \pm \|-4,-2\rangle)]$ |
| $\|\psi^{(4)}_{43,44}\rangle = \frac{1}{\sqrt{8}}[(\|1,3\rangle \pm \|-1,-3\rangle) - (\|2,4\rangle \pm \|-2,-4\rangle) - (\|3,1\rangle \pm \|-3,-1\rangle) + (\|4,2\rangle \pm \|-4,-2\rangle)]$ |
| $\|\psi^{(4)}_{45,46}\rangle = \frac{1}{\sqrt{8}}[(\|1,3\rangle \pm \|-1,-3\rangle) - (\|2,4\rangle \pm \|-2,-4\rangle) + (\|3,1\rangle \pm \|-3,-1\rangle) - (\|4,2\rangle \pm \|-4,-2\rangle)]$ |
| $\|\psi^{(4)}_{47,48}\rangle = \frac{1}{\sqrt{8}}[(\|1,3\rangle \pm \|-1,-3\rangle) + (\|2,4\rangle \pm \|-2,-4\rangle) - (\|3,1\rangle \pm \|-3,-1\rangle) - (\|4,2\rangle \pm \|-4,-2\rangle)]$ |
| $\|\psi^{(4)}_{49,50}\rangle = \frac{1}{\sqrt{8}}[(\|1,-4\rangle \pm \|-1,4\rangle) + (\|2,-1\rangle \pm \|-2,1\rangle) + (\|3,-2\rangle \pm \|-3,2\rangle) + (\|4,-3\rangle \pm \|-4,3\rangle)]$ |
| $\|\psi^{(4)}_{51,52}\rangle = \frac{1}{\sqrt{8}}[(\|1,-4\rangle \pm \|-1,4\rangle) - (\|2,-1\rangle \pm \|-2,1\rangle) - (\|3,-2\rangle \pm \|-3,2\rangle) + (\|4,-3\rangle \pm \|-4,3\rangle)]$ |
| $\|\psi^{(4)}_{53,54}\rangle = \frac{1}{\sqrt{8}}[(\|1,-4\rangle \pm \|-1,4\rangle) - (\|2,-1\rangle \pm \|-2,1\rangle) + (\|3,-2\rangle \pm \|-3,2\rangle) - (\|4,-3\rangle \pm \|-4,3\rangle)]$ |
| $\|\psi^{(4)}_{55,56}\rangle = \frac{1}{\sqrt{8}}[(\|1,-4\rangle \pm \|-1,4\rangle) + (\|2,-1\rangle \pm \|-2,1\rangle) - (\|3,-2\rangle \pm \|-3,2\rangle) - (\|4,-3\rangle \pm \|-4,3\rangle)]$ |
| $\|\psi^{(4)}_{57,58}\rangle = \frac{1}{\sqrt{8}}[(\|1,4\rangle \pm \|-1,-4\rangle) + (\|2,1\rangle \pm \|-2,-1\rangle) + (\|3,2\rangle \pm \|-3,-2\rangle) + (\|4,3\rangle \pm \|-4,-3\rangle)]$ |
| $\|\psi^{(4)}_{59,60}\rangle = \frac{1}{\sqrt{8}}[(\|1,4\rangle \pm \|-1,-4\rangle) - (\|2,1\rangle \pm \|-2,-1\rangle) - (\|3,2\rangle \pm \|-3,-2\rangle) + (\|4,3\rangle \pm \|-4,-3\rangle)]$ |
| $\|\psi^{(4)}_{61,62}\rangle = \frac{1}{\sqrt{8}}[(\|1,4\rangle \pm \|-1,-4\rangle) - (\|2,1\rangle \pm \|-2,-1\rangle) + (\|3,2\rangle \pm \|-3,-2\rangle) - (\|4,3\rangle \pm \|-4,-3\rangle)]$ |
| $\|\psi^{(4)}_{63,64}\rangle = \frac{1}{\sqrt{8}}[(\|1,4\rangle \pm \|-1,-4\rangle) + (\|2,1\rangle \pm \|-2,-1\rangle) - (\|3,2\rangle \pm \|-3,-2\rangle) - (\|4,3\rangle \pm \|-4,-3\rangle)]$ |

*Tab. B.7:* Alice's preparation process for $N = 4$.

| Transformation | New state |
|---|---|
| $(I \otimes I)\|\psi_1^{(4)}\rangle$ | $\|\psi_1^{(4)}\rangle$ |
| $(N_1N_2N_3N_4 \otimes I)\|\psi_1^{(4)}\rangle$ | $\|\psi_2^{(4)}\rangle$ |
| $(P_2N_2P_2N_2P_3N_3P_3N_3 \otimes I)\|\psi_1^{(4)}\rangle$ | $\|\psi_3^{(4)}\rangle$ |
| $(N_1N_4P_2N_2P_2P_3N_3P_3 \otimes I)\|\psi_1^{(4)}\rangle$ | $\|\psi_4^{(4)}\rangle$ |
| $(P_2N_2P_2N_2P_4N_4P_4N_4 \otimes I)\|\psi_1^{(4)}\rangle$ | $\|\psi_5^{(4)}\rangle$ |
| $(N_1N_3P_2N_2P_2P_4N_4P_4 \otimes I)\|\psi_1^{(4)}\rangle$ | $\|\psi_6^{(4)}\rangle$ |
| $(P_3N_3P_3N_3P_4N_4P_4N_4 \otimes I)\|\psi_1^{(4)}\rangle$ | $\|\psi_7^{(4)}\rangle$ |
| $(N_1N_2P_3N_3P_3P_4N_4P_4 \otimes I)\|\psi_1^{(4)}\rangle$ | $\|\psi_8^{(4)}\rangle$ |
| $(P_1P_2P_3P_4 \otimes I)\|\psi_1^{(4)}\rangle$ | $\|\psi_9^{(4)}\rangle$ |
| $(N_1N_2N_3N_4P_1P_2P_3P_4 \otimes I)\|\psi_1^{(4)}\rangle$ | $\|\psi_{10}^{(4)}\rangle$ |
| $(P_2N_2P_2N_2P_3N_3P_3N_3P_1P_2P_3P_4 \otimes I)\|\psi_1^{(4)}\rangle$ | $\|\psi_{11}^{(4)}\rangle$ |
| $(N_1N_4P_2N_2P_2P_3N_3P_3P_1P_2P_3P_4 \otimes I)\|\psi_1^{(4)}\rangle$ | $\|\psi_{12}^{(4)}\rangle$ |
| $(P_2N_2P_2N_2P_4N_4P_4N_4P_1P_2P_3P_4 \otimes I)\|\psi_1^{(4)}\rangle$ | $\|\psi_{13}^{(4)}\rangle$ |
| $(N_1N_3P_2N_2P_2P_4N_4P_4P_1P_2P_3P_4 \otimes I)\|\psi_1^{(4)}\rangle$ | $\|\psi_{14}^{(4)}\rangle$ |
| $(P_3N_3P_3N_3P_4N_4P_4N_4P_1P_2P_3P_4 \otimes I)\|\psi_1^{(4)}\rangle$ | $\|\psi_{15}^{(4)}\rangle$ |
| $(N_1N_2P_3N_3P_3P_4N_4P_4P_1P_2P_3P_4 \otimes I)\|\psi_1^{(4)}\rangle$ | $\|\psi_{16}^{(4)}\rangle$ |
| $(L_+^3 \otimes I)\|\psi_1^{(4)}\rangle$ | $\|\psi_{17}^{(4)}\rangle$ |
| $(N_1N_2N_3N_4L_+^3 \otimes I)\|\psi_1^{(4)}\rangle$ | $\|\psi_{18}^{(4)}\rangle$ |
| $(P_2N_2P_2N_2P_3N_3P_3N_3L_+^3 \otimes I)\|\psi_1^{(4)}\rangle$ | $\|\psi_{19}^{(4)}\rangle$ |
| $(N_1N_4P_2N_2P_2P_3N_3P_3L_+^3 \otimes I)\|\psi_1^{(4)}\rangle$ | $\|\psi_{20}^{(4)}\rangle$ |
| $(P_2N_2P_2N_2P_4N_4P_4N_4L_+^3 \otimes I)\|\psi_1^{(4)}\rangle$ | $\|\psi_{21}^{(4)}\rangle$ |
| $(N_1N_3P_2N_2P_2P_4N_4P_4L_+^3 \otimes I)\|\psi_1^{(4)}\rangle$ | $\|\psi_{22}^{(4)}\rangle$ |
| $(P_3N_3P_3N_3P_4N_4P_4N_4L_+^3 \otimes I)\|\psi_1^{(4)}\rangle$ | $\|\psi_{23}^{(4)}\rangle$ |
| $(N_1N_2P_3N_3P_3P_4N_4P_4L_+^3 \otimes I)\|\psi_1^{(4)}\rangle$ | $\|\psi_{24}^{(4)}\rangle$ |
| $(L_+^3P_1P_2P_3P_4 \otimes I)\|\psi_1^{(4)}\rangle$ | $\|\psi_{25}^{(4)}\rangle$ |
| $(N_1N_2N_3N_4L_+^3P_1P_2P_3P_4 \otimes I)\|\psi_1^{(4)}\rangle$ | $\|\psi_{26}^{(4)}\rangle$ |
| $(P_2N_2P_2N_2P_3N_3P_3N_3L_+^3P_1P_2P_3P_4 \otimes I)\|\psi_1^{(4)}\rangle$ | $\|\psi_{27}^{(4)}\rangle$ |
| $(N_1N_4P_2N_2P_2P_3N_3P_3L_+^3P_1P_2P_3P_4 \otimes I)\|\psi_1^{(4)}\rangle$ | $\|\psi_{28}^{(4)}\rangle$ |
| $(P_2N_2P_2N_2P_4N_4P_4N_4L_+^3P_1P_2P_3P_4 \otimes I)\|\psi_1^{(4)}\rangle$ | $\|\psi_{29}^{(4)}\rangle$ |
| $(N_1N_3P_2N_2P_2P_4N_4P_4L_+^3P_1P_2P_3P_4 \otimes I)\|\psi_1^{(4)}\rangle$ | $\|\psi_{30}^{(4)}\rangle$ |
| $(P_3N_3P_3N_3P_4N_4P_4N_4L_+^3P_1P_2P_3P_4 \otimes I)\|\psi_1^{(4)}\rangle$ | $\|\psi_{31}^{(4)}\rangle$ |
| $(N_1N_2P_3N_3P_3P_4N_4P_4L_+^3P_1P_2P_3P_4 \otimes I)\|\psi_1^{(4)}\rangle$ | $\|\psi_{32}^{(4)}\rangle$ |
| $(L_+^2 \otimes I)\|\psi_1^{(4)}\rangle$ | $\|\psi_{33}^{(4)}\rangle$ |
| $(N_1N_2N_3N_4L_+^2 \otimes I)\|\psi_1^{(4)}\rangle$ | $\|\psi_{34}^{(4)}\rangle$ |
| $(P_2N_2P_2N_2P_3N_3P_3N_3L_+^2 \otimes I)\|\psi_1^{(4)}\rangle$ | $\|\psi_{35}^{(4)}\rangle$ |
| $(N_1N_4P_2N_2P_2P_3N_3P_3L_+^2 \otimes I)\|\psi_1^{(4)}\rangle$ | $\|\psi_{36}^{(4)}\rangle$ |
| $(P_2N_2P_2N_2P_4N_4P_4N_4L_+^2 \otimes I)\|\psi_1^{(4)}\rangle$ | $\|\psi_{37}^{(4)}\rangle$ |
| $(N_1N_3P_2N_2P_2P_4N_4P_4L_+^2 \otimes I)\|\psi_1^{(4)}\rangle$ | $\|\psi_{38}^{(4)}\rangle$ |
| $(P_3N_3P_3N_3P_4N_4P_4N_4L_+^2 \otimes I)\|\psi_1^{(4)}\rangle$ | $\|\psi_{39}^{(4)}\rangle$ |
| $(N_1N_2P_3N_3P_3P_4N_4P_4L_+^2 \otimes I)\|\psi_1^{(4)}\rangle$ | $\|\psi_{40}^{(4)}\rangle$ |
| $(L_+^2P_1P_2P_3P_4 \otimes I)\|\psi_1^{(4)}\rangle$ | $\|\psi_{41}^{(4)}\rangle$ |
| $(N_1N_2N_3N_4L_+^2P_1P_2P_3P_4 \otimes I)\|\psi_1^{(4)}\rangle$ | $\|\psi_{42}^{(4)}\rangle$ |
| $(P_2N_2P_2N_2P_3N_3P_3N_3L_+^2P_1P_2P_3P_4 \otimes I)\|\psi_1^{(4)}\rangle$ | $\|\psi_{43}^{(4)}\rangle$ |
| $(N_1N_4P_2N_2P_2P_3N_3P_3L_+^2P_1P_2P_3P_4 \otimes I)\|\psi_1^{(4)}\rangle$ | $\|\psi_{44}^{(4)}\rangle$ |
| $(P_2N_2P_2N_2P_4N_4P_4N_4L_+^2P_1P_2P_3P_4 \otimes I)\|\psi_1^{(4)}\rangle$ | $\|\psi_{45}^{(4)}\rangle$ |
| $(N_1N_3P_2N_2P_2P_4N_4P_4L_+^2P_1P_2P_3P_4 \otimes I)\|\psi_1^{(4)}\rangle$ | $\|\psi_{46}^{(4)}\rangle$ |
| $(P_3N_3P_3N_3P_4N_4P_4N_4L_+^2P_1P_2P_3P_4 \otimes I)\|\psi_1^{(4)}\rangle$ | $\|\psi_{47}^{(4)}\rangle$ |
| $(N_1N_2P_3N_3P_3P_4N_4P_4L_+^2P_1P_2P_3P_4 \otimes I)\|\psi_1^{(4)}\rangle$ | $\|\psi_{48}^{(4)}\rangle$ |
| $(L_+ \otimes I)\|\psi_1^{(4)}\rangle$ | $\|\psi_{49}^{(4)}\rangle$ |
| $(N_1N_2N_3N_4L_+ \otimes I)\|\psi_1^{(4)}\rangle$ | $\|\psi_{50}^{(4)}\rangle$ |
| $(P_2N_2P_2N_2P_3N_3P_3N_3L_+ \otimes I)\|\psi_1^{(4)}\rangle$ | $\|\psi_{51}^{(4)}\rangle$ |
| $(N_1N_4P_2N_2P_2P_3N_3P_3L_+ \otimes I)\|\psi_1^{(4)}\rangle$ | $\|\psi_{52}^{(4)}\rangle$ |
| $(P_2N_2P_2N_2P_4N_4P_4N_4L_+ \otimes I)\|\psi_1^{(4)}\rangle$ | $\|\psi_{53}^{(4)}\rangle$ |
| $(N_1N_3P_2N_2P_2P_4N_4P_4L_+ \otimes I)\|\psi_1^{(4)}\rangle$ | $\|\psi_{54}^{(4)}\rangle$ |
| $(P_3N_3P_3N_3P_4N_4P_4N_4L_+ \otimes I)\|\psi_1^{(4)}\rangle$ | $\|\psi_{55}^{(4)}\rangle$ |
| $(N_1N_2P_3N_3P_3P_4N_4P_4L_+ \otimes I)\|\psi_1^{(4)}\rangle$ | $\|\psi_{56}^{(4)}\rangle$ |
| $(L_+P_1P_2P_3P_4 \otimes I)\|\psi_1^{(4)}\rangle$ | $\|\psi_{57}^{(4)}\rangle$ |
| $(N_1N_2N_3N_4L_+P_1P_2P_3P_4 \otimes I)\|\psi_1^{(4)}\rangle$ | $\|\psi_{58}^{(4)}\rangle$ |
| $(P_2N_2P_2N_2P_3N_3P_3N_3L_+P_1P_2P_3P_4 \otimes I)\|\psi_1^{(4)}\rangle$ | $\|\psi_{59}^{(4)}\rangle$ |
| $(N_1N_4P_2N_2P_2P_3N_3P_3L_+P_1P_2P_3P_4 \otimes I)\|\psi_1^{(4)}\rangle$ | $\|\psi_{60}^{(4)}\rangle$ |
| $(P_2N_2P_2N_2P_4N_4P_4N_4L_+P_1P_2P_3P_4 \otimes I)\|\psi_1^{(4)}\rangle$ | $\|\psi_{61}^{(4)}\rangle$ |
| $(N_1N_3P_2N_2P_2P_4N_4P_4L_+P_1P_2P_3P_4 \otimes I)\|\psi_1^{(4)}\rangle$ | $\|\psi_{62}^{(4)}\rangle$ |
| $(P_3N_3P_3N_3P_4N_4P_4N_4L_+P_1P_2P_3P_4 \otimes I)\|\psi_1^{(4)}\rangle$ | $\|\psi_{63}^{(4)}\rangle$ |
| $(N_1N_2P_3N_3P_3P_4N_4P_4L_+P_1P_2P_3P_4 \otimes I)\|\psi_1^{(4)}\rangle$ | $\|\psi_{64}^{(4)}\rangle$ |

*Tab. B.8:* Bob's measurement process for $N = 4$.

| Initial state | $(H_{x_1}H_{x_2}H_{x_3}H_{x_4} \otimes I)$PCS | $U_{(4)}$ | Renamed |
|---|---|---|---|
| $\|\psi_1^{(4)}\rangle_x$ | $\frac{1}{2}[\|1,-1\rangle + \|2,-2\rangle + \|3,-3\rangle + \|4,-4\rangle]$ | $\|1,-1\rangle$ | 0 ⊔ ⊔ ⊔ 1 ⊔ ⊔⊔ |
| $\|\psi_2^{(4)}\rangle_x$ | $\frac{1}{2}[\|-1,-1\rangle + \|-2,-2\rangle + \|-3,-3\rangle + \|-4,-4\rangle]$ | $\|-1,-1\rangle$ | 1 ⊔ ⊔ ⊔ 1 ⊔ ⊔⊔ |
| $\|\psi_3^{(4)}\rangle_x$ | $\frac{1}{2}[\|1,-1\rangle - \|2,-2\rangle - \|3,-3\rangle + \|4,-4\rangle]$ | $\|4,-4\rangle$ | ⊔ ⊔ ⊔0 ⊔ ⊔ ⊔ 1 |
| $\|\psi_4^{(4)}\rangle_x$ | $\frac{1}{2}[\|-1,-1\rangle - \|-2,-2\rangle - \|-3,-3\rangle + \|-4,-4\rangle]$ | $\|-4,-4\rangle$ | ⊔ ⊔ ⊔1 ⊔ ⊔ ⊔ 1 |
| $\|\psi_5^{(4)}\rangle_x$ | $\frac{1}{2}[\|1,-1\rangle - \|2,-2\rangle + \|3,-3\rangle - \|4,-4\rangle]$ | $\|2,-2\rangle$ | ⊔0 ⊔ ⊔ ⊔ 1 ⊔ ⊔ |
| $\|\psi_6^{(4)}\rangle_x$ | $\frac{1}{2}[\|-1-,1\rangle - \|-2,-2\rangle + \|-3,-3\rangle - \|-4,-4\rangle]$ | $\|-2,-2\rangle$ | ⊔1 ⊔ ⊔ ⊔ 1 ⊔ ⊔ |
| $\|\psi_7^{(4)}\rangle_x$ | $\frac{1}{2}[\|1,-1\rangle + \|2,-2\rangle - \|3,-3\rangle - \|4,-4\rangle]$ | $\|3,-3\rangle$ | ⊔ ⊔ 0 ⊔ ⊔ ⊔ 1⊔ |
| $\|\psi_8^{(4)}\rangle_x$ | $\frac{1}{2}[\|-1,-1\rangle + \|-2,-2\rangle - \|-3,-3\rangle - \|-4,-4\rangle]$ | $\|-3,-3\rangle$ | ⊔ ⊔ 1 ⊔ ⊔ ⊔ 1⊔ |
| $\|\psi_9^{(4)}\rangle_x$ | $\frac{1}{2}[\|1,1\rangle + \|2,2\rangle + \|3,3\rangle + \|4,4\rangle]$ | $\|2,2\rangle$ | ⊔0 ⊔ ⊔ ⊔ 0 ⊔ ⊔ |
| $\|\psi_{10}^{(4)}\rangle_x$ | $\frac{1}{2}[\|-1,1\rangle + \|-2,2\rangle + \|-3,3\rangle + \|-4,4\rangle]$ | $\|-2,2\rangle$ | ⊔1 ⊔ ⊔ ⊔ 0 ⊔ ⊔ |
| $\|\psi_{11}^{(4)}\rangle_x$ | $\frac{1}{2}[\|1,1\rangle - \|2,2\rangle - \|3,3\rangle + \|4,4\rangle]$ | $\|3,3\rangle$ | ⊔ ⊔ 0 ⊔ ⊔ ⊔ 0⊔ |
| $\|\psi_{12}^{(4)}\rangle_x$ | $\frac{1}{2}[\|-1,1\rangle - \|-2,2\rangle - \|-3,3\rangle + \|-4,4\rangle]$ | $\|-3,3\rangle$ | ⊔ ⊔ 1 ⊔ ⊔ ⊔ 0⊔ |
| $\|\psi_{13}^{(4)}\rangle_x$ | $\frac{1}{2}[\|1,1\rangle - \|2,2\rangle + \|3,3\rangle - \|4,4\rangle]$ | $\|1,1\rangle$ | 0 ⊔ ⊔ ⊔ 0 ⊔ ⊔⊔ |
| $\|\psi_{14}^{(4)}\rangle_x$ | $\frac{1}{2}[\|-1,1\rangle - \|-2,2\rangle + \|-3,3\rangle - \|-4,4\rangle]$ | $\|-1,1\rangle$ | 1 ⊔ ⊔ ⊔ 0 ⊔ ⊔⊔ |
| $\|\psi_{15}^{(4)}\rangle_x$ | $\frac{1}{2}[\|1,1\rangle + \|2,2\rangle - \|3,3\rangle - \|4,4\rangle]$ | $\|4,4\rangle$ | ⊔ ⊔ ⊔0 ⊔ ⊔ ⊔ 0 |
| $\|\psi_{16}^{(4)}\rangle_x$ | $\frac{1}{2}[\|-1,1\rangle + \|-2,2\rangle - \|-3,3\rangle - \|-4,4\rangle]$ | $\|-4,4\rangle$ | ⊔ ⊔ ⊔1 ⊔ ⊔ ⊔ 0 |
| $\|\psi_{17}^{(4)}\rangle_x$ | $\frac{1}{2}[\|1,-2\rangle + \|2,-3\rangle + \|3,-4\rangle + \|4,-1\rangle]$ | $\|4,-1\rangle$ | ⊔ ⊔ ⊔01 ⊔ ⊔⊔ |
| $\|\psi_{18}^{(4)}\rangle_x$ | $\frac{1}{2}[\|-1,-2\rangle + \|-2,-3\rangle + \|-3,-4\rangle + \|-4,-1\rangle]$ | $\|-4,-1\rangle$ | ⊔ ⊔ ⊔11 ⊔ ⊔⊔ |
| $\|\psi_{19}^{(4)}\rangle_x$ | $\frac{1}{2}[\|1,-2\rangle - \|2,-3\rangle - \|3,-4\rangle + \|4,-1\rangle]$ | $\|2,-3\rangle$ | ⊔0 ⊔ ⊔ ⊔ ⊔1⊔ |
| $\|\psi_{20}^{(4)}\rangle_x$ | $\frac{1}{2}[\|-1,-2\rangle - \|-2,-3\rangle - \|-3,-4\rangle + \|-4,-1\rangle]$ | $\|-2,-3\rangle$ | ⊔1 ⊔ ⊔ ⊔ ⊔1⊔ |
| $\|\psi_{21}^{(4)}\rangle_x$ | $\frac{1}{2}[\|1,-2\rangle - \|2,-3\rangle + \|3,-4\rangle - \|4,-1\rangle]$ | $\|1,-2\rangle$ | 0 ⊔ ⊔ ⊔ ⊔1 ⊔ ⊔ |
| $\|\psi_{22}^{(4)}\rangle_x$ | $\frac{1}{2}[\|-1,-2\rangle - \|-2,-3\rangle + \|-3,-4\rangle - \|-4,-1\rangle]$ | $\|-1,-2\rangle$ | 1 ⊔ ⊔ ⊔ ⊔1 ⊔ ⊔ |
| $\|\psi_{23}^{(4)}\rangle_x$ | $\frac{1}{2}[\|1,-2\rangle + \|2,-3\rangle - \|3,-4\rangle - \|4,-1\rangle]$ | $\|3,-4\rangle$ | ⊔ ⊔ 0 ⊔ ⊔ ⊔ ⊔1 |
| $\|\psi_{24}^{(4)}\rangle_x$ | $\frac{1}{2}[\|-1,-2\rangle + \|-2,-3\rangle - \|-3,-4\rangle - \|-4,-1\rangle]$ | $\|-3,-4\rangle$ | ⊔ ⊔ 1 ⊔ ⊔ ⊔ ⊔1 |
| $\|\psi_{25}^{(4)}\rangle_x$ | $\frac{1}{2}[\|1,2\rangle + \|2,3\rangle + \|3,4\rangle + \|4,1\rangle]$ | $\|1,2\rangle$ | 0 ⊔ ⊔ ⊔ ⊔0 ⊔ ⊔ |
| $\|\psi_{26}^{(4)}\rangle_x$ | $\frac{1}{2}[\|-1,2\rangle + \|-2,3\rangle + \|-3,4\rangle + \|-4,1\rangle]$ | $\|-1,2\rangle$ | 1 ⊔ ⊔ ⊔ ⊔0 ⊔ ⊔ |
| $\|\psi_{27}^{(4)}\rangle_x$ | $\frac{1}{2}[\|1,2\rangle - \|2,3\rangle - \|3,4\rangle + \|4,1\rangle]$ | $\|3,4\rangle$ | ⊔ ⊔ 0 ⊔ ⊔ ⊔ ⊔0 |
| $\|\psi_{28}^{(4)}\rangle_x$ | $\frac{1}{2}[\|-1,2\rangle - \|-2,3\rangle - \|-3,4\rangle + \|-4,1\rangle]$ | $\|-3,4\rangle$ | ⊔ ⊔ 1 ⊔ ⊔ ⊔ ⊔0 |
| $\|\psi_{29}^{(4)}\rangle_x$ | $\frac{1}{2}[\|1,2\rangle - \|2,3\rangle + \|3,4\rangle - \|4,1\rangle]$ | $\|4,1\rangle$ | ⊔ ⊔ ⊔00 ⊔ ⊔⊔ |
| $\|\psi_{30}^{(4)}\rangle_x$ | $\frac{1}{2}[\|-1,2\rangle - \|-2,3\rangle + \|-3,4\rangle - \|-4,1\rangle]$ | $\|-4,1\rangle$ | ⊔ ⊔ ⊔10 ⊔ ⊔⊔ |
| $\|\psi_{31}^{(4)}\rangle_x$ | $\frac{1}{2}[\|1,2\rangle + \|2,3\rangle - \|3,4\rangle - \|4,1\rangle]$ | $\|2,3\rangle$ | ⊔0 ⊔ ⊔ ⊔ ⊔0⊔ |
| $\|\psi_{32}^{(4)}\rangle_x$ | $\frac{1}{2}[\|-1,2\rangle + \|-2,3\rangle - \|-3,4\rangle - \|-4,1\rangle]$ | $\|-2,3\rangle$ | ⊔1 ⊔ ⊔ ⊔ ⊔0⊔ |
| $\|\psi_{33}^{(4)}\rangle_x$ | $\frac{1}{2}[\|1,-3\rangle + \|2,-4\rangle + \|3,-1\rangle + \|4,-2\rangle]$ | $\|3,-1\rangle$ | ⊔ ⊔ 0 ⊔ 1 ⊔ ⊔⊔ |
| $\|\psi_{34}^{(4)}\rangle_x$ | $\frac{1}{2}[\|-1,-3\rangle + \|-2,-4\rangle + \|-3,-1\rangle + \|-4,-2\rangle]$ | $\|-3,1\rangle$ | ⊔ ⊔ 1 ⊔ 1 ⊔ ⊔⊔ |
| $\|\psi_{35}^{(4)}\rangle_x$ | $\frac{1}{2}[\|1,-3\rangle - \|2,-4\rangle - \|3,-1\rangle + \|4,-2\rangle]$ | $\|2,-4\rangle$ | ⊔0 ⊔ ⊔ ⊔ ⊔ ⊔ 1 |
| $\|\psi_{36}^{(4)}\rangle_x$ | $\frac{1}{2}[\|-1,-3\rangle - \|-2,-4\rangle - \|-3,-1\rangle + \|-4,-2\rangle]$ | $\|-2,-4\rangle$ | ⊔1 ⊔ ⊔ ⊔ ⊔ ⊔ 1 |
| $\|\psi_{37}^{(4)}\rangle_x$ | $\frac{1}{2}[\|1,-3\rangle - \|2,-4\rangle + \|3,-1\rangle - \|4,-2\rangle]$ | $\|4,-2\rangle$ | ⊔ ⊔ ⊔0 ⊔ 1 ⊔ ⊔ |
| $\|\psi_{38}^{(4)}\rangle_x$ | $\frac{1}{2}[\|-1,-3\rangle - \|-2,-4\rangle + \|-3,-1\rangle - \|-4,-2\rangle]$ | $\|-4,-2\rangle$ | ⊔ ⊔ ⊔1 ⊔ 1 ⊔ ⊔ |
| $\|\psi_{39}^{(4)}\rangle_x$ | $\frac{1}{2}[\|1,-3\rangle + \|2,-4\rangle - \|3,-1\rangle - \|4,-2\rangle]$ | $\|1,-3\rangle$ | 0 ⊔ ⊔ ⊔ ⊔ ⊔ 1⊔ |
| $\|\psi_{40}^{(4)}\rangle_x$ | $\frac{1}{2}[\|-1,-3\rangle + \|-2,-4\rangle - \|-3,-1\rangle - \|-4,-2\rangle]$ | $\|-1,-3\rangle$ | 1 ⊔ ⊔ ⊔ ⊔ ⊔ 1⊔ |
| $\|\psi_{41}^{(4)}\rangle_x$ | $\frac{1}{2}[\|1,3\rangle + \|2,4\rangle + \|3,1\rangle + \|4,2\rangle]$ | $\|4,2\rangle$ | ⊔ ⊔ ⊔0 ⊔ 0 ⊔ ⊔ |
| $\|\psi_{42}^{(4)}\rangle_x$ | $\frac{1}{2}[\|-1,3\rangle + \|-2,4\rangle + \|-3,1\rangle + \|-4,2\rangle]$ | $\|-4,2\rangle$ | ⊔ ⊔ ⊔1 ⊔ 0 ⊔ ⊔ |
| $\|\psi_{43}^{(4)}\rangle_x$ | $\frac{1}{2}[\|1,3\rangle - \|2,4\rangle - \|3,1\rangle + \|4,2\rangle]$ | $\|1,3\rangle$ | 0 ⊔ ⊔ ⊔ ⊔ ⊔ 0⊔ |
| $\|\psi_{44}^{(4)}\rangle_x$ | $\frac{1}{2}[\|-1,3\rangle - \|-2,4\rangle - \|-3,1\rangle + \|-4,2\rangle]$ | $\|-1,3\rangle$ | 1 ⊔ ⊔ ⊔ ⊔ ⊔ 0⊔ |
| $\|\psi_{45}^{(4)}\rangle_x$ | $\frac{1}{2}[\|1,3\rangle - \|2,4\rangle + \|3,1\rangle - \|4,2\rangle]$ | $\|3,1\rangle$ | ⊔ ⊔ 0 ⊔ 0 ⊔ ⊔⊔ |
| $\|\psi_{46}^{(4)}\rangle_x$ | $\frac{1}{2}[\|-1,3\rangle - \|-2,4\rangle + \|-3,1\rangle - \|-4,2\rangle]$ | $\|-3,1\rangle$ | ⊔ ⊔ 1 ⊔ 0 ⊔ ⊔⊔ |
| $\|\psi_{47}^{(4)}\rangle_x$ | $\frac{1}{2}[\|1,3\rangle + \|2,4\rangle - \|3,1\rangle - \|4,2\rangle]$ | $\|2,4\rangle$ | ⊔0 ⊔ ⊔ ⊔ ⊔ ⊔ 0 |
| $\|\psi_{48}^{(4)}\rangle_x$ | $\frac{1}{2}[\|-1,3\rangle + \|-2,4\rangle - \|-3,1\rangle - \|-4,2\rangle]$ | $\|-2,4\rangle$ | ⊔1 ⊔ ⊔ ⊔ ⊔ ⊔ 0 |
| $\|\psi_{49}^{(4)}\rangle_x$ | $\frac{1}{2}[\|1,-4\rangle + \|2,-1\rangle + \|3,-2\rangle + \|4,-3\rangle]$ | $\|2,-1\rangle$ | ⊔0 ⊔ ⊔1 ⊔ ⊔⊔ |
| $\|\psi_{50}^{(4)}\rangle_x$ | $\frac{1}{2}[\|-1,-4\rangle + \|-2,-1\rangle + \|-3,-2\rangle + \|-4,-3\rangle]$ | $\|-2,-1\rangle$ | ⊔1 ⊔ ⊔1 ⊔ ⊔⊔ |
| $\|\psi_{51}^{(4)}\rangle_x$ | $\frac{1}{2}[\|1,-4\rangle - \|2,-1\rangle - \|3,-2\rangle + \|4,-3\rangle]$ | $\|4,-3\rangle$ | ⊔ ⊔ ⊔0 ⊔ ⊔1⊔ |
| $\|\psi_{52}^{(4)}\rangle_x$ | $\frac{1}{2}[\|-1,-4\rangle - \|-2,-1\rangle - \|-3,-2\rangle + \|-4,-3\rangle]$ | $\|-4,-3\rangle$ | ⊔ ⊔ ⊔1 ⊔ ⊔1⊔ |
| $\|\psi_{53}^{(4)}\rangle_x$ | $\frac{1}{2}[\|1,-4\rangle - \|2,-1\rangle + \|3,-2\rangle - \|4,-3\rangle]$ | $\|3,-2\rangle$ | ⊔ ⊔ 0 ⊔ ⊔1 ⊔ ⊔ |
| $\|\psi_{54}^{(4)}\rangle_x$ | $\frac{1}{2}[\|-1,-4\rangle - \|-2,-1\rangle + \|-3,-2\rangle - \|-4,-3\rangle]$ | $\|-3,-2\rangle$ | ⊔ ⊔ 1 ⊔ ⊔1 ⊔ ⊔ |
| $\|\psi_{55}^{(4)}\rangle_x$ | $\frac{1}{2}[\|1,-4\rangle + \|2,-1\rangle - \|3,-2\rangle - \|4,-3\rangle]$ | $\|1,-4\rangle$ | 0 ⊔ ⊔ ⊔ ⊔ ⊔ ⊔1 |
| $\|\psi_{56}^{(4)}\rangle_x$ | $\frac{1}{2}[\|-1,-4\rangle + \|-2,-1\rangle - \|-3,-2\rangle - \|-4,-3\rangle]$ | $\|-1,-4\rangle$ | 1 ⊔ ⊔ ⊔ ⊔ ⊔ ⊔1 |
| $\|\psi_{57}^{(4)}\rangle_x$ | $\frac{1}{2}[\|1,4\rangle + \|2,1\rangle + \|3,2\rangle + \|4,3\rangle]$ | $\|3,2\rangle$ | ⊔ ⊔ 0 ⊔ ⊔0 ⊔ ⊔ |
| $\|\psi_{58}^{(4)}\rangle_x$ | $\frac{1}{2}[\|-1,4\rangle + \|-2,1\rangle + \|-3,2\rangle + \|-4,3\rangle]$ | $\|-3,2\rangle$ | ⊔ ⊔ 1 ⊔ ⊔0 ⊔ ⊔ |
| $\|\psi_{59}^{(4)}\rangle_x$ | $\frac{1}{2}[\|1,4\rangle - \|2,1\rangle - \|3,2\rangle + \|4,3\rangle]$ | $\|1,4\rangle$ | 0 ⊔ ⊔ ⊔ ⊔ ⊔ ⊔0 |
| $\|\psi_{60}^{(4)}\rangle_x$ | $\frac{1}{2}[\|-1,4\rangle - \|-2,1\rangle - \|-3,2\rangle + \|-4,3\rangle]$ | $\|-1,4\rangle$ | 1 ⊔ ⊔ ⊔ ⊔ ⊔ ⊔0 |
| $\|\psi_{61}^{(4)}\rangle_x$ | $\frac{1}{2}[\|1,4\rangle - \|2,1\rangle + \|3,2\rangle - \|4,3\rangle]$ | $\|2,1\rangle$ | ⊔0 ⊔ ⊔0 ⊔ ⊔⊔ |
| $\|\psi_{62}^{(4)}\rangle_x$ | $\frac{1}{2}[\|-1,4\rangle - \|-2,1\rangle + \|-3,2\rangle - \|-4,3\rangle]$ | $\|-2,1\rangle$ | ⊔1 ⊔ ⊔0 ⊔ ⊔⊔ |
| $\|\psi_{63}^{(4)}\rangle_x$ | $\frac{1}{2}[\|1,4\rangle + \|2,1\rangle - \|3,2\rangle - \|4,3\rangle]$ | $\|4,3\rangle$ | ⊔ ⊔ ⊔0 ⊔ ⊔0⊔ |

the primitive assumption about the existence of such initial entangled orthonormal states is not true and we do not have enough initial orthonormal states for the procedure. Thus, there does not exist any related efficient dense coding scheme for the $N = 6$ case. It can be seen that a similar argument also works for the case of $N = 3$, which was explained before. The important point is that one can see by a similar method as above that this scheme dose not work for odd $N$'s and $N$'s which have odd halves. This inability to extend the scheme for the mentioned $N$'s is in accordance with the known fact concerning Hadamard matrices, which says that the order of a Hadamard matrix is 1, 2 or $4k$, with $k$ being a positive integer [79, 80]. At this stage, we understood that there might be a kind of extension that takes any Hadamard matrix design and turns it into a general framework for representing Bell states and the required unitary operators which are used in the preparation and measurement processes. For example, using the coefficients of kets for each family represented in Tables. B.3 and B.6, one can obtain 4 and 8-dimensional normalized Hadamard matrices, respectively. Hence, it is possible to represent the Bell bases of each family and also the unitary operators which transform them together using this kind of Hadamard matrices, as we have shown generally in Eqs. (5.4) and (5.24).

To confirm our results as well as to obtain an explicit form for $U_{(N)}$ operator, we have checked our procedure, step by step, for $N = 8$ and 12 cases. Here, just, we have shown the results obtained for $U_{(N)}$ operator at these two cases. For the $N = 8$ case, for example, we have obtained

$$\begin{aligned}
U_{(8)} = \ & \frac{1}{\sqrt{8}} \ [(I \otimes P_1P_3P_6P_8N_1 \ldots N_8P_1P_3P_6P_8) \\
+ \ & (L_+ \otimes P_5N_5P_5N_5P_6N_6P_6N_6P_8N_8P_8N_8L_+) \\
+ \ & (L_+^2 \otimes P_1P_3P_6P_7N_1 \ldots N_8P_1P_3P_6P_7L_+^2) \\
+ \ & (L_+^3 \otimes P_3N_3P_3N_3P_6N_6P_6N_6P_7N_7P_7N_7L_+^3) \\
+ \ & (L_+^4 \otimes P_1P_4P_5P_8N_1 \ldots N_8P_1P_4P_5P_8L_+^4) \\
+ \ & (L_+^5 \otimes P_3N_3P_3N_3P_4N_4P_4N_4P_8N_8P_8N_8L_+^5) \\
+ \ & (L_+^6 \otimes P_1P_4P_5P_7N_1 \ldots N_8P_1P_4P_5P_7L_+^6) \\
+ \ & (L_+^7 \otimes P_4N_4P_4N_4P_5N_5P_5N_5P_7N_7P_7N_7L_+^7)]
\end{aligned} \tag{B.9}$$

where $N_1 \ldots N_8 = N_1N_2N_3N_4N_5N_6N_7N_8$. Meantime, for $N = 12$, the $U_{(N)}$ operator can be written in the following form

$$\begin{aligned}
& U_{(12)} = \\
& \frac{1}{\sqrt{12}}[(I \otimes P_1P_6P_7P_8P_{11}P_{12}N_1 \ldots N_{12}P_1P_6P_7P_8P_{11}P_{12}) \\
& +(L_+ \otimes P_6N_6P_6N_6P_7N_7P_7N_7P_9N_9P_9N_9P_{11}N_{11}P_{11}N_{11}P_{12}N_{12}P_{12}N_{12}L_+) \\
& +(L_+^2 \otimes P_1P_3P_5P_9P_{10}N_1 \ldots N_{12}P_1P_3P_5P_9P_{10}L_+^2) \\
& +(L_+^3 \otimes P_6N_6P_6N_6P_7N_7P_7N_7P_8N_8P_8N_8P_{10}N_{10}P_{10}N_{10}P_{12}N_{12}P_{12}N_{12}L_+^3) \\
& +(L_+^4 \otimes P_1P_3P_4P_5P_8P_9P_{10}N_1 \ldots N_{12}P_1P_3P_4P_5P_8P_9P_{10}L_+^4) \\
& +(L_+^5 \otimes P_3N_3P_3N_3P_4N_4P_4N_4P_5N_5P_5N_5P_8N_8P_8N_8P_9N_9P_9N_9L_+^5)
\end{aligned}$$

$$
\begin{aligned}
&+(L_+^6 \otimes P_1P_4P_6P_7P_{10}P_{12}N_1\ldots N_{12}P_1P_4P_6P_7P_{10}P_{12}L_+^6)\\
&+(L_+^7 \otimes P_3N_3P_3N_3P_4N_4P_4N_4P_{10}N_{10}P_{10}N_{10}P_{11}N_{11}P_{11}N_{11}P_{12}N_{12}P_{12}N_{12}L_+^7)\\
&+(L_+^8 \otimes P_1P_4P_5P_6P_9P_{11}P_{12}N_1\ldots N_{12}P_1P_4P_5P_6P_9P_{11}P_{12}L_+^8)\\
&+(L_+^9 \otimes P_3N_3P_3N_3P_4N_4P_4N_4P_5N_5P_5N_5P_7N_7P_7N_7P_9N_9P_9N_9L_+^9)\\
&+(L_+^{10} \otimes P_1P_3P_7P_8P_{11}N_1\ldots N_{12}P_1P_3P_7P_8P_{11}L_+^{10})\\
&+(L_+^{11} \otimes P_5N_5P_5N_5P_6N_6P_6N_6P_8N_8P_8N_8P_{10}N_{10}P_{10}N_{10}P_{11}N_{11}P_{11}N_{11}L_+^{11})].
\end{aligned}
\tag{B.10}
$$

In fact, these representations, in addition to the previous ones, as examples, help one to obtain and understand the general form presented for the $U_{(N)}$ operator in Eq. (5.39).

# C. A COMMENT ON DENSE CODING IN PAIRWISE ENTANGLED CASE

Recently, Lee *et al.* [87] have considered two different multiqubit schemes for dense coding and compared their efficiency. In the pairwise scheme, Alice and Bob share $N$ separated maximally entangled pairs, so this scheme is equivalent to $N$ separate dense coding schemes. On the other hand, in the maximally entangled scheme, Alice has $N$ qubits and Bob possesses one, which are prepared in $N+1$ maximally entangled qubits. The number of different unitary operators which can be constructed in the first scheme is $4^N$ corresponding to the number of different messages which can be sent. Since if we have $M$ different messages we can codify them in $\log_2 M$ bits of information, then $\log_2(4^N) = 2N$ bits of information can be transferred by using the pairwise scheme form Alice to Bob, by sending $N$ particles. However, Lee *et al.* [87] performed an essential mistake just here by which they have claimed that number of bits is as much as $2^N$, which is unfounded.[1] This, of course, is inconsistent with the Holevo bound [99, 100], which states that $N$ qubits can at most carry $N$ bits of information. Furthermore, another criticism on their work can be found in [101].

Now, the rates of information gain (bits per unit time) that they have deduced in their Eqs. (6) and (7) for the pairwise and maximally entangled schemes must change into

$$\begin{aligned} r_p &= \frac{2N}{N(t_h + t_c)} \\ r_m &= \frac{N+1}{t_h + N t_c} \end{aligned} \tag{C.1}$$

where $t_c$ and $t_h$ are the operation times for CNOT and Hadamard gates, respectively. If $t_c$ and $t_h$ are assumed equal, it is seen that $r_p = r_m = \frac{1}{t_c}$, and consequently, there is no exponential efficiency in the pairwise scheme. Therefore, considering $N$ different Alices or combining them as a sole Alice (which changes the number of parties involved but not the number of sent particles) cannot lead into a more efficient protocol as they have claimed. This is in accordance with the result of Bose *et al.* [69].

Meantime, it should be noted that a more reasonable measure of efficiency of the scheme can be the number of bits transferred in the protocol per needed

[1] It is curious to note, though, that they still claim that the $2^N$ result is correct [96].

time per sent particles. By considering this we have

$$r_p = \frac{2}{N(t_h + t_c)}$$
$$r_m = \frac{N+1}{N(t_h + Nt_c)}. \tag{C.2}$$

Now these quantities give more better sense of efficiency of the protocols, though, the sent particles are the same here. This concept of efficiency for these schemes has been also applied by Bose *et al.* [69].

# D. CURRICULUM VITAE

## Omid Akhavan

| | |
|---|---|
| 1974 | Born in (Vanak) Tehran-Iran on July 8. |
| 1980-1985 | Elementary school (Ferdows-Barin, Zafar) in Tehran. |
| 1985-1988 | Middle school (Razi, Mirdamad) in Tehran. |
| 1988-1992 | High school (Razi, Mirdamad) in Tehran. |
| 1992-1996 | Studied physics at Uroumieh University, First Department Honors, Uroumieh, Iran. |
| 1996 | Summer student at Institute for Advanced Studies in Basic Sciences, Zanjan, Iran. |
| 1996 | National Physics Olympiad Competition, Ranking 12th, Iran. |
| 1996-1998 | Obtained M. Sc. degree in physics under supervising of Prof. A.Z. Moshfegh at Department of Physics, Sharif University of Technology, Tehran. Experimental work on physics of thin films. |
| 1998-2003 | PhD student in Physics under supervising of Prof. M. Golshani at Department of Physics, Sharif University of Technology. Theoretical work on the quantum theory of motion and hidden variables, quantum communication schemes such as teleportation, dense coding and key distribution. Participation in Workshop on "Realizability of Quantum Computation: Entanglement at the Nanoscale", ICTP, Italy. |
| 2003 | Invited to Physics Department of Sharif University of Technology for a postdoc position. |

## *Papers and Manuscripts by the Author*

14. *The growth of $CoSi_2$ thin films by Co/W/Si(100) multilayer structure*, A.Z. Moshfegh, S.J. Hashemifar, and O. Akhavan, *Solid State Communication* **128**, 239 (2003).

15. *A scheme for spatial wave function teleportation in three dimensions*, O. Akhavan, A.T. Rezakhani, and M. Golshani, *J. Quant. Inf. Comp.*, submitted.